# Material Platforms for Defect Qubits and Single Photon Emitters

Gang Zhang[1], Yuan Cheng[1], Jyh-Pin Chou[2,3,*], Adam Gali[4,5,*]

[1] Institute of High Performance Computing, A*STAR, 1 Fusionopolis Way, Singapore 138632, Singapore.
[2] Department of Physics, National Changhua University of Education, Changhua 50007, Taiwan.
[3] Department of Mechanical Engineering, City University of Hong Kong, 83 Tat Chee Avenue, Hong Kong SAR 999077, China.
[4] Institute for Solid State Physics and Optics, Wigner Research Centre for Physics, P.O. Box 49, H-1525 Budapest, Hungary.
[5] Department of Atomic Physics, Budapest University of Technology and Economics, Budafoki út 8, H-1111 Budapest, Hungary.

Author to whom correspondence should be addressed: jpchou@cc.ncue.edu.tw (Jyh-Pin Chou) and agali@eik.bme.hu (Adam Gali)

**Abstract**

Quantum technology has grown out of quantum information theory and now provides a valuable tool that researchers from numerous fields can add to their toolbox of research methods. To date, various systems have been exploited to promote the application of quantum information processing. The systems that can be used for quantum technology include superconducting circuits, ultra-cold atoms, trapped ions, semiconductor quantum dots, and solid-state spins and emitters. In this review, we will discuss the state of the art on material platforms for spin-based quantum technology, with a focus on the progress in solid-state spins and emitters in several leading host materials, including diamond, silicon carbide, boron nitride, silicon, two-dimensional semiconductors, and other materials. We will highlight how first-principles calculations can serve as an exceptionally robust tool for finding the novel defect qubits and single photon emitters in solids, through detailed predictions of the electronic, magnetic and optical properties.

## 1. Introduction

Quantum technology is a platform that exploits the basic rules of quantum mechanics to realize certain techniques and functionality more efficiently than classic technologies[1]. The development of quantum technology opens new arenas of previously-impossible technologies due to the nature of quantum mechanics. In recent years, the interest in quantum technology has grown rapidly[2]. Quantum technology has been flourished from quantum information theory which relies on *controllable* quantum bits (qubits) as the most elementary unit of quantum information[3]. Quantum information theory predicted to solve such complex problems by algorithms based on qubits that are otherwise intractable by classic digital computers [3–8] as well as led to the foundation of inherently secure communication based on the no-clone theorem of quantum states[9]. The latter is the fundament of quantum communication applications that are now spreading commercially, in particular, by the use of quantum key distribution[10]. The simulation of quantum systems and quantum computation is still its infancy but quantum computers[11] are now publicly available with producing the first important results[12]. The key of the development of reliable quantum devices is to realize physical qubits that have sufficiently long *coherence times* for doing quantum operations on them with *high fidelities*. In practice, the fidelity of the operation with physical qubits is not perfect and *quantum error correction* is necessary; this correction can be realized by multiple physical qubits for each *logical qubit* of the quantum algorithm. An exception would be the physical realization of the Majorana qubit[13–16], which is intrinsically a logical qubit.

One may distinguish two types of physical qubits in terms of their role in quantum information processing: local qubits are the units for quantum operations at the quantum nodes whereas the quantum information is delivered by flying qubits between the distant quantum nodes. A natural physical realization of the flying qubits are the photons that can connect the nodes. However, photons are not able to store quantum information at a given site, thus local qubits are typically realized on materials basis with sufficiently long coherence times of the quantum state. In this review, we focus on the physical realization of materials qubits. We note that the quantum nodes may be connected to each other by realizing an interface between the materials qubit and the photon.

The materials platforms of local physical qubits are numerous such as atoms in optical lattices[17], trapped ions[18,19], nuclear spin[20–22], superconducting circuits[23–28], spins in semiconductor quantum dots[29–35], and atomic sized defects in solids[36–38] (see Fig. 1). Innovative areas of device application will arise once the physical properties of these materials are fully understood. From a physical perspective, materials with highly controllable quantum states will help to improve the key performance regarding the dynamical properties of quantum devices in general. Practical quantum devices will likely exploit different real materials for certain purpose. For example, a solid-state spin qubit[36] may be used in a *quantum memory* because of the long coherence time, while a superconducting Josephson junction[23–25] has an advantage as a computational qubit due to the fast processing capability. Since quantum technology is a subject of interest to a broad audience, the literature on this topic is now sizable. Obviously, given the breadth of the topics touched by quantum technology and defects in solids, not all technical details can be provided in one article. Interested readers are directed to the relevant references. For a brief overview of quantum simulators, they can refer to Ref.[39]; for specialized reviews on trapped atoms and ions, they can refer to Refs.[40–43]. For reviews on superconducting circuits, they should refer to Ref.[26]. This review attempts to provide a self-contained description of the current status of research on materials development for solid state defect quantum technologies.

A key representative of solid state defect qubits is the nitrogen-vacancy (NV) center in diamond, which is a paramagnetic color center[44,45]. The electron spin of single defect centers can be initialized and read out through spin-dependent luminescent measurement by optical pumping (see Section 3.1). Importantly, the electron spin state can be coherently manipulated with an alternating magnetic field. In addition, the electron spin of the defect center exhibits long coherence times (~ms) even at room temperature[46,47], which makes it promising for room-temperature quantum information processing. The single defect manipulation can be achieved by using a confocal microscopy setup for excitation and optical readout of the spin state in such diamond samples where the defect concentration is sufficiently low to excite a single defect within the confocal spot. The NV center then acts as a single photon emitter which can be the source of quantum communication protocols or may be applied as fluorescent agent in biomarker and other applications. We note that the nuclear spin of the nitrogen and $^{13}C$ isotopes around the NV defect may be applied as qubits to realize quantum memory[47–49], quantum error correction protocols[50], and bear a great potential in quantum simulation and quantum computation applications. The magneto-optical parameters and the electron spin coherence time of the NV center in diamond are relatively sensitive to the environment inside the diamond lattice or external to diamond, in particular, for NV centers residing close to the diamond surface. The sensitive dependence of spin to strain[51], magnetic[52,53], electric[54], and temperature fields[55–57] combined with its atomic-scale resolution make the defect center in a solid a versatile quantum sensor. These excellent and versatile properties make the defect in a solid a promising candidate for use in a wide range of applications, including quantum information processing and quantum sensing of biological systems[58–64] (see Fig. 2).

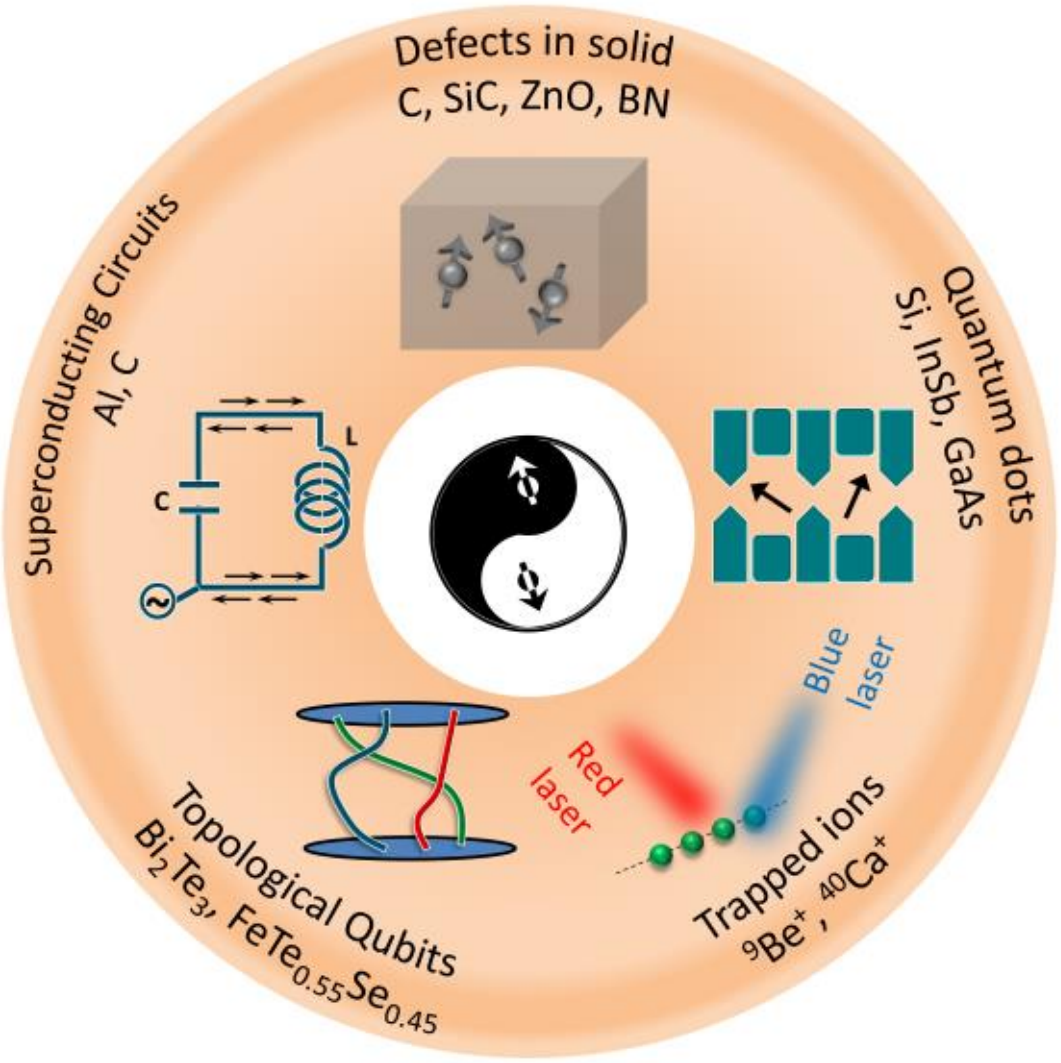

Figure 1. The illustration of five pursuable quantum technologies and the corresponding manufacture materials, including quantum dots of Si[65], InSb[66], and GaAs[67]; trapped ions/ultracold atoms of $^{9}Be^{+}$[68] and $^{40}Ca^{+}$[69]; topological qubits of $Bi_2Te_3$[15] and $FeTe_{0.55}Se_{0.45}$[16] with realizing Majorana bound states; superconducting circuits of Al[70] and C[71]; and solid-state spins/defects of diamond (C)[72], SiC[73], ZnO[74], and BN[75]. These materials can be applied to quantum computers, quantum communication, and quantum sensing. This review focuses on (spin) defects in solids. The famous Chinese symbol in the center, "Tai Chi", is represented by a circle divided into light and dark, or “Yang” and “Yin”, just as is the case with the components of qubits of $S = \pm 1/2$.

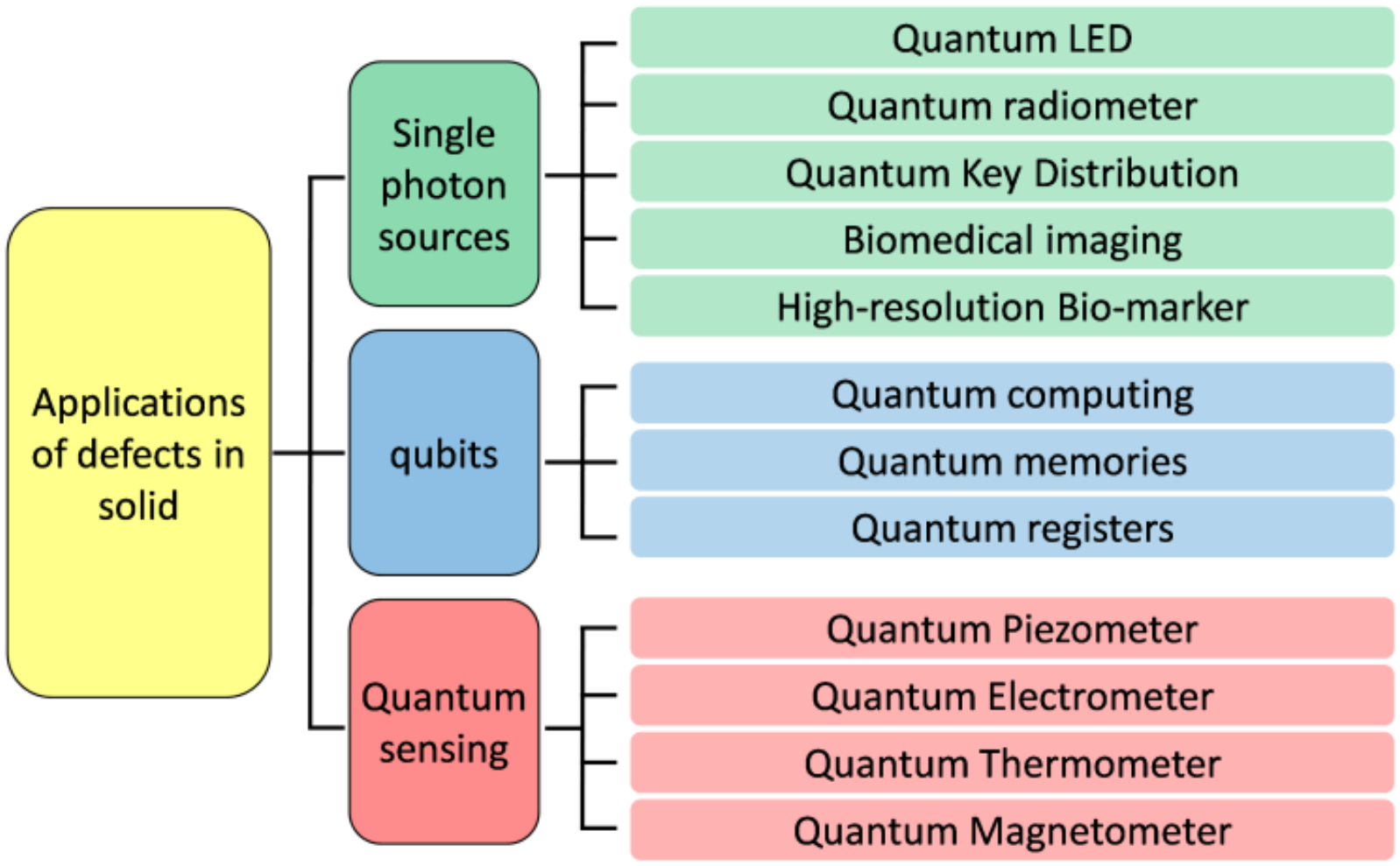

Figure 2. The major applications of defects in solids include single photon sources, qubits, and quantum sensors due to the coupling of spin to strain, magnetic, electric fields, and temperature. For each application, diverse magneto-optical parameters are expected for the defect center. For details, please refer to the main text.

One advantage in the solid-state qubit is its capability of being integrated into the traditional microelectronic structures that have revolutionized our world once already. Diamond has favorable optical properties with room temperature defect qubit but it is difficult to grow diamond at wafer scale and fabricate a semiconductor device from diamond. This was one motivation to seek alternative materials and potential atomic-scale defects that are both robust and easily controlled, similarly to the favorable coherence and readout properties of the diamond NV center. These conditions pose restrictions on the choice of host materials and defect states as explained in detail in Ref.[61]. We briefly emphasize here that optical readout and the need of well-confined atomic-scale defect states favor wide band gap materials as hosts, nevertheless, moderate or small band gap materials can also have great potential, in particular, with low-temperature operation.

The list of possible applications of defect qubits in Fig. 2 already hints that a single defect qubit cannot cover the diverging criteria of various quantum applications. This is another reason to search for alternative solid state defect qubits. We briefly demonstrate this issue on the most successful defect qubit again. The diamond NV center has very broad phonon-assisted fluorescence in the region of 637–900 nm under green laser illumination, where the contribution of the zero-phonon-line (ZPL) emission to the total emission is only ~3%, which is called the Debye–Waller (DW) factor. The electron spin resonance (ESR) occurs typically in the microwave region (~3 GHz) when no external magnetic field is applied; i.e., ESR resonance at the zero-field-splitting (ZFS)[76]. Biological studies need room temperature operation that diamond NV center satisfies. On the other hand, *in vivo* biological applications also favor near-infrared (NIR) optical excitation and emission where the absorption and the auto-fluorescence of the living cells is minimal (see the definition of biological windows in Fig. 9 in the Summary and Outlook section). Furthermore, it is favorable to avoid the use of intense microwave spin excitation at 2–3 GHz in biological systems because the latter can heat the organism. As a consequence, alternative room temperature defect qubits with these magneto-optical properties are in the focus of intense research. On the other hand, room temperature operation for quantum communication application is not demanding but rather the optical properties should suit to the technicalities of the present communication technologies. The dominant and bright emission should come at the ZPL with coherent photons, and the ZPL should be stable against the stray electric fields to avoid spectral diffusion[77]. Furthermore, the ZPL wavelength should ideally be in the region of 1300–1550 nm to be compatible with the maximum transmission of current optical fibers available nowadays (see Fig. 9). Thus, it is evident that new solid-state defect qubits should be found with these favorable magneto-optical properties.

These issues drive the field in searching for new solid-state qubits that operate akin to the NV center in terms of initialization and readout[61,78] but have the aforementioned magneto-optical properties for quantum sensing and communication. It is very difficult to experimentally perform systematic studies on various point defects embedded in numerous host materials. The search for new candidates can be seriously accelerated by highly predictive *ab initio* studies that can first identify the target defect that can be realized experimentally. Alternatively, experimentation might find a new, prospective quantum emitter by chance, but the interpretation of the signals and improvement on the quantum protocols is hindered by the fact that the microscopic origin of the quantum emitters is unknown. Often, it is very difficult to obtain knowledge about the nature of the quantum emitters by experimental methods, in particular, for such emitters that rarely occur in the host material. Again, *ab initio* modeling can be extremely powerful in the identification of quantum emitters and qubits. As a consequence, this review summarizes the recent efforts and results from *ab initio* atomistic simulations, i.e. theoretical spectroscopy, and experiments hand-in-hand. We will provide a brief overview about the *ab initio* modeling and show the physical concept of the key magneto-optical parameters of solid state defect qubits in Sec. 2.

We here review the most fundamental magneto-optical properties of the observed defect qubits and single photon sources with focusing on wide band gap materials that can host room temperature qubits but moderate or small band gap materials are also briefly discussed in Sec. 6.2. We attempt to provide a full coverage of the already identified qubits or single photon sources. We note that other review papers have been appeared related to this topic, and probably numerous review papers have been written parallel to this work or are now being in the preparation phase. We realized that previous review papers either specific to a given material like diamond[79,80], silicon carbide (SiC)[81,82], or the hexagonal boron nitride (*h*-BN)[83,84], or only focusing on the issues of specific quantum applications of selected color centers (e.g., Ref.[85]). Our review is not limited either to specific materials or applications. The full list of the fundamental magneto-optical properties of the already observed single photon sources in various three-dimensional and low dimensional materials orients the readers to pick up those qubits or single photon emitters in their research that might suit the target quantum application. We will discuss briefly the ongoing research associated with the given qubits or the host materials in this regard. Accordingly, we show the physical concept of the key magneto-optical parameters of solid state defect qubits and its intimate connection to the theoretical spectroscopy from *ab initio* modeling in Sec. 2. In sections III–IV, we will discuss shortly the defect qubits and single photon emitters in diamond, SiC, cubic BN, gallium nitride (GaN), aluminum nitride (AlN), zinc oxide (ZnO), zinc sulfide (ZnS), titanium dioxide ($TiO_2$), silicon (Si), wide band gap materials hosting rare-earth ions and two-dimensional materials such as *h*-BN and low-dimensional materials including carbon nanotubes. Finally, we conclude this review paper in the last section.

## 2. Defect qubits with various magneto-optical parameters

In the example of diamond NV center, we already listed the key magneto-optical parameters of qubits in the Introduction. These parameters and their sensitivity on external effects (electric and magnetic fields, strain, temperature, etc.) determine their applicability for a given quantum application. The physical concept of these parameters will be discussed in Sec. 2.1, also in the context of *ab initio* calculations. In a recent review[45], a full coverage of *ab initio* methods was provided for the diamond NV center. These *ab initio* methods heavily rely on the application of Kohn-Sham density functional theory plane wave supercell calculation to defects in solids where the basic equations and advanced functionals were summarized in an earlier general review paper[86] and other recent review papers focused on specific defect qubits[87]. We therefore describe the basic first principles methods very briefly in Sec. 2.1, in order to make a connection between the theoretical and experimental spectroscopy without extensively deferring the readers from the context of the present review. We then describe in Sec. 2.2 how these parameters are determined in the experiments, and provide a very comprehensive list of the magneto-optical properties of the known single defect qubits in various wide band gap materials in Table I which is a central point of this review paper.

### 2.1 Basic first principles methods in a nutshell and theoretical spectroscopy

Characterization of point defects in solids is the key task for identification candidates for qubits. Computation of the electronic structure of the point defects is the first inevitable step to this end. Having the electronic structure in hand, the key magneto-optical properties can be calculated which may be called theoretical spectroscopy of point defects in solids. We briefly summarize the typical *ab initio* methods employed to this end.

Born–Oppenheimer approximation is usually applied for the electron-nuclei system of point defects, in which the ions are treated as charged particles with atomic masses, and the electrons move fast in the adiabatic potential created by the ions. The total energy of the electronic states should be calculated as a function of coordinates of ions of the system so that the adiabatic potential energy surface (APES) of the system can be mapped. The global energy minimum of the APES for a given electronic configuration can be found by minimizing the quantum mechanical forces acting on the ions. The quasi-harmonic vibration modes can also be calculated by interpolating a parabola around the global minimum by moving the ions out of the equilibrium positions and solving the Hessian equation. To this end, the total energy of the electronic system should be determined. Two widespread basic methods can be used: (1) quantum chemistry codes with linear combination of atomic orbitals (LCAO) and Gaussian-type orbital (GTO) based molecular clusters; and (2) density functional theory (DFT) supercell methods. Quantum chemistry codes with coupled cluster or configuration interaction (CI) methods can treat highly correlated orbitals, but only small systems can be calculated, which is problematic when the band edges converge slowly to the infinite cluster's values because of the quantum confinement effect, which opens the gap between empty and occupied defect levels. DFT supercell methods are often applied together with a plane waves basis set, as it is a natural basis for periodic boundary conditions (nevertheless, it can also be a GTO basis) which requires pseudopotentials[88] or projected augmented wave (PAW)[89] methods. The charged defects in the supercell methods can be calculated by neutralizing the supercell with an opposite charge delocalized in the entire supercell, which results in artificial interaction between the periodic images of these charges. This can be compensated for usually by Lany–Zunger correction[90] based on the Makov–Payne theory[91] or Freysoldt correction[92] in 3D materials. For 2D materials, new schemes that should be applied have been developed [93,94], but doing so can be painstaking. The application of charge correction or donor-acceptor pair models is highly important, for the value of the charge correction could go beyond 1 eV in 2D materials. We mention another method to treat charged 2D systems. We note here that Kaviani and co-workers transferred this troublesome problem of charged slab supercell calculation into another but readily solvable problem[95]: A neutral diamond surface model is used for the negatively charged NV defect, NV(−), at the expense of a substitutional, positively charged nitrogen defect $N_S$ entering the diamond slab. Due to the valence electron of a nitrogen atom being one more than that of a carbon atom, this $N_S$ donor will naturally donate its additional electron to the neutral NV acceptor defect by creating a pair comprising an NV center and a positively charged $N_S$. If this defect pair are placed into the same layer of the slab and the slab has a cubic-like shape, then the dipole–dipole interaction between the periodic images of the defect pairs can be minimized (also see Ref.[96]).

In most of the defect calculations, the charged defects are embedded in the supercell models. By using the correction energy ($E_{\mathrm{corr}}$) for charged defective supercells, one can calculate the formation energy ($E_q^f[d]$) of a given point defect ($d$) in a charged state $q$, which provides thermodynamic properties. For a binary compound material (XY), this can be written as[97]

$$E_q^f[d] = E_{\mathrm{tot}}[d] - n_X\mu_X - n_Y\mu_Y - n_d\mu_d + q(E_F + E_V) + E_{\mathrm{corr}} \quad (1)$$

where $\mu_i$ are the chemical potential of ion ($i$) with $n_i$ number in the supercell, and $E_F$ is the Fermi-energy with respect to the calculated valence band maximum $E_V$, which is usually aligned to zero energy in the corresponding band diagrams, and then $E_F$ varies between zero and the fundamental bandgap ($E_g$) of the host material. The adiabatic charge transition levels between the $q$ and $q$+1 charged states can be calculated as

$$E(q|q+1) = \left(E_{\mathrm{tot}}^{q+1}[d] + E_{\mathrm{corr}}^{q+1}\right) - \left(E_{\mathrm{tot}}^{q}[d] + E_{\mathrm{corr}}^{q}\right) + E_V \quad (2)$$

which defines the position of the Fermi-level where the defect adiabatically switches between the $q$ and $q$+1 charge states with respect to $E_V$. The calculated charge transition levels provide the ionization thresholds of the defect qubit as explained in Fig. 3a. These quantities are very important to understand the photostability of the qubit. Photoexcitation of the defect above the ionization threshold leads to temporary or complete loss of the qubit that are often called dark state in the literature (no fluorescence from the defect qubit). If the qubit defect was photoionized then the qubit state may be restored by photoexcitation of the dark state. The photoionization threshold needed to restore the qubit can be larger than the ionization threshold of the qubit (see divacancy defect qubit in 4H-SiC[98,99]).

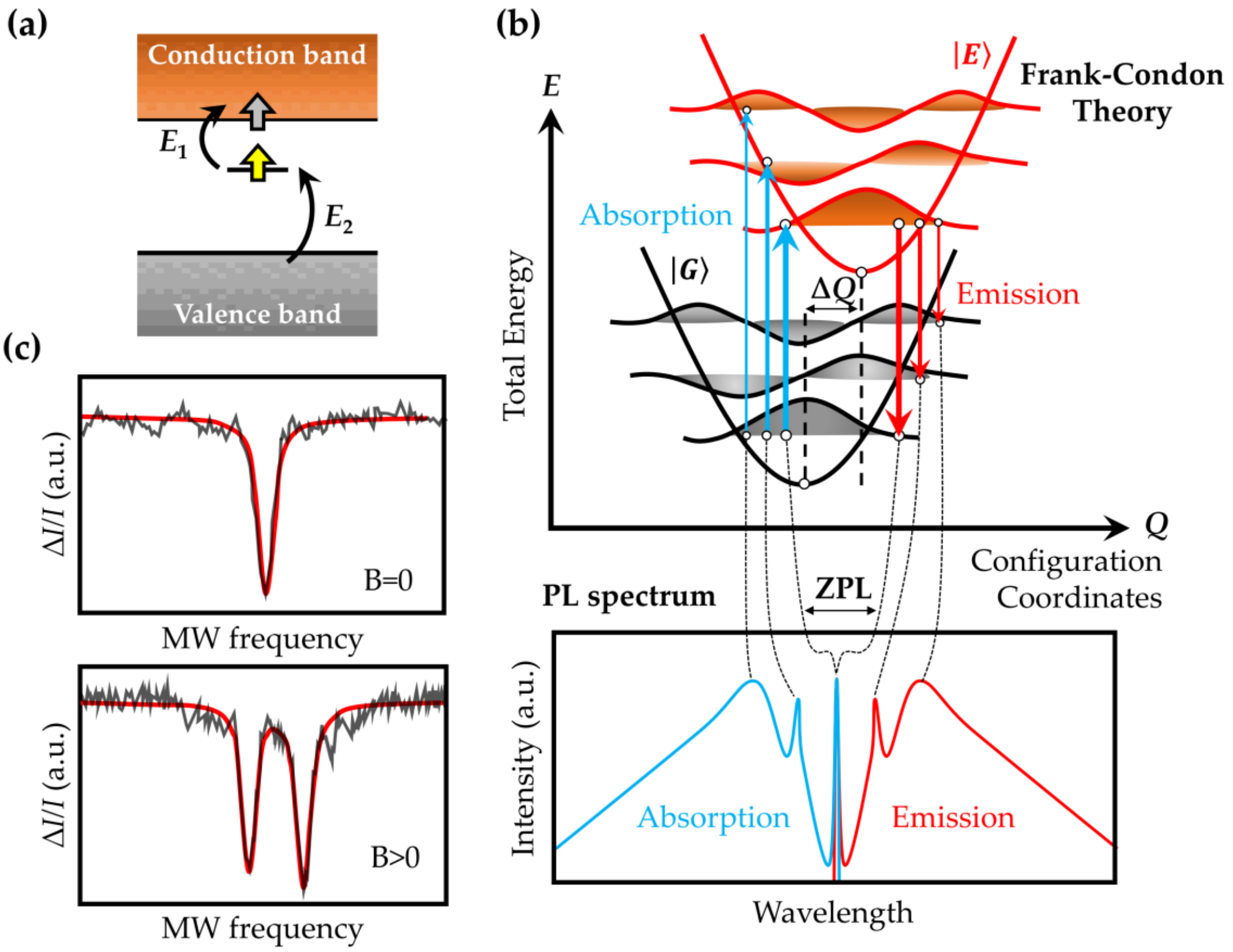

Figure 3. Illustrations of key magneto-optical parameters. (a) Ionization threshold. The defect has a deep acceptor level in the band gap. Assuming that the negatively charged defect is the qubit one can see that $E_1$ energy is needed to neutralize the defect by photo-excitation of the electron from the defect to the conduction band edge, whereas $E_2$ energy is needed to restore the qubit by ionizing the defect from the valence band. (b) Theory of emission spectrum. Adiabatic potential energy surface (APES) of the ground state and excited state along the configuration coordinate. In the Huang-Rhys approximation it is assumed that the APES of the excited state has the same shape as that of the ground state. The thin lines in the APES represent the energy levels of the effective phonon. In reality, many phonons may contribute to the optical transition which may result in a broad phonon sideband. (c) Illustration of typical optically detected magnetic resonance (ODMR) spectrum. The intensity of the emission changes as microwave frequency sweeps that is the signature of electron spin transition. At B=0 zero magnetic field the position of the dip is the ZFS parameter.

We note that the given charge and electronic configuration should be calculated as a function of the spin state too, and the lowest energy spin configuration belongs to the ground state spin of the system at a given charge state. This is highly important, for the qubit state is often associated with the high-spin ground state. The spin levels may split due to spin–orbit or dipolar spin–spin interactions at zero magnetic field that previously called ZFS, which can be calculated from first principles (see Ref.[100]). The ZFS can be measured by either alternating magnetic fields (microwave transition to rotate the spin state) or resonant optical transitions (see Fig. 3c).

The calculation of the key optical quantities, such as the ZPL and DW factors, requires the computation of the optically allowed excited state and the vibration modes. In the Huang–Rhys approximation of the optical transition, it is sufficient to calculate the vibration modes only in the ground state[101]. Nevertheless, the calculation of the ZPL energy requires computing the excited state and finding the global minimum of the APES in this electronic configuration; i.e., computation of the quantum mechanical forces acting on the ions. The calculation of the total energy in the excited state is not trivial for Kohn–Sham DFT. Nevertheless, it may work by constraint occupation of Kohn–Sham states, also known as the ΔSCF method (see application to the NV center in diamond in Ref.[102]). We will discuss below situations when this simple formalism fails and the possible methods to calculate this quantity. By combining these calculations, the DW factor ($w$) can be derived as $S = -ln\, w$, where $S$ is the Huang–Rhys factor obtained in the *ab initio* calculations[103,104] (see Fig. 3b).

## 2.2 List of defect qubits in semiconductors and wide band gap materials

Over the past decades, qubits have been implemented in a wide range of materials, including atoms[105], superconductors[106], semiconductors[36], and rare-earth materials[107]. Among these possible candidates, point defects in semiconductors and wide band gap materials, also called "color centers", show a rich spin and optoelectronic physics that can be exploited to fabricate quantum information devices, including qubits for quantum computing and quantum sensing technology and single-photon emitters in both quantum communication and quantum computation. Color centers are point defects in a semiconductor or insulator that bind electrons to an extremely localized region on the scale of Å. Thus, color centers behave as atoms embedded in the host crystal. These atom-like states make the color centers ideal candidates for solid-state qubits, for many deep centers exhibit a nonzero spin magnetic moment in the ground state and might be optically polarized, manipulated into energetically excited states by illumination and radiation of microwave fields.

At a low concentration of point defects in a solid, the photo-excitation source can be focused on the target point defect with a simultaneously induced ESR transition by sweeping alternating magnetic fields. If the fluorescence intensity is spin dependent, then the electron spin resonance can be observed by monitoring the fluorescence intensity as a function of the microwave frequency. This technique is called optically detected magnetic resonance (ODMR) measurement[108]. The spin resonance may be detected at zero magnetic field for non-isotropic defects where the low symmetry crystal field will split the spin levels known as ZFS. Here, we list these most basic key optical (ZPL, DW) and magnetic properties (ZFS in the ground and excited states at the given spin state) for the materials of prominent qubits and single photon emitters that have been experimentally identified to date, including three-dimensional materials such as diamond, SiC polymorphs (4H, 3C, and 6H), ZnO, GaN, cubic BN (*c*-BN), 2H-$MoS_2$, yttrium aluminum garnet ($Y_3Al_5O_{12}$ or YAG), yttrium orthosilicate ($Y_2SiO_5$ or YSO), and yttrium orthovanadate ($YVO_4$ or YVO); and two-dimensional materials such as hexagonal BN (*h*-BN) and transition metal dichalcogenides ($WSe_2$ and $WS_2$). In sections III–IV, we will discuss these materials in detail.

Table I. Summary of the key parameters in experiment, such as zero-phonon-line (ZPL), zero-field-splitting (ZFS), and the Debye–Waller (DW) factor, for emerging quantum-coherent materials. The ZPL weight, DW factor, and the electron–phonon coupling parameter, the Huang–Rhys (HR) factor, are deduced from the relationship $w = e^{-S}$, where $w$ is the DW factor and $S$ is the HR factor. DAP labels donor-acceptor pair defect and REI for rare-earth ions. V(4+) represent vanadium impurity substituting Si in the SiC lattice that should not be read as a vacancy. Most of the listed defects were observed as single photon emitters; PbV(−) in diamond, Cr(4+) in 4H SiC and GaN, VB(−) in *h*-BN, all the defects in *c*-BN, all the defects in ZnO except for $V_{Zn}$ were observed as ensembles.

| | ZPL (eV) | ZFS (GHz) | DW |
|---|---|---|---|
| **Diamond** | NV(0) 2.156[109] | | NV(0) 0.14[121] |
| | NV(−) 1.945[110] | NV(−) $^2A_2$, 2.88[76], $^3E$, 1.43[118] | NV(−) 0.04[62] |
| | SiV(0) 1.310[111] | SiV(0) 1.00[119] | SiV(0) 0.90±0.10[122] |
| | SiV(−) 1.681[112] | SiV(−) $^2E_g$, 50[120], $^2E_u$, 260[120] | SiV(−) 0.75–0.79[123] |
| | GeV(−) 2.06[113] | GeV(−) $^2E_g$, 181[113], $^2E_u$, 1120[113] | GeV(−) 0.61[113] |
| | SnV(−) 2.00[114] | SnV(−) $^2E_g$, 850[114], $^2E_u$, 3000[114] | SnV(−) 0.41[114] |
| | $NiN_4$ 1.546[115] | | |
| | PbV(−) 2.385[116] | | |
| | ST1 2.255[117] | ST1 1.274±0.139[117] | ST1 <0.1[117] |
| **4H-SiC** | $V_{Si}$-$V_C$(0) 1.094–1.150[124] | $V_{Si}$-$V_C$(0) $^3A_2$, 1.22–1.34[124] | $V_{Si}$-$V_C$(0) ~0.05[132] |
| | PL5 divacancy 1.190[125] | PL5 divacancy 1.353[125] | |
| | PL6 divacancy 1.193[125] | PL6 divacancy 1.365[125] | |
| | $V_{Si}$-$N_C$(−) 0.99–1.06[126–128] | $V_{Si}$-$N_C$(−)$^3A_2$, 1.282–1.331[128] | |
| | $V_{Si}$(V1) 1.438[129] | $V_{Si}$(V1) $^4A_2$, 0.004[129] | $V_{Si}$(V1) 0.08[133] |
| | $V_{Si}$(V2) 1.352[129] | $V_{Si}$(V2) $^4A_2$, 0.035[129] | $V_{Si}$(V2) 0.09[133] |
| | Cr(4+) 1.158, 1.190[130] | Cr(4+) $^3A_2$, 1–6.46[130] | Cr(4+) 0.75[130] |
| | V(4+) 0.970, 0.929[131] | V(4+) $^2E$, 529, 43[131] | V(4+) 0.25, 0.5[131] |
| **3C-SiC** | $V_{Si}$-$V_C$(0) 1.121[132] | $V_{Si}$-$V_C$(0) $^3A_2$, 1.336[132] | $V_{Si}$-$V_C$(0) ~0.05[132] |
| | $V_{Si}$-$N_C$(−) 0.845[134] | $V_{Si}$-$N_C$(−) $^3A_2$, 1.303[134] | |
| | 1.27[a][135] | | |
| **6H-SiC** | $V_{Si}$-$V_C$(0) 1.094–1.134[136] | $V_{Si}$-$V_C$(0) $^3A_2$, 1.236–1.347[124] | |
| | $V_{Si}$-$N_C$(−) 0.960–1.000[127] | | |
| | $V_{Si}$(V1) 1.433[129] | $V_{Si}$(V1) $^4A_2$, 0.028[137] | |
| | $V_{Si}$(V2) 1.398[129] | $V_{Si}$(V2) $^4A_2$, 0.128[137] | |
| | $V_{Si}$(V3) 1.366[129] | $V_{Si}$(V3) $^4A_2$, 0.028[137] | |
| | V(4+) 0.948, 0.917, 0.893[131] | V(4+) $^2E$, 524, 25, 16[131] | V(4+) <0.5 [131] |
| ***h*-BN nanotube** | 1.941[138] | | |
| | 2.172 and 2.179[139] | | |
| ***h*-BN** | ~ 2.0[d], 4.1[e], 5.3[f] | | 0.82[75], 0.59[141] |
| | $V_B$(−) ~1.6[g][140] | $V_B$(−) $^3A_{2g}$, 3.4[140] | ~0.030[g] |
| **$WSe_2$** | 0.936[142] | | |
| **$WS_2$** | $V_S$ 1.72 and 1.98[143] | | 0.30 and 0.50[143] |
| **2H-$MoS_2$** | 1.174[144] | | |
| ***c*-BN** | $O_N$-$V_B$ ~1.63[145] | | |
| | $B_i$ 3.3[146] | | |
| | $B_i$-$V_N$ 3.57[146] | | $B_i$-$V_N$ 0.017[148] |
| | $C_N$ 4.09[147] | | |
| | Si impurity 4.94[148] | | Si impurity 0.007[148] |
| **ZnO** | $V_{Zn}$ 2.331[149] | | |
| | $Cu_{Zn}$ 2.859[150] | | $Cu_{Zn}$ 0.0015[153] |
| | $V_{Zn}$-$Cl_O$ 2.365[151] | | |
| | DAP 3.333[152] | | DAP 0.996[154] |
| **GaN** | 0.855[a][155] | | 0.71[156] |
| | 3.33[b], 2.594[c], 1.82[d] | | 0.63[c] |
| | Cr(4+) 1.193[130] | | Cr(4+) 0.73[130] |
| **Silicon** | G-centre $C_iC_s$ 0.969[157] | | G-centre $C_iC_s$ 0.30[159] |
| | Er(3+) 0.805[h][158] | | |
| **REI** | YAG:Pr(3+) 4.122[160] | | |
| | Ce(3+) 2.536[161] | | YAG:Ce(3+) 0.002[165] |
| | YSO: Er(3+) 0.807[162] | | |
| | YVO:Yb(3+) 1.280[163] | | |
| | $LaFe_3$:Pr(3+) 2.594[164] | | |

[a] The possible candidates are $C_NO_NH_i$[166], $C_N$-$H_i$[166], or $C_N$-$Si_{Ga}$[167].

[b] Theoretical studies predicted that the possible candidates are $C_N$[168] or other complexes[169], such as $V_{Ga}$-3H or $V_{Ga}O_N$-2H. Experimental studies proposed that the possible candidates are $C_N$-$O_N$[170,171] and $C_N$-$Si_{Ga}$[171].
[c] Point defects near cubic inclusions within hexagonal lattice of GaN were proposed [172].
[d] Theoretical study predicted $C_B$-$V_N$[173] and experimental study proposed $N_B$-$V_N$[75].
[e] DAP[174] and $C_N$[175] were proposed originally but a recent theoretical study predicts $C_N$-$C_B$ dimer defect[75].
[f] Theoretical study predicted $O_N$[176], and experimental studies proposed $V_{NB}$[177] and DAP[178].
[g] As the reported measurements[140] were performed at room temperature with no observable ZPL energy, we show the estimated value from *ab initio* calculations for ZPL and DW factor[179].
[h] Co-doping oxygen with erbium in silicon has been reported to enhance emission and improve luminescence at high temperature[180].

## 3. Diamond-hosted qubits

Diamond has a band gap of about 5.5 eV, thus it can host numerous color centers with emission wavelengths in a wide region from NIR through visible to UV[181]. Diamond has strong and short carbon-carbon bonds which makes it a dense and hard material. This may be considered as disadvantageous from doping point of view but also results in high frequency phonon modes. It is worth emphasizing that phonons play critically important roles in quantum information[182,183]. The phonon, the quantized excitation of lattice vibrational modes in crystals, can carry heat and information, which has broad applications for heat energy control and management. Unlike in bulk materials, in low-dimensional nano structures, the confined phonon mode changes the nature of phonon transport, leading to anomalous thermal conductivity and heat energy diffusion[184,185]. In addition to the key role in thermal conductivity of solid state systems[186–188], the interactions of phonons with other carriers, such as electrons and photons, also have important influences on the performance of nano electronics[189], optoelectronic devices[190], and thermoelectric devices[191]. In particular, phonons affect significantly the coherence time of defect spins. For example, the spin coherence time of an NV center can be about a few seconds at cryogenic temperatures but is remarkably reduced to 1.8 ms at room temperature[46,47,192]. We note here that many magneto-optical properties of NV centers depend on the phonon density of states, which highlights the advantage of diamond as a host material; i.e., its high Debye temperature of ∼2200 K. Hence, room temperature can be considered as a relatively low temperature for diamond, which results in a relatively low phonon population even under ambient conditions that generally suppresses the possibility of spin–lattice relaxation. Therefore, other materials with high Debye temperatures may also be promising for hosting defects with long spin–lattice relaxation times, such as boron-based materials[193,194].

### 3.1 NV center

Among the many color centers in diamond, the NV center stands out because of its desirable spin coherence and readout properties[195,196,100]. The NV center is particularly attractive due to its ability to interface with a variety of external degrees of freedom. While different charge states of the NV defect have been reported, the negatively-charged NV defect (NV(−) center) has the most attractive quantum properties and is usually denoted as the NV center. Structurally, the NV center is a point defect in diamond in which a substitutional nitrogen atom is adjacent to a vacancy (Fig. 4a), which can capture an extra electron to form an NV(−) defect. A simplified schematic of the electronic structure of an NV center is shown in Fig. 4a. It is clear that the ground state is a spin triplet ($^3A_2$), the excited state is also a spin triplet state ($^3E$), and two metastable singlet spin states lie between them ($^1A_1$ and $^1E$).

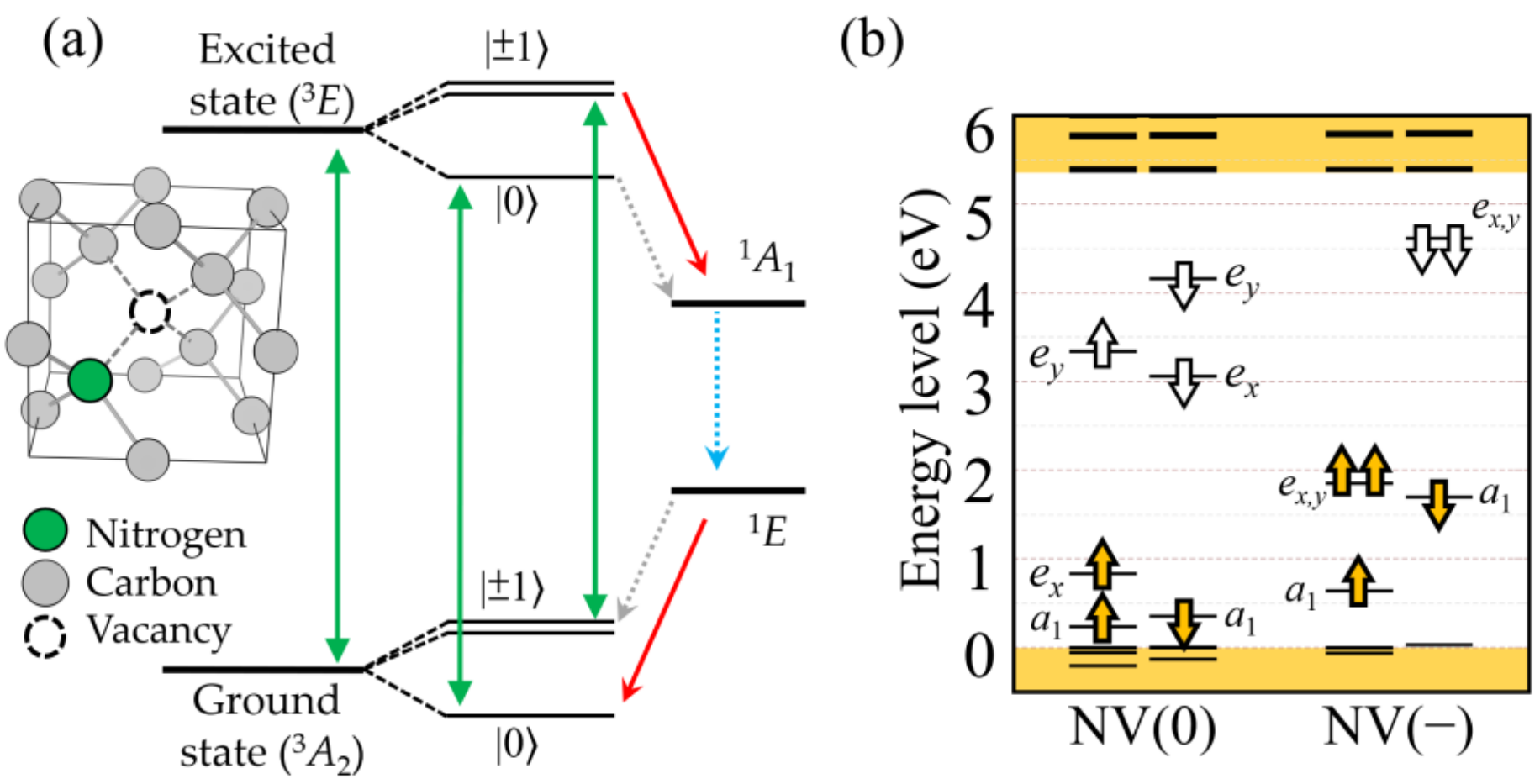

Figure 4. (a) Structure of an NV center in diamond and its energy level diagram in an optical polarization cycle. The green arrows represent spin-conserving absorption and emission, and the grey dotted and red straight arrows represent the weak and efficient intersystem crossings, respectively, and the cyan dotted arrow represents a weak near infrared emission competing with the direct non-radiative decay. The excited state level diagram is simplified to the case of elevated temperatures. (b) The ground state of single particle NV defect energy levels in the fundamental bandgap of diamond. Filled and empty arrows depict occupied (electron) and unoccupied (hole) states, respectively, as found in spin-polarized density functional theory calculations.

The working mechanism of a diamond NV center as a qubit is shown in Fig. 4a. First, a spin-conserving optical excitation is applied between the spin triplet ground state ($^3A_2$) and spin triplet excited state ($^3E$), with an excitation energy of 1.945 eV (637 nm) when no phonons participate in the optical transition [195,196,100]. At low temperatures (< 20 K), a fine structure emerges in the $^3E$ manifold because of spin–orbit and spin–spin interactions (not shown). At room temperature, electron–phonon interactions will rapidly mix the orbital levels, leading to an effective orbital singlet with a fine structure that strongly resembles the ground state[195,196,100], as shown in Fig. 4a. The radiative decay from the triplet excited state to the ground state is spin-conserving. In addition, an efficient nonradiative decay pathway via intermediate dark states $^1A_1$ and $^1E$ exists between the $m_S = \pm 1$ sublevels of the triplet excited state and the $m_S = 0$ sublevel of the triplet ground state, respectively, which is called intersystem crossing. Rogers *et al.*[197] applied uniaxial stress to demonstrate the ZPL of the infrared transition between these two singlet states that was found at 1.19 eV. The energy gap between $^3E$ and $^1A_1$ was estimated at around 0.4 eV from the simulation and measurement of the intersystem crossing rate[198,199]. The combination of radiative and spin selective nonradiative pathways makes the NV center optically polarized into the ground state upon illumination. Once initialized, the ground state spin can then be coherently manipulated by microwave radiation, and its state can be read out by the observed fluorescence intensity upon illumination.

The electronic structure of the NV center is shown in Fig. 4b. The bound states of this deep center are multiparticle states composed of six electrons, which come from the four dangling bonds surrounding the vacancy (five electrons) and one electron from the environment. The double degenerate $e$ ($e_x$, $e_y$) states are completely occupied in the spin majority channel and are empty in the spin minority channel, whereas the $a_1$ orbital is fully occupied in the electronic ground state. The spin-conserving optical excitation from the $^3A_2$ to the $^3E$ excited triplet can be described as promoting an electron from the $a_1$ state to the empty $e_x/e_y$ state in the defect level diagram[200]. We note that the $^1A_1$ state is a strongly correlated state that can only be described by multireference methods beyond the Kohn–Sham DFT methods[201–203].

At room temperature, extremely long spin coherence times of up to 1.8 ms are reported for the ground state of the NV center [46], which is close to the regime needed for quantum error correction. To realize the long electron spin coherence time, the deep center should reside in a wide bandgap host material that allows large energy spacing between adjacent bound states, and it should have small spin–orbit coupling. The remarkable features of the NV center are largely afforded by the host material, diamond. Therefore, the electronic bound states of the NV center can reside deep within the bandgap, resulting in weak interaction with states in the valence and conduction bands. This may imply that wide band gap materials are suitable hosts for color centers akin to the NV center in diamond. It is worth noting that due to the quantum confinement effect, the semiconductor bandgap increases with reductions the dimensions of the crystals[204,205]. Therefore, for some narrow bandgap semiconductors, despite their not being ideal to host deep centers, their nano-sized counterparts may become promising candidates for elevated temperature operation.

The temperature dependence of the optical transition linewidth and excited states relaxation of a single NV center was associated with strong electron–phonon coupling[206]. It was shown that, at low temperature, the electron–phonon coupling caused by the dynamic Jahn–Teller effect in the $^3E$ excited state is the dominant mechanism for the optical dephasing. The Jahn–Teller effect was first reported in 1937 by Hermann Arthur Jahn and Edward Teller[207]. It states that a system with a spatially degenerate electronic state will undergo a spontaneous geometrical distortion, resulting in symmetry breaking, because the removal of degeneracy can lower the overall energy of the system. Using a combination of first-principles calculations and the Jahn–Teller type of electron–phonon interaction model analysis, Abtew *et al.* studied the dynamic Jahn–Teller effect in the $^3E$ excited state of a diamond NV(−) center[208]. To understand the Jahn–Teller effect, the APES was obtained by carrying out constrained DFT calculations. For the NV center, there are six symmetrized displacements with two $A_1$ and two $E$ vibrations induced by the motion of the immediate neighbor atoms of the vacancy. The totally symmetric $A_1$ vibrational modes do not lower the symmetry, but the $E$ vibrational modes will lower the symmetry and split the defect levels. The combination of theoretical modelling[208] and experimental measurement[206] clearly suggested that the dephasing of the ZPL of the NV center is dominated by the coupling between the double degenerate $E$ state and the $E$ phonon modes. This results in a $T^5$ dependence of the ZPL linewidth. First principles calculations from Thiering and Gali have shown that the strong electron–phonon coupling, i.e., the dynamic Jahn–Teller effect, with higher energy phonons plays a crucial role in determining the fine structure of the $^3E$ excited state at cryogenic temperatures and affects the intersystem crossing rate and the shape of the PL spectrum[209,210].

Recent advances in quantum technology make it possible to create proximate qubits that form a quantum network in solids. Via electron spin–spin interaction, the proximate qubits might interact with each other. It was found that in diamond with extremely high concentrations of NV centers (about 45 ppm), the electron spin coherence time was significantly reduced to about 67 μs[211], in striking contrast to the 1.8 ms coherence time of the isolated NV centers[46]. The decoherence is caused by the charge fluctuation between neighboring NV(−) and NV(0) defects. To obtain deep knowledge about the underlying physical mechanism, very recently, Chou *et al.*, using first principles calculations, explored the tunneling-mediated charge diffusion between point defects in diamond[212]. Based on the quantum mechanical tunneling of the electron of NV(−) and a proximate acceptor defect in diamond, they proposed a physical model for the source of decoherence in NV qubits. Although the wave functions decay fast and the electron probability is only about 0.1% at a distance of 1 nm away from the center of the defect, strong interaction and charge transfer still exist even when the distance between neighboring NV(−) and NV(0) centers is up to 4.4 nm. Importantly, Chou *et al.* theoretically suggested that, to maintain the coherence time (≈1 ms) of the isolated NV qubit, a distance of 9 nm between the NV sensor and the acceptor defect is required. This is of high importance for both quantum network and quantum sensor applications of the solid-state qubits[213,214,85].

The most promising area of application of the NV center is the ultrasensitive nanoscale sensors[95,215,216]. The spin levels and coherence time of the NV center is sensitive to external fields, such as an electromagnetic field, pressure and temperature, making it an atomic-scale quantum sensor capable of detecting changes in the surrounding environment, and in particular, it is very attractive for biological or biomolecule sensing applications. In sensor applications, the NV centers should be engineered relatively close to the surface of the diamond. However, compared to that of deeply buried NV centers in diamond, the coherence time of near-surface NV centers is significantly reduced. One possible approach to improvement is proper surface functionalization with different molecule/atom groups[95,216,217]. So far, not all the possible sources of the noise causing this effect are fully understood, and the foundation of a precise strategy to keep the long coherence time as well as high sensitivity is still under intense research.

We note that diamond NV center has a great potential to realize room temperature quantum computing, in strike contrast to the present ultralow temperature operation of superconductor based quantum computers. It has been recently shown that single diamond NV center can coherently address 27 $^{13}$C nuclear spins[218]. Furthermore, a high fidelity of >99% has been demonstrated for the gate operation for single- (99.99%) and two-qubit gates (99.2%)[50,219]. By combination of present advances in the formation and activation of NV centers by high precision ion implantation (e.g., Ref.[220] and references therein), the photocurrent based spin readout technique with going below the diffraction limit in the readout process[221,222], an array of NV centers can be envisioned with addressing ~30 qubits around each NV center with sufficiently close distance between NV centers to produce a scalable room temperature quantum processor unit. *Ab initio* theory can provide a great asset to locate the individual $^{13}$C spins around the NV center[223,224] but further theoretical efforts are needed to explore the complex spin-spin interactions between the cluster of $^{13}$C spins and the central electron spin, in terms of optimizing the quantum control and maintaining the favorable coherence time of the NV electron spin.

### 3.2 Silicon-vacancy center

Recently, the silicon-vacancy (SiV) defect in diamond, which forms a structure with inversion symmetry, has been of great interest[123]. The inversion symmetry of defects is one strategy to suppress the influence of local electric field noise to signal; i.e., the spectral diffusion. In particular, near transform-limited photons could be generated by the negatively charged SiV(−) centers because of the strong ZPL emission[225–227]. The SiV defect consists of an interstitial silicon atom in a split-vacancy configuration belonging to symmetry group of $D_{3d}$, which gives rise to an electronic level structure consisting of ground ($^2E_g$) and excited ($^2E_u$) states that both have double orbital degeneracy[120,228,229]. In the SiV(−) center of diamond, these degenerate orbitals are occupied by a single hole with spin S = 1/2, leading to both orbital and spin degrees of freedom. Thus, spin projection and orbital angular momentum are good quantum numbers for this system. The energy spacing between the spin levels in the electronic ground and excited state is caused by the spin–orbit coupling reduced by the electron–phonon interaction[79], which are about 50 GHz and 250 GHz, respectively. The interaction of the external magnetic field and strain to the spin and electronic orbitals was developed based on group theory principles[120] and later combined with extended theory of the Jahn–Teller effect from first principles calculations[79].

By employing group theory and an advanced *ab initio* method with a realistic band gap of bulk diamond, Gali and Maze determined the electronic structure of the SiV defect in diamond[229]. In this structure, the Si atom goes automatically into a bond center position between two adjacent vacancies, i.e., split-vacancy configuration, which is the inversion center of the diamond lattice. Thus, this defect is also called the V-Si-V defect in the literature, but the SiV label is common in the quantum optics community, and we will use this label in this context. Since carbon atoms are more electronegative than silicon atoms, the charge is transferred from the silicon atom to the nearby carbon atoms. The Si-related four $sp^3$ states should form $a_{1g}$, $e_u$, and $a_{2u}$ orbitals, while the carbon dangling bonds form $a_{1g}$, $a_{2u}$, $e_u$, and $e_g$ orbitals, and all of them form the defect states in diamond. For the neutral SiV defect, only the $e_g$ orbital occurs in the bandgap, with 0.3 eV above the valence-band maximum (VBM). Unlike the positively charged state, the neutral and negatively charged states are found to be stable in a large range of Fermi levels. In the SiV(−) center, the ground state electron configuration is $e_u^4 e_g^3$, where the excited state can be described as promoting an electron from the $e_u$ to the $e_g$ level. The calculated ZPL energy is 1.72 eV without spin–orbit interaction, which agrees well with the experimental data (1.682 eV[230]). It is worth noting that the ground ($^2E_g$) and excited ($^2E_u$) states have similar charge densities; therefore, only small charge redistribution occurs upon optical excitation of the SiV(−) center, explaining the relatively large Debye–Waller factor of about 0.7 or 70%. This is quite different from the case of the NV(−) center, where the charge densities of the ground state and excited state significantly differ, and results in a large motion of ions upon illumination[102]. The relatively sharp and stable ZPL emission with tiny spectral diffusion made it possible to demonstrate indistinguishable photon production from two single SiV(−) centers without any external tuning[231].

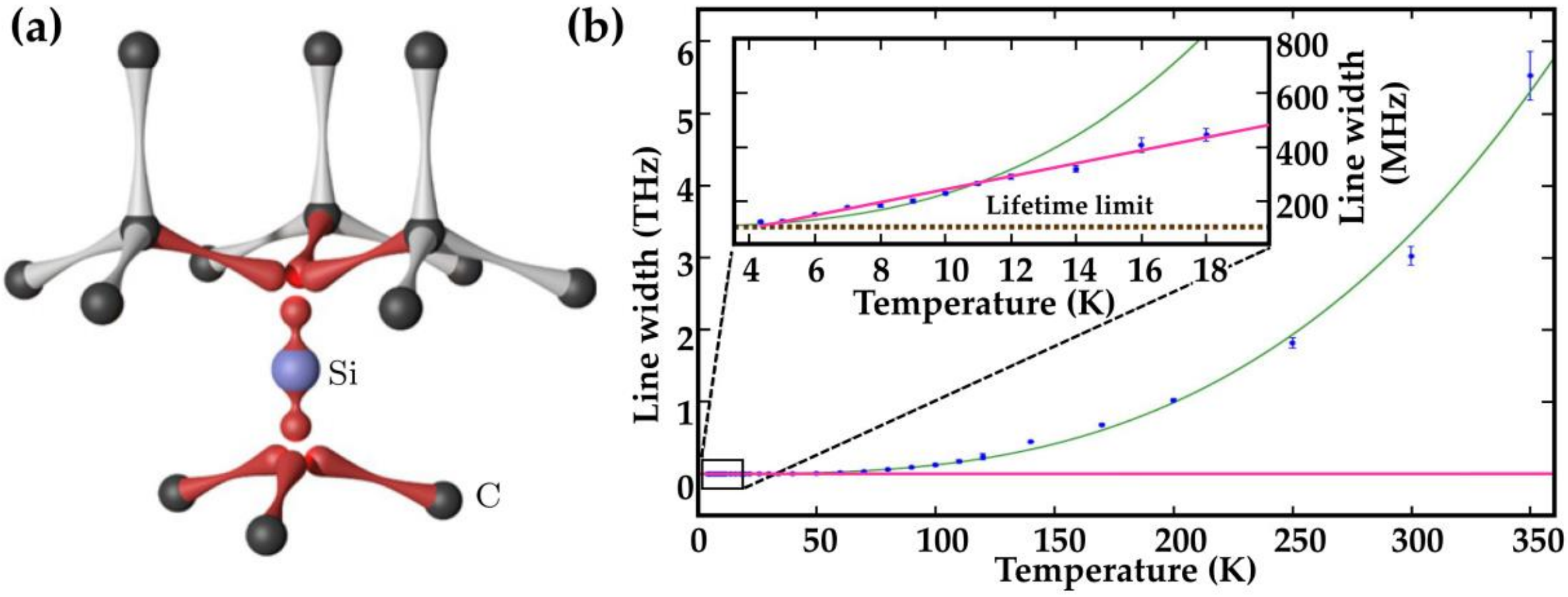

Figure 5. (a) The SiV center consists of a silicon atom centered between two neighboring vacant lattice sites (red small balls). Si ion sits in the inversion center of the diamond lattice which makes the optical transition less dependent against the stray electric fields than that for NV center in diamond. (b) Line width of the SiV center for different temperatures. The green curve corresponds to cubic scaling in the high temperature range (>70 K). At low temperatures (< 20 K), the pink line represents linear scaling, as shown in the inset. Reproduced with permission from Jahnke *et al.*, New J. Phys. **17**, 043011 (2015). Copyright 2015 license under a Creative Commons Attribution 3.0 (CC BY 3.0) International License.

We note that both the ground $^2E_\mathrm{g}$ and excited $^2E_\mathrm{u}$ states are subject to the Jahn–Teller effect. As a consequence, the electron–phonon interactions have a significant impact on the photon emission. Fig. 5 shows the full width at half maximum line widths for the single SiV(−) center[227]. It is obvious that the line width increases with temperature, but with different scaling laws at different ranges of temperature. From 70K to 350K, the line width scales as the cube of the temperature. However, at low temperature (< 20K), there is linear dependence. Temperature can introduce vibronic coupling between different orbitals due to the dynamic Jahn–Teller effect, resulting in population mixture.

Jahnke *et al.* developed a microscopic model of the electron–phonon processes within the ground and excited electronic levels to describe the observed temperature dependent optical line width of the SiV(−) center [227]. In this model, the electron–phonon coupling is a consequence of the linear Jahn–Teller interaction between the $E$-symmetric electronic states and corresponding acoustic phonon modes. At low temperature, the first-order transition between the orbital states involves the absorption or emission of a single phonon. The frequency of the phonon is resonant with the spin–orbit splitting energy $\hbar\Delta$. Within the frame of time-dependent perturbation, the transition rates are approximately described as:

$$\gamma \approx \frac{2\pi}{\hbar}\chi\rho\Delta^2 k_B T \quad (3)$$

where $\chi$ is the interaction frequency for phonons, $\rho$ is the proportionality constant, and $T$ is temperature. Hence the single phonon mediated transition results in the relaxation of the population between the ground and excited states as well as the dephasing of the states. At low temperature, it leads to a linear dependence.

On the other side, at high temperature, the line broadening deviates from the linear relationship with temperature, suggesting the involvement of higher order phonon coupling in the relaxation process. It is worth emphasizing that for NV(−) centers, the inelastic Raman-type scattering process dominates; however, for SiV(−) centers, the inelastic Raman scattering is suppressed and the elastic Raman-type scattering process is dominated. Thus, the two-phonon elastic scattering dominated transition rate is:

$$\gamma \approx \frac{2\pi^3}{3\hbar}\chi^2\rho^2\Delta^2 k_B^3 T^3 \quad (4)$$

Therefore, the two-phonon scattering dominates the dephasing of the orbital states at high temperature, which agrees well with the experimental results.

Understanding the electronic structure and coherence properties of the SiV(−) center contributed to the development of a coherent population trapping experiment[225,226] wherein the spin state of the center could be optically addressed by applying resonant lasers at cryogenic temperatures. However, SiV(−) has short coherence times of about 100 ns at cryogenic temperatures because of the inelastic Raman processes[232]. By cooling the system down to the millikelvin region, the phonon density of states can be reduced and millisecond coherence times can be achieved[233,234]. In special diamond structures at millikelvin temperatures, coupling of identical SiV(−) spins has been achieved[235,236] and coherent manipulation of a single SiV(−) center spin by acoustic phonons has been recently demonstrated[237].

We note a recent breakthrough on creating stable and individual neutral SiV defects[122] in which the zero-phonon line is at 946 nm (see Table I) and the spin state is S = 1. Unlike in the SiV(−) center, the ground state spin does not strongly couple to phonons similarly to the case of the NV(−) ground state, so it has a long spin-relaxation time and presumably long coherence times at cryogenic temperatures. Optical spin polarization of SiV(0) was achieved[122] that could be partially understood from the known electronic structure of the defect[111,119,229,238–241]. Very recently, ODMR signal from single SiV(0) has been reported[242], which is, in particular, spectacular because it has been achieved via excitation of the Rydberg states. The Rydberg excited states have been analyzed by the combination of group theory and *ab initio* calculations[242], where Rydberg excited states are bound exciton states in which the valence band hole is attracted by Coulomb forces of the negatively charged defect. Interestingly, similar charge correction is required for the description of the Rydberg excited states like for the fully ionized defect up to a certain cutoff radius from the center of the defect that is associated with the Bohr radius of the Rydberg excited state[242]. The underlying mechanism of the ODMR contrast has not yet been fully understood, and further research is required for optimizing the corresponding signal. Nevertheless, SiV(0) certainly opens a novel avenue for quantum telecommunication applications.

### 3.3 Other types of split-vacancy complex color centers

The inversion-symmetry based group-IV vacancy color centers XV(−), where X = Si, Ge, Sn, and Pb, in diamond are fast emerging qubits that can be harnessed in quantum communication and sensor applications[79]. As mentioned above, SiV(−) possesses several advantages, such as a narrow inhomogeneous linewidth[243], negligible spectra diffusion[235], and immunity to electric field noise to first order[235]; however, it suffers from low quantum efficiencies of ∼10%[244] and short coherence times at cryogenic temperatures[232]. These limitations urged researchers to seek alternative quantum emitters that potentially have larger spin–orbit splitting that may raise the spin-coherence times. The heavier group-IV vacancy centers of GeV, SnV, and PbV are characterized by similar geometric structures and optical properties but have increased energy spacing between the spin levels because of the expected larger spin–orbit coupling. Indeed, these centers could be formed in diamond. The GeV(−) center has a strong photoluminescence band with a zero-phonon line at 602 nm[113,245–247]. Siyushev *et al.* used two optical excitation wavelengths with resonant matching with the different electronic transitions between the ground and excited states, and they observed coherent population trapping[247]. This demonstrates that the GeV color center in diamond has excellent optical spectral stability and controllable spin states. The SnV(−) center revealed a narrow emission linewidth of ∼232 MHz[248] and 17-fold greater ground state splitting of ∼ 850 GHz[79], implying it has potential for a long spin coherence time. It has also been found to have large quantum efficiency (∼ 80%) and a long ZPL wavelength of 2.0 eV[114]. Novel color centers have been associated with PbV centers. Cryogenic photoluminescence measurements revealed several transitions, including a prominent doublet near 520 nm, and the ground state splitting of 5.7 THz far exceeds that reported for other group-IV split-vacancy centers[249,116]. We emphasize that the degenerate orbital in the ground and excited states result in strong Jahn-Teller interaction for these defects, i.e., a strong electron-phonon coupling well beyond the Born-Oppenheimer approximation[79], therefore the electron-phonon or vibronic spectrum should be determined in *ab initio* calculations. It is spectacular that theory predicted similar order of magnitude the strength of electron-phonon coupling and spin-orbit coupling for the PbV(−) center which results in an effective phonon-spin coupling[79]. Although, similar effect has been demonstrated for SiV(−) center at ultralow temperatures with the use of complex materials engineering to produce acoustic waves in diamond[237] but the significantly enlarged energy spacing between the spin sublevel of PbV(−) center should enable to demonstrate this coupling at elevated temperatures with laser excitation of high energy acoustic phonons, thus the complex materials processing can be avoided. Pb is a huge ion, thus it may create many unwanted defects after ion implantation. Presently, the optical[250] and spin properties[248,251] of SnV(−) center is rather under intense research to apply as a convenient alternative to SiV(−) center in terms of operation temperature, where high temperature annealing or chemical vapor deposition (CVD) after-growth of diamond after Sn implantation resulted in high quality diamond[250,252]. Very recently, GeV(−) and SnV(−) centers have been grown into diamond via microwave plasma CVD process[253] as alternative methods for smooth introduction of dopants. This has the advantage of high quality diamond crystal but the positioning of the individual color centers is not controlled.

We note that the sister defects of SiV(0) were studied by first principles calculations[240], where the complex nature of the triplet excited states, i.e., a product of the Jahn–Teller effect, was studied in detail. We note here that the three triplet excited states are strongly coupled by phonons much well beyond the Born-Oppenheimer approximation. The three triplet excited states are separated due to the exchange interaction between electrons which cannot be accurately described by Kohn-Sham DFT, and multi-reference electronic structure calculations are required. Nevertheless, deep insight to the nature of the exchange interaction makes it feasible to have a fair estimate for the energy splitting between the triplet excited states by using Kohn-Sham DFT[240]. It is predicted that GeV(0), SnV(0), and PbV(0) have ZPL transitions at around 1.80 eV, 1.82 eV, and 2.21 eV, respectively[240]. The corresponding optical centers have not yet been reported, although it is likely that they could appear in diamond beside their negatively charged counterparts because they do not require high p-type doping, in contrast to the case of SiV(0) defects[79]. It is likely that GeV(0) and SnV(0) centers will be observed in diamond in the near future that can be prospective as visible emitter alternative to the NIR emitter SiV(0).

### 3.4 Other selected single color centers

Two other color centers in diamond were selected and listed in Table I which were observed as quantum emitters. The $NiN_4$ defect has NIR emission with inversion symmetry, therefore it is an interesting alternative to other color centers in diamond. However, after the early observation of its quantum emission[115] no progress has be reported so far. Further exploration of this color center may be difficult because its controlled formation (clustering of nitrogen atoms around Ni impurity) is a very challenging task. The ST1 color center was also picked up because it has a unique property: the optical emission occurs between the singlet states whereas the qubit state is a metastable S=1 state[117]. The defect exhibits a high room temperature readout contrast at 45%. The electron spin could be coherently coupled to proximate $^{13}C$ nuclear spins. It was demonstrated that nuclei coupled to single metastable electron spins are useful quantum systems with long memory times, in spite of electronic relaxation processes[117]. This qubit has not yet fully exploited as the microscopic origin is yet unknown which makes its controlled formation difficult. *Ab initio* calculations are needed for identification of this solid state defect qubit which may boost the quantum memory application of this qubit. Identification of the microscopic structure would help to do symmetry analysis on the excited state and metastable state which can principally lead to the optimization of the quantum optics protocol.

## 4. Defect centers in SiC

Inspired by the color centers in diamond, deep color centers have been considered in other host crystals[78,254]. Weber *et al.* proposed a list of physical criteria that a candidate color center and its host material should meet[61], and silicon carbide (SiC) was identified as a potential candidate because of its large band gap and low concentration of nuclear spins, which was in line with a previous theoretical study that proposed, in particular, a divacancy defect in SiC as an alternative to the NV center in diamond[254]. SiC is a wide bandgap semiconductor with extensive applications. High quality 4-inch wafers are available, and this availability contributes to the success of SiC in industrial applications. Defect centers in SiC have emerged as promising candidates for quantum technology applications due to their lower cost, the availability of mature microfabrication technologies, and their favorable optical properties. SiC is composed of silicon and carbon atoms in a hexagonal lattice arrangement in the plane, which can be stacked as quasihexagonal ($h$) or quasicubic ($k$) Si-C bilayers in the sequence. Resulting from the stacking of these bilayers, over 200 polytypes exist in SiC, including 3C, 4H and 6H phases in Ramsdell notation with band gaps of 2.4 eV, 3.2 eV, and 3.0 eV, respectively, where C and H refer to cubic and hexagonal crystals, respectively, and the number represents the instances of Si-C bilayers in the unit cell. Among these polytypes or polymorphs of SiC, 4H-SiC is commonly used as everyday semiconductors for electronic devices, which makes this material a unique platform for the integration of classical semiconductor technology with quantum technology. In 4H-SiC, $h$ and $k$ bilayers alter each other, creating two distinct configurations for monovacancy defects and four different configurations for pair-like defects such as two adjacent vacancies, called divacancies. This multiplies the complexity of the identification of defects in this material but also provides opportunities to produce multiple qubits with distinct but similar properties at the same time.

As a compound semiconductor, SiC contains intrinsic defects of carbon vacancies ($V_C$), silicon vacancies ($V_{Si}$), anti-site type defects ($Si_C$ and $C_{Si}$), the carbon antisite-vacancy pair defect ($C_{Si}$-$V_C$), and divacancy ($V_{Si}$-$V_C$). The identification of the microscopic configuration of point defects is a key step in the advance of quantum information materials. First-principles calculation combined with magneto-optical spectra were widely employed to study the electronic and spin properties of these defect centers in SiC (e.g., recent *ab initio* results about native defects in 4H-SiC in Ref.[255] and a previous result in Ref.[256]). It was expected that the optically active native point defects might act as single color centers and qubits in well-engineered SiC materials.

Indeed, one of the brightest solid state quantum emitters in the visible region (ZPLs around 600 nm) was reported in 4H-SiC, and it was identified by *ab initio* calculations as configurations of the positively charged $C_{Si}$-$V_C$ defect, which has an S=1/2 electron spin[257]. The manipulation of the spin state of these emitters has not yet been reported.

Using optical and microwave techniques similar to those used with diamond qubits, Koehl *et al.* demonstrated that several point defects in 4H-SiC are optically active and coherently controlled with a range of temperature from 20 K to room temperature[125]. They were mostly identified as divacancy configurations[256], which they called PL1-4 centers. Neutral divacancies in 4H-SiC have S = 1 ground state spin and ZPL of around 1.1 eV with a DW factor of about 0.03. The coherence times of $V_{Si}$-$V_C$ are 1.2 ms with ODMR readout contrast of 15% in 4H-SiC and 0.9 ms in 3C-SiC[132]. Compared to the NV center in diamond, the divacancy with correlated states in SiC can provide competitive advantages, such as that the emission wavelength in the NIR region would suit biological studies because it can effectively penetrate organic tissue, the ODMR and ZPL signals can be resolved at room temperature, and ZFS is at lower frequencies (~1.3 GHz) than that of the NV center in diamond, which is also preferential for biological systems [258–260]. Beside PL1-4 centers, other ODMR centers with similar ZPL and ZFS energies were also found and called PL5-6 centers. Very recently, these centers have been identified as neutral divacancies inside the stacking faults of 4H-SiC[260], which shows the principle that stacking faults or short polytype inclusions may generate atomic-scale defect qubits distinct from the configurations in the bulk counterpart. These defect spins could be fabricated and manipulated at the single defect level[132,261]. Recently, the ability to control the specific charge states of divacancy spin defects in 4H-SiC has been realized experimentally, providing enhanced spin-dependent readout and long-term charge stability[99].

The divacancy spin levels and coherence times are sensitive to the presence of strain[73,262,263], electric fields[73], magnetic field[264], and temperature[136] and can potentially be harnessed in sensor applications or used to drive the electron spins mechanically[265] or electrically[259,266]. The variety of divacancy qubits and the Stark-shift tuning of the optical and spin levels makes possible the production of interference between different types of divacancy qubits in a controlled fashion when engineered into SiC electronic devices[267]. It was experimentally proven that the spin polarization can be efficiently transferred from the electron spin toward nearby nuclear spins in the SiC lattice[136], which was understood by combined *ab initio* and effective spin Hamiltonian study[268], with prediction of the flipping of nuclear spins by optical means and small constant magnetic fields[269].

Silicon-vacancy related ODMR centers were well known in hexagonal SiC polytypes and even in a rhombohedral SiC polytype. These centers were labeled as V1 and V2 in 4H-SiC[270,271] and as V1, V2, and V3 in 6H-SiC[272], where the corresponding emission comes at the near infrared region (see Table I) with a broad phonon sideband. Very recently, the DW factor was found to be around 6%[133,273]. The corresponding electron paramagnetic resonance (EPR) signals showed S=3/2 spin states with a relatively small ZFS in the ground state between the sublevels caused by the crystal field of the hexagonal crystal, which slightly deviates from the quasi-tetrahedral symmetry and results in $C_{3v}$ symmetry[272]. The origin of these ODMR emitters was debated in the literature. Soltamov and co-workers interpreted the observed hyperfine signatures for a V2 center so that the V2 center should be a complex of $V_{Si}$ near a second or farther neighbor $V_C$ along the crystal axis[274]. Using first-principles calculations and electron spin resonance measurements, Ivády *et al.* have studied the isolated silicon vacancies, the negatively charged silicon vacancy, and a proximate neutral carbon vacancy [$V_{Si}(-)$ + $V_C(0)$], as shown in Fig. 6[271]. The calculated DFT energies indicated that the pair vacancy [$V_{Si}(-)$ + $V_C(0)$] is a metastable configuration. Furthermore, ZPL energy and hyperfine tensor were calculated via the Heyd–Scuseria–Ernzerhof (HSE06) hybrid exchange-correlation functional[275,276], and the zero-field splitting calculations were conducted using the Perdew–Burke–Ernzerhof (PBE) functional[277]. These high-precision first-principles calculations showed that the pair vacancy [$V_{Si}(-)$ + $V_C(0)$] has a spin-1/2 ground state without any zero-field splitting. By theoretical simulations and high-resolution EPR measurements, Ivády *et al.* demonstrated that the isolated silicon-vacancy [$V_{Si}(-)$] accounts for the majority of the experimentally observed magneto-optical properties. The molecular orbitals of the isolated silicon-vacancy are constructed from symmetry-adapted linear combinations of the three equivalent *sp*$^3$-orbitals from the basal-plane carbons and the *sp*$^3$-orbital belonging to the carbon atom on the crystalline c-axis[278]; thus, it is not a part of a nearby carbon-vacancy, as was proposed by Soltamov *et al*. The conclusion is that these ODMR centers are the configurations of isolated $V_{Si}(-)$ defects, where V1($h$) and V2($k$) in 4H-SiC and V1($h$), V2($k_2$), V3($k_1$) in 6H-SiC were identified by first principles calculations[278–280] which can be engineered as single photon emitters[281]. An individual $V_{Si}(-)$ qubit at room temperature was coherently controlled with a coherence time of about 1.5 ms[282]. The ODMR contrast does not exceed 1% for any configuration. $V_{Si}(-)$ qubits could be engineered into SiC electronic devices, where the charge state of single defect qubits could be stabilized[283] and electroluminescence from ensemble $V_{Si}(-)$ qubits could be observed[284] with a potential to produce masers[285].

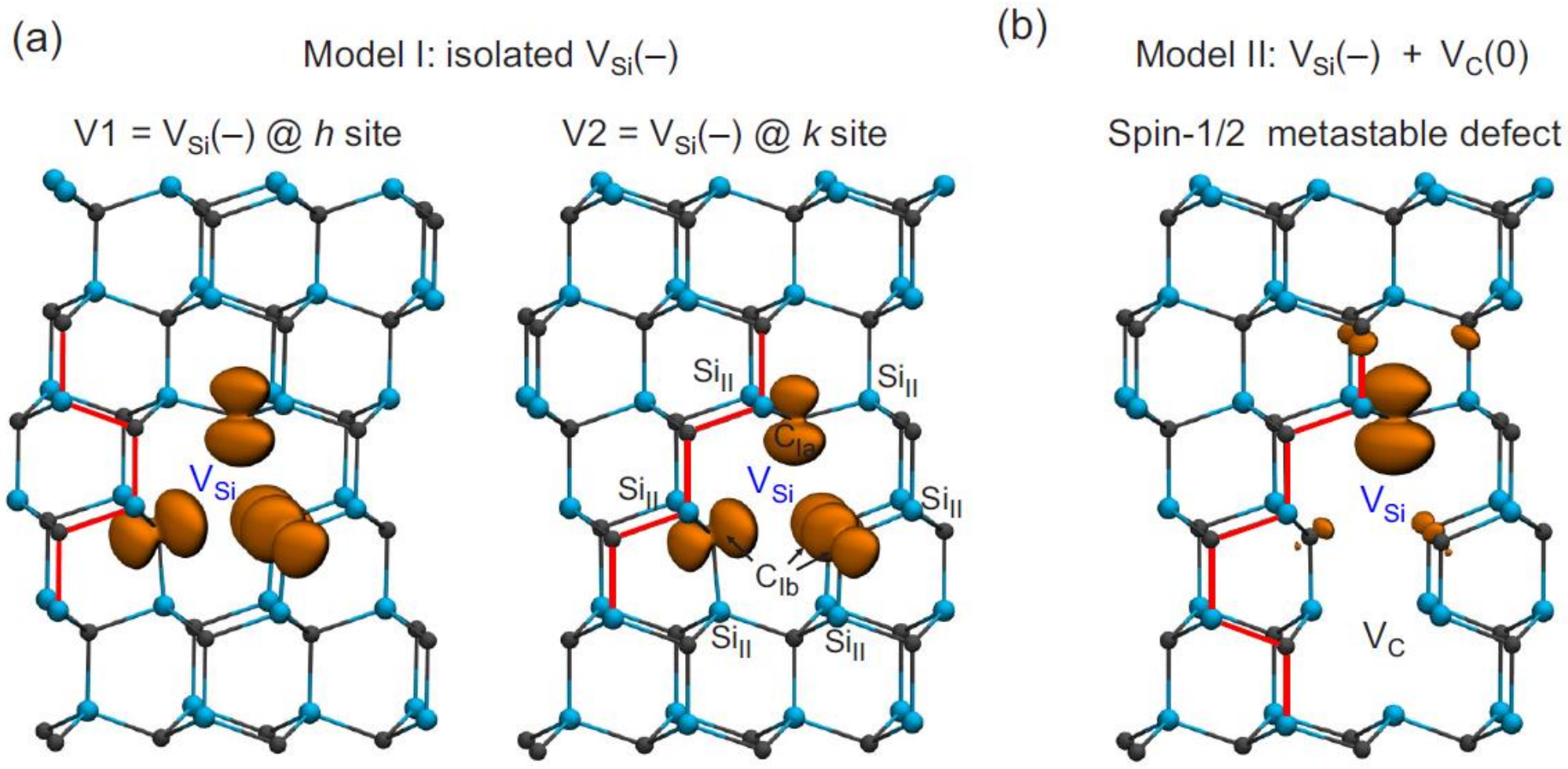

Figure 6. Models of silicon-vacancy related qubits in 4H-SiC. (a) Isolated Si-vacancy model and the alignment of V1-V2 centers to different silicon-vacancy configurations in 4H-SiC; (b) vacancy pair model of V2 center in 4H-SiC. Here, the red lines highlight the stacking of the SiC double layers to identify the different configurations of silicon and carbon vacancies. The highlighted orange lobes demonstrate the spin density of the defects. It can be seen that, based on the calculations shown in [Ivády *et al.* Phy. Rev. B **96**, 161114 (2017)], the isolated vacancy model in (b) is assigned to the V1 and V2 centers in 4H-SiC. Reprinted figure with permission from Ivády *et al.* Phy. Rev. B **96**, 161114 (2017). Copyright 2017 American Physical Society.

Group theory considerations together with first principles calculations have already revealed the basic electronic structure of $V_{Si}(-)$ qubits, and the corresponding selection rules in the optical excitation of these defects[286,287]: The four carbon dangling bonds (see Fig. 6a) create an $a_1$ and $t_2$ orbital in $a_1^{(2)}t_2^{(3)}$ electronic configuration in the negatively charged state in 3C SiC, where the $t_2$ orbital splits to $a_1$ and $e$ orbitals in the hexagonal SiC polytypes. The lowest excited state may be described as promoting an electron from the lower $a_1$ level to the upper $a_1$ level with forming $^4A_2$ ground and excited states. Soykal *et al.* further elaborated the group theory by revealing the fine spin level structures and states for both the optically active spin quartet states and the dark spin doublet states[278]. First principles calculations revealed that a strong Jahn–Teller effect appears in the excited state, causing the appearance of a polaronic V1' level in the PL spectrum and the temperature dependence of the spin dephasing[133]. In addition, the V1 center shows negligible spectral diffusion[288] because of the small change in the charge density between the quartet lowest energy excited state and the ground state[289]. The favorable magneto-optical properties of $V_{Si}(-)$ qubits could be harnessed to realize spin-controlled generation of indistinguishable and distinguishable photons from silicon vacancy centers in 4H-SiC[290].

Nitrogen-vacancy pair defects ($N_CV_{Si}$) have recently been observed in N-doped 3C, 4H, and 6H-SiC polytype crystals[128,291]. The negatively charged $N_CV_{Si}$ show broad photoluminescence spectra and NIR emission arising from the transition between the $^3E$ excited and $^3A_2$ ground states, both of which exhibit S = 1 spin states[292,293]. The ODMR measurement of the single nitrogen-vacancy centers in 4H-SiC at room temperature has been very recently achieved in N-doped SiC[294], which presumably shows similar properties to their NV counterpart in diamond. The photoluminescence signals from transition metal point defects were also observed in SiC, e.g., chromium, vanadium, niobium, and molybdenum, which possess relatively high DW factors. Chromium defects in hexagonal SiC have about 0.73 DW factor and S = 1 spin ground state, which could be spin polarized upon illumination at the ensemble level[130]. The emission wavelength is at 1.158 eV with long optical decay rates of $\sim$100 $\mu$s, where the latter is due to the intra-configurational spin-flip transition between the spin singlet excited and spin triplet ground states[295]. The neutral vanadium defects have two configurations in 4H-SiC with optical emission at 0.969 (α) and 0.929 (β) eV[296], and three centers with similar NIR emission were found in 6H-SiC[297]. Individual vanadium centers created by vanadium implantation were observed in 4H- and 6H-SiC, in which the observed coherence time was roughly 1 $\mu$s at cryogenic temperatures[131]. The high-quality molybdenum doped SiC (ZPL is at 1.106 eV and 1.152 eV in 6H and 4H polytypes, respectively) resulted in a coherence time of 0.32 $\mu$s for the S = ½ electron spin at cryogenic temperatures, behaving as doublets with highly anisotropic Landé $g$-factor[298] largely deviated from the value of the free electron, which implies a contribution of the spin–orbit interaction and should result in a relatively large zero-field splitting. In these defects, the $d$ orbitals play a crucial role by carrying angular momentum and can interact with the external magnetic fields. However, these systems are also Jahn–Teller active, and the electron–phonon coupling will significantly affect the strength of interaction of the system with external magnetic fields, explaining the observations on vanadium and molybdenum qubits[299].

Overall, SiC has obvious advantages in applications of quantum technology. High quality 4-inch 4H-SiC wafer is widely exploited in semiconductor industry with well controlled doping technologies, which is promising to reduce the production cost of the future SiC based quantum devices. Indeed, single divacancy[256] and Si-vacancy spins[300] have been integrated and controlled in SiC diode structures (see Fig. 7). Furthermore, single SiC spins could be integrated into photonics structures which is a promising route towards future quantum optoelectronics devices[300]. 4H-SiC has a smaller than diamond's but still considerably large band gap at 3.3 eV which can host mostly NIR quantum emitters. NIR emission is generally favorable for quantum sensors for biology and quantum communication. On the other hand, the relatively low small energy spacing between the defect levels and about twice smaller Debye temperature of 4H-SiC (about 1000K)[301] than diamond's make the defect qubits' key magneto-optical properties sensitive to temperature. For instance, the ODMR readout contrast of single divacancy spins upon off-resonant excitation is well above 10% close to cryogenic temperatures but significantly reduces at room temperature. We refrain here that the room temperature ODMR readout contrast of Si-vacancy V2 center is around 1%. Compared to the ~30% room temperature ODMR readout contrast of diamond NV center, the low ODMR contrast of 4H-SiC spins is disappointing as the high readout contrast is an important factor in the overall sensitivity and temporal resolution of quantum sensors. Very recently, a breakthrough has been achieved in 4H-SiC[302], where ~20% off-resonant room temperature ODMR readout contrast is demonstrated for single PL6 spins. Theory implies that the ODMR readout contrast starts to decline at about 150 K for divacancy defect qubits in 4H-SiC[302], in contrast to diamond NV center's data at around 600 K[303] because of the smaller energy spacing between the defect levels in 4H-SiC with respect to those of NV diamond[302]. Nevertheless, PL6 divacancy qubit's room temperature ODMR contrast still persists at relatively high value. This opens the door for SiC quantum technology in the field of room temperature quantum NMR measurements and other biology applications. The tight control of the surface and interface of SiC and silicon dioxide ($SiO_2$) is far from being solved (e.g., Ref.[304]) which will be the next important task in the development of sensitive quantum sensors from SiC.

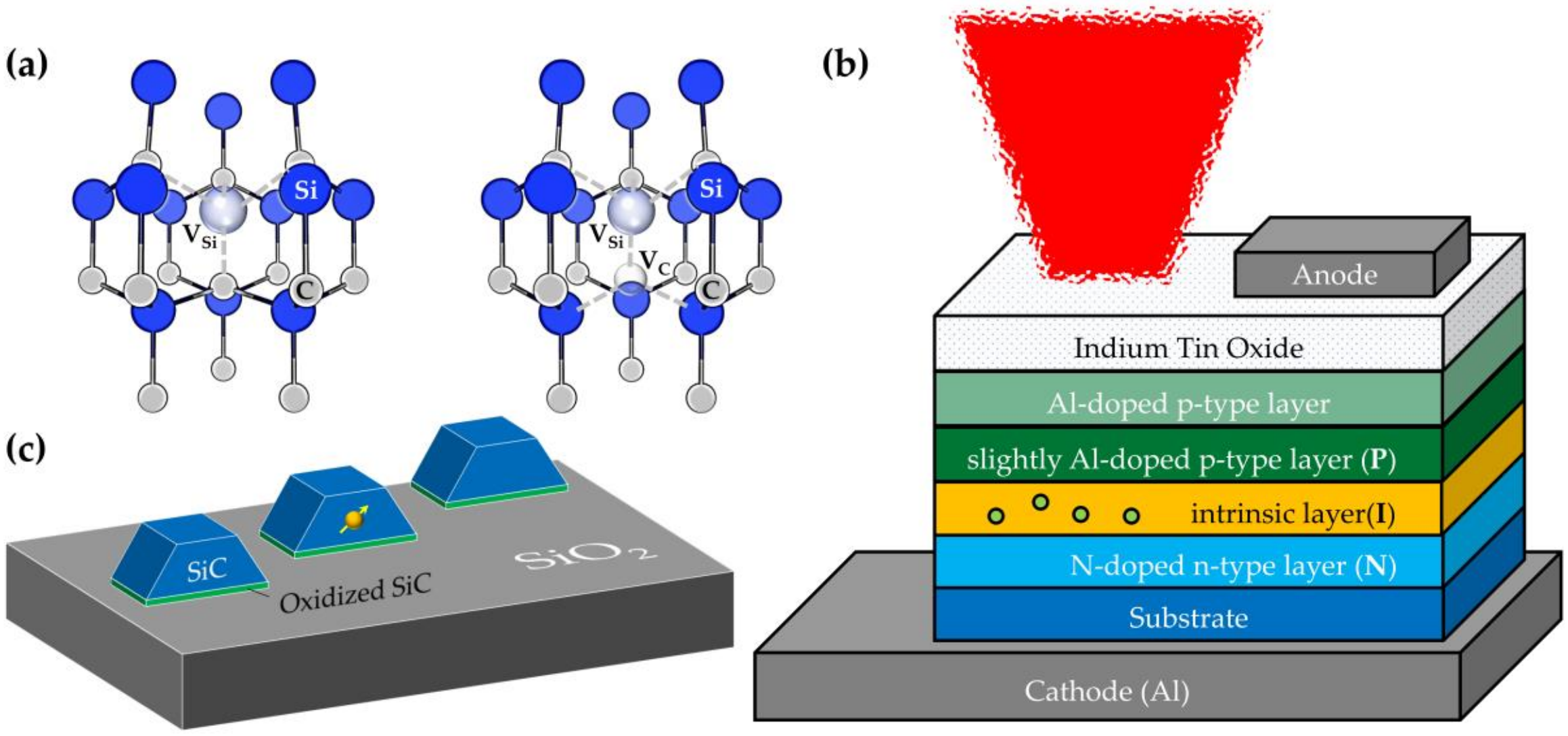

Figure 7. As discussed here, SiC offers certain advantages in applications in quantum technology. High quality 4-inch 4H-SiC wafer is exploited widely in semiconductor industry with well controlled doping technologies, which can reduce the production cost of future SiC-based quantum devices. Single divacancy and Si-vacancy spins have been integrated and controlled in SiC diode structures. (a) Si-vacancy and divacancy structures in 4H-SiC. (b) Defect qubits integrated into SiC diodes. (c) Defect qubits integrated into SiC photonics structures.

## 5. Defect centers hosted in 2D materials

Two-dimensional (2D) materials have been attracting extensive research interest in the last few decades. Recently, room-temperature quantum emitters have been reported in two-dimensional (2D) wide bandgap materials[75,305–308], creating great interest on quantum emitters and potential quantum bits in 2D materials. The big advantage of 2D materials is that the creation of the defects can be well controlled with the present experimental techniques with the almost deterministic localization of the defects that is needed for scaling-up quantum bits for quantum computers. In addition, strain and electric fields can be engineered on these 2D materials, and optical cavities can be formed around them[309]. Another playground for 2D materials is the stacking of different 2D materials on top of each other, and one might imagine sandwiching an atom or molecule between these 2D layers as qubits. Beside quantum computing and quantum communication purposes, it is noted that 2D materials are surfaces *per se*, so they can be good hosts for quantum sensing. The single-layer $MoS_2$ layers, delta-doped diamond slabs, and Si thin slabs were predicted as promising hosts for spin qubits[310]. Isotopic purification is much more effective in 2D materials that leading to an exceptionally long spin coherence time of the order of 30 ms. Apparently, there is great potential for using 2D materials for quantum technology. However, the first quantum emitters found in hexagonal boron nitride (*h*-BN) have not been unambiguously identified, and the origin of these emitters is still under intense research[307,75,311,173,312–314,176,148,315,316]. For instance, some quantum emitters have been tentatively assigned to specific native point defects in *h*-BN[176]. If these quantum emitters are identified and then engineered into a nanocavity (e.g., for $WSe_2$ in Ref.[317]), then room-temperature photon blockades, which are an ultimate demonstration of nonlinearity at a single photon level, might be realized for the first time at room temperature.

### 5.1 Hexagonal boron nitride

Hexagonal boron nitride (*h*-BN) is one representative 2D material. It is a wide-bandgap (~6 eV) 2D material with the potential to host color centers that are promising candidates for quantum technology. Many emitters in *h*-BN are bright with narrow linewidths, are tunable, and have high stability[75,306,318,319]. In 2016, room-temperature, polarized and ultrabright single-photon emission from color centers in 2D *h*-BN was first experimentally demonstrated[75]. These advantages have sparked strong research interest in *h*-BN defects[75,173,306,312,313,318–324].

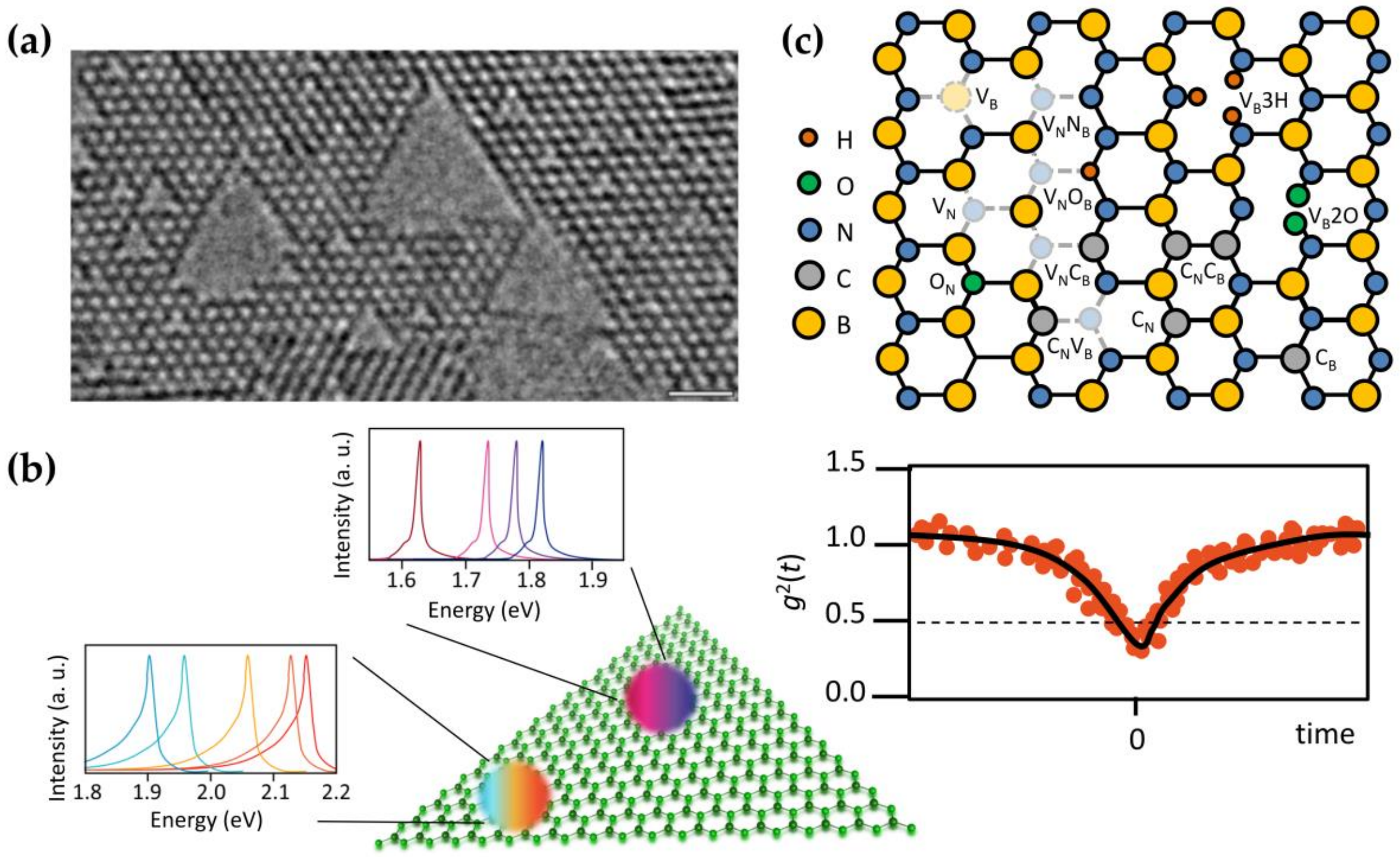

Figure 8. Atomic defects in in *h*-BN. (a) High resolution transmission electron microscope image showing lattice defects in *h*-BN such as single vacancies and larger vacancies respectively. These can be seen as triangle shapes with the same orientation. The scalebar in this figure is 1 nm. Figure has been reprinted with permission from Jin *et al.* Phy. Rev. Lett. **102**, 195505 (2009)[325]. Copyright 2009 American Physical Society. (b) Typical visible and ultraviolet PL spectra from quantum emitters. The photon correlation spectrum ($g^2$) reveals the quantum nature of the emitters. (c) Geometry of various possible defects in *h*-BN responsible for the quantum emitters, including $V_B$, $V_N$, $O_N$, $C_B$, $C_N$, $V_N N_B$, $V_N O_B$, $V_N C_B$, $C_N V_B$, $V_B 3H$, $V_B 2O$, and $C_N C_B$ defects.

In *h*-BN, a narrow emission band has been observed experimentally with a ZPL transition at a range of 1.6 eV to 2.2 eV[75,306], but *h*-BN has also shown luminescence with a ZPL at 4.1 eV[174] and 5.3 eV[177]. Tran *et al.*[75] experimentally demonstrated room-temperature and ultrabright emission from *h*-BN with ZPL energy at 1.95 eV. A plasma-treatment of *h*-BN removes this emitter and produces another emitter with ZPL emission at ~1.7 eV that was associated with the presence of oxygen in the quantum emitters [314]. Observations suggest that some emitters may absorb at and emit from distinct electronic states[319]. The origin of these emitters is still unclear. A broad spectral distribution, spanning ~1 eV of the group of 2-eV ZPL emitters also observed in *h*-BN, implies that a number of distinct defect structures may exist, and they are tentatively attributed to the structural composition variations[307], the defect charge states uncertainty[326], as well as the local strain and dielectric environment variations[308]. The broad window of ZPLs that are generated by numerous color centers makes the identification of these quantum emitters very challenging. Indeed, it has been shown that Stark-shift on the order of 10 nm can be observed upon 1 GV/m electric field on the ZPL energy of single emitters in *h*-BN[327], which demonstrates the strong coupling of these emitters to external fields.

Inspired by the similar phenomenon in NV centers in diamond, it is naturally suggested that the origin of these quantum emitters is the combination of substitutional and vacancy defects. *h*-BN has a wide variety of possible intrinsic defects, antisite defects, and unintentional impurities (such as H, O, C, and Si) in its lattice structure. Several DFT computational works have attributed these emissions to charge neutral native and substitutional defects with deep bandgap states[173,312,313,176,322]. The plausible defect structures proposed by theoretical studies and experimental observations are shown in Fig. 8, including boron vacancy $V_B$[140,325], nitrogen vacancy $V_N$[325], carbon substitutional $C_B$ and $C_N$[328,329], oxygen substituting for nitrogen $O_N$[328], silicon substituting for boron $Si_B$[330]; a vacancy next to a substitutional atom, e.g., nitrogen $V_N N_B$[75,313], oxygen $V_N O_B$[327], carbon $V_N C_B$[313,331] and $C_N V_B$[322]; boron vacancy passivated by oxygen $V_B 2O$[314] and hydrogen $V_B 3H$[176,311]; and a double substitutional carbon defect $C_N C_B$[332].

The intrinsic defects $V_B$ and $V_N$ in a freestanding *h*-BN have been observed in a TEM experiment [325]. The defect $V_N$ possesses $D_{3h}$ symmetry, which has two defect orbitals in the bandgap that enable optical excitation below the band gap energy. The negatively charged $V_N$ shows a closed-shell singlet ground state, which is excluded as the source of the observed EPR signal[322]. Room-temperature EPR and ODMR observations demonstrated a triplet ground state in electron irradiated and annealed *h*-BN crystal and exfoliated flakes[140], which was identified as the negatively charged $V_B$ [313,179]. Substitutional impurity defects such as $O_N$, $C_N$, $C_B$, and $C_N C_B$ carbon pairs were directly observed by annular dark-field imaging in a STEM[328]; that study demonstrated that substitutional carbon atoms may occur mostly in pair-form. The 4.1 and 5.3 eV ZPLs were plausibly attributed to the deep donor $C_B$ and deep acceptor $C_N$[329], which has not yet been confirmed either from experiment or theory, though these defects have indeed low formation energies in *h*-BN[176]. Recent experiments have recorded ODMR signals on single or a few defects at room temperature for color centers in *h*-BN [140,333] that were tentatively assigned to $C_N$ defects based on the detected hyperfine signatures compared to the calculated ones[322]. On the other hand, the calculated ZPL energy of $C_N$ defects do not agree with the observed ones[176], thus further investigations are needed to identify the origin of the close-to-single ODMR centers. McDougall *et al.*[311] combined a X-ray Absorption Near-Edge Structures (XANES) experiment and DFT calculation on the shift of the core levels of the host atoms to investigate the nature of point defects in *h*-BN. They were able to identify the presence of $O_N$ and $V_B 3H$ defects, where the latter was a boron vacancy with three hydrogen atoms saturating the dangling bonds. As the *h*-BN samples exhibited 4.1 eV emission, this was tentatively assigned to the $V_B 3H$ defect based on DFT calculations that underestimated the band gap by about 1.5 eV. Recent theoretical investigations indicated that, based on the calculated ZPL energy, the lifetime of the excited state and DW factor[334] of carbon pair defect gives rise to a relatively narrow luminescence band, which can be associated with the 4.1-eV emitter. Mark *et al.*[335] found by first principles calculations that the internal transition from the boron dangling bonds to a boron $p_z$ state has a ZPL at 2.06 eV and emission with a Huang–Rhys factor of 2.3, which was suggested for the origin of the group of 2-eV quantum emitters in *h*-BN. In this model, the boron dangling bond was accompanied by hydrogen atoms saturating the other dangling bonds of the vacancies[335], but the presence of hydrogen has not yet been proven in experiments. In addition, the calculated ionization threshold of the negatively charged dangling bond was smaller in energy than the calculated ZPL energy at 2.06 eV[335], which makes this proposition tentative at the moment.

Advanced first-principles calculations in combination with group theory analyses would be a powerful approach to analyze the defects' magneto-optical properties, as shown in Ref.[313]. Abdi *et al.* predicted that $V_B(-)$ can exist and produce an S = 1 ground state[313], and this phenomenon was later observed by ODMR[140]. On the other hand, prediction of the excited states from DFT calculations has limitations for the highly correlated orbitals or multireference states that typically appear in vacancy-type defects. As an example, Cheng *et al.* predicted that a $V_N C_B$ defect has an S = 1 ground state in the neutral charge state, which is stable in a relatively wide range of Fermi levels[312]. The estimated ZPL energy between the triplet states is about 1.6 eV where emitters with such ZPL energy are often observed in $h$-BN samples. In contrast to this result, Sajid and co-authors found that the ground state of $V_N C_B$ is $^1A_1$ singlet with $C_{2v}$ symmetry. They predicted 2.08 eV ZPL energy for the emission, where the phonon sideband should be broad[173]. This may explain the observed 1.95 eV emitters in $h$-BN[75]. If the intersystem crossing is very fast from the singlet excited state towards the triplet state, then the predicted ZPL energy is about 1.58 eV between the triplets[322], which basically agrees with the previous DFT result in Ref.[312].

A multireference character of the closed-shell and open-shell ground states of the defect-induced strong electron correlation effect would be another challenge in computational methods to accurately describe the energy levels of point defects[336]. As an example of $V_N C_B$ defects in $h$-BN, hybrid functionals such as HSE06 work very well for excitations within the triplet manifold of the defects; however, they underestimate significantly the triplet-state energies by about 1 eV. The estimation of the inaccuracy was based on an $h$-BN flake cluster model of a few tens of atoms terminated by hydrogen atoms by comparing the DFT results with Hartree–Fock based multireference methods. The estimated DFT error might be overestimated in Ref.[336] because of the limitations of the small models. Recently, a multireference method was applied on a supercell model of about 200 atoms for the calculation of the $V_B(-)$ states in $h$-BN based on a density matrix renormalization group algorithm, and the error in HSE06 DFT was not so severe[179].

It is worth noting that phonons play a critical role in determining the magneto-optical properties of the quantum emitters in $h$-BN. Strong electron–phonon interaction activates the emission of $V_N N_B$ defects[337], which also causes a strong coupling of strain to the ZPL emission. In this particular case, the out-of-plane or membrane phonon modes are coupled strongly to the defect, which moves the atoms out of the plane. Experiments combined with first principles calculations imply that this motion of ions can explain the polarization of the emitted photons from such emitters[327]. The strong electron–phonon coupling, manifested as the Jahn–Teller effect in the excited state, is responsible for the weakly allowed optical transition in $V_B(-)$ states in $h$-BN, and membrane phonons may play a crucial role in the intersystem crossing responsible for the ODMR contrast[179].

## 5.2 Transition metal dichalcogenides

The family of two-dimensional materials, including graphene, $h$-BN, transition metal dichalcogenides (TMDs) (e.g., $MoS_2$, $WS_2$, etc.), and phosphorene, have attracted much attention due to their extraordinary physical, chemical and mechanical properties, as well as their promising applications in nano electronics, thermoelectric devices, wearable electronics, flexible displays and smart health diagnostics[338–347]. In addition to $h$-BN, TMDs have semiconducting characteristics, with direct bandgaps in their monolayer structures. For a $MoS_2$ monolayer, there exist several types of intrinsic defects[348,349]. The defect based single-photon sources were realized in $WSe_2$[350–358]. The potential advantage of a 2D single-photon source is obvious, for it offers the possibility of integrating single-photon sources with van der Waals heterostructures. Most observed optical emission from $WSe_2$ in the band between 1.63 eV and 1.72 eV, with narrow linewidths of 120–130 μeV, and strong photon anti-bunching was reported, unambiguously establishing the single-photon nature[350–353]. This feature is attributed to spatially localized exciton states and the electron–hole interaction in the presence of anisotropy. The photon emission properties can be controlled via the application of external electric and magnetic fields. The decay time of these emissions is around 10 ns, which is ten-fold longer than that of the broad localized excitons. This reveals that the exciton bound to deep-level defects is the underlying mechanism for the observed single photon emissions[356]. In 2017, large-scale deterministic creation of quantum emitters was demonstrated experimentally[357,358]. Although research on defect centers in TMDs is currently intensifying, this is a new field that addresses many open questions, such the major defects dominating the single-photon emission, their atomic configurations and electronic structures, which still need to be determined. Manipulating desired point defects at a predetermined set of locations lies at the heart of 2D quantum sensing engineering, and it can be technically realized[359].

## 6. Other materials as hosts

### 6.1. Other wide bandgap materials

Similar to the aforementioned host materials, other wide bandgap materials also host optically active defects that can principally emit single photons. Compared with NV center in diamond, the research of color centers in these wide band gap materials is still in its infancy. Preliminary results show a great advance in these materials, however, further efforts are still required to obtain a comprehensive understanding, such as the correlation between defect orbitals with the pristine orbitals of host material, the effects of strain and phonon coupling on electron coherence. For the completeness of this review article, here we briefly introduce the basic properties, performance and open questions of these wide bandgap semiconductors for application in quantum information technology.

#### 6.1.1. Zinc oxide

Point defects in ZnO were explored for potential application as single-photon sources[149,360,361]. Numerous theoretical and experimental studies have investigated the electrical and optical properties of ZnO with a band gap of around 3.4 eV, and the role of defects, including oxygen vacancies ($V_O$), zinc vacancies ($V_{Zn}$), the zinc interstitials ($Zn_i$), and the oxygen interstitials ($O_i$)[362]. Using hybrid density functional calculations, the native vacancies, interstitials, and dangling bonds in ZnO have been investigated[363]. The oxygen vacancy ($V_O$) was found to be a neutral defect, and the highest-energy photoluminescence peak associated with $V_O$ is at 0.62 eV. Under realistic growth conditions, the zinc vacancy ($V_{Zn}$) is the lowest-energy native defect in n-type ZnO, acting as an acceptor. $V_{Zn}$ gives rise to multiple transition levels and emission between 1 and 2 eV. Compared with isolated $V_{Zn}$, hydrogen atoms form highly stable complexes with $V_{Zn}$, shifting the acceptor levels closer to the valence band edge. Hydrogenated $V_{Zn}$ has optical transitions similar to those of isolated $V_{Zn}$, both in good agreement with recent experimental results, supporting them as the source of photon emission. It is suggested that Zn dangling bonds-related emission may be an intrinsic source of green luminescence in ZnO. The first report on the room temperature quantum emitters with broad phonon sideband (see Table I) was associated with the different charge states of the Zn-vacancy[149]. Room temperature single photon emission was also found in ZnO nanoparticles[364,365]. The application of ZnO in quantum technology would benefit from detailed knowledge of its optical properties that should be further explored. We note that the exchange interaction between the bands of Zn *d* orbitals and O *p* orbitals makes the accurate DFT calculation of pristine ZnO and defective ZnO complicated with respect to the case of other semiconductor materials. Hybrid DFT functionals resulted in a great advance as listed above, however, further efforts are required to study the localized and correlated orbitals of the vacancies and related defects.

#### 6.1.2. Zinc sulfide

A room temperature quantum emitter was also observed in cubic ZnS particles of ∼100 nm in diameter[366]. The zinc blende ZnS exhibits a room temperature band gap of ∼3.6 eV which is somewhat larger than that of ZnO. The fluorescence intensity of the quantum emitter emerges at 600 nm, it has a maximum at around 640 nm, and it decays to zero at around 750 nm upon 532-nm excitation. The ZPL peak is not visible at room temperature. The observed lifetime is at ∼2.2 ns. Since the emitter was observed without any treatment it was tentatively associated with an intrinsic defect, the Zn-vacancy. *Ab initio* predicts for ZPL energy of about 2.4 eV (516 nm) for Zn-vacancy[367], which is a bit far from the observed shortest wavelength of the emission. Further efforts are needed for identification of this single photon emitter. High quality ZnS material with large band gap has a good potential in hosting visible quantum emitters.

#### 6.1.3. Titanium oxide

Recently, defects in $TiO_2$ thin films and nanopowders exhibited single-photon emission have been found[368]. The excited-state and non-radiative lifetimes were found to be within the range of several nanoseconds and tens of nanoseconds, respectively. The fluorescence occurred in the red emission band. The three types of room temperature quantum emitters show relatively broad luminescence with some peaks between 600 and 700 nm. Low temperature measurements for optical characterization and *ab initio* calculations may reveal the origin of these single photon emitters in this prospective material with a band gap at ~3.0 eV.

### 6.1.4. Gallium nitride

Room temperature quantum emitters in gallium nitride (GaN) were reported to exhibit narrowband luminescence[172]. In semiconductor quantum dots, because of the high densities of defects and charge traps, the rapid charging/de-charging can result in an obvious broadening of the emission linewidth. To overcome this issue, interface fluctuation GaN quantum dots have been developed[369]. In the interface-fluctuation quantum dots, because the formation of charge localization centers is located at positions of thickness fluctuation in quantum wells, they emit in the near ultraviolet at wavelength of ~350 nm and exhibit narrow linewidths as compared to typical QDs[369–371]. Recently, near-infrared emitters in GaN have been found; they exhibit both excellent photon purity and a record-high brightness exceeding $10^6$ counts/sec[155]. The origin of the emitter remains unknown. The substitutional chromium ($Cr^{4+}$) impurity is a qubit candidate in GaN that possesses small phonon sidebands, so most of the fluorescence can be harnessed for quantum communication[130].

### 6.1.5. Cubic BN and hexagonal BN nanotube

In addition to *h*-BN 2D crystals, the zinc-blende allotropic form of boron nitride, so-called cubic boron nitride (*c*-BN), has shown important application as a wide-band-gap semiconductor. Recently, Tararan *et al.*[148] provided a literature survey and systematically investigated the optical properties of *c*-BN, including the optical gap and the luminescence of intragap defects. Using EELS, a large optical gap exceeding 10 eV was reported. A number of defect luminescence centers were demonstrated in the visible and UV spectral range. In the UV spectral range, possible single-photon emission was observed[148], which may motivate further investigation on this material.

Another possible form of BN is the hexagonal boron nitride nanotubes (*h*-BNNT). Unlike carbon nanotubes, the polarized B-N bonds create a large band gap of ~6 eV independent from the chirality of the nanotube. The first room temperature quantum emitters were reported in about 5-nm-diameter *h*-BNNT[138]. Recently, two similar room temperature quantum emitters have been observed in 50-nm diameter *h*-BNNT[139]. The peaks for the two emitters are found at 571 nm and 569 nm, respectively, presumably attributed to ZPL of the defects. Both emitters have broad features and asymmetric line shapes in the spectra. Similar quantum emitters in the visible wavelength region were reported in 2D *h*-BN[75]. For comparison, they also conducted experiments on BNNT samples with an average diameter of about 5 nm under the same condition, on different substrates[139]. It is much more difficult to find a quantum emitter and the emissions are very unstable under the 532 nm illumination. Most of them bleach within 10 seconds. This is consistent with former results[138] and is likely due to a much larger curvature in 5-nm-diameter BNNTs.

### 6.1.6. Wurtzite aluminum nitride

The wurtzite phase of aluminum nitride (w-AlN) has $C_{6v}$ symmetry. It is a wide band gap (6.03–6.12 eV) semiconductor that has many applications[372]. The very large band gap can suppresses coupling between band gap levels and bulk states. Moreover, small spin-orbit splitting of 19 meV was reported[373], which can enhance qubit-state lifetime. These two unique advantages make AlN an attractive host material for scalable solid-state qubits analogous to diamond and SiC. Furthermore, growth of high-quality crystal w-AlN has been reported[374]. The spin states of neutral nitrogen vacancy in AlN has been experimentally detected using electron paramagnetic resonance[375].

AlN has large band gap, this meets the criteria for a host material for addressable single photon emitter. However, in AlN, the defect levels induced by anion vacancy are too close to the band edge, results in strong resonance with the bulk band edges. This hinders the application of AlN in scalable quantum technologies. To overcome this issue, Varley *et al.* proposed that alloying AlN with transition-metal dopants can push the defect levels deeper into the band gap[376]. For example, Ti and Zr atoms substitute on the Al site can lead to the formation of desired electronic and spin states.

Seo *et al.* proposed an alternative by applying strain[377]. It was found that in the stress-free negatively charged nitrogen vacancy ($V_N^-$), the S = 1 state is slightly higher in energy than the S = 0 state, reveals that the two spin states are approximately degenerate in energy. On the other side, these two spin states are associated with two distinct Jahn-Teller distortions configurations. This suggests that these two spin states can be separated in energy by applying strain. Their DFT calculation results demonstrate this expectation, i.e., even at a small compressive strain of −3% along the [1120] direction, the S = 1 state is significantly lower by about 250 meV than the S = 0 state. In addition to the difference between energy of these two spin states, the hyperfine tensors are also sensitive to loading strain.

Very recently, Xue *et al.* have observed numerous room temperature single photon emitters in w-AlN films[378]. At low temperatures, the PL spectra exhibit relatively narrow and strong ZPL peaks with positions varying from the visible (543 nm) till the NIR (~980 nm) region. The full PL spectra thus the corresponding Debye-Waller factors have not reported. They also applied DFT calculations with such functionals that do not well reproduce the band gap or not well tested, and only the vertical excitation energies were determined[378]. Based on these calculations they concluded that some NIR quantum emitters in the AlN film originate from the antisite nitrogen vacancy complexes ($N_{Al}V_N$) and divacancy complexes ($V_{Al}V_N$). Further characterization is required at both the theoretical and experimental fronts in this highly prospective host material.

### 6.1.7. Rare-earth ions in garnets, silicate and vanadate

Rare-earth ions (REI) embedded into solid state matrix are prospective building blocks for quantum memory and quantum communication applications[379]. In particular, it has been demonstrated for an ensemble of europium doped into $Y_2SiO_5$ or YSO that the quantum information can be stored for six hours[380]. In REI systems, the *4f* electrons split in the low symmetry crystal field, and the optical transition basically originates from these split atomic states. The atomic-like states have the advantage of producing narrow optical emission lines, and the large ion has relatively large hyperfine fine structure in the ground state with the long coherence time. However, the disadvantage is the long optical lifetimes because of the optical transition dipole between the atomic states is often forbidden in the first order. This limitation has been partly circumvented in the case of $Pr^{3+}$ doped $Y_3Al_5O_{12}$ or YAG by using a upconversion process that has the mutual benefits of accessing a short-lived excited state and providing background-free optical images[160], which was the first demonstration of REI quantum emission at He flow temperatures. Quantum emission of $Pr^{3+}$ doped $LaF_3$ at 1.5 K was also demonstrated in a following study[164]. Later, spin-to-photon interface and single spin readout were observed in $Ce^{3+}$ doped YAG[161,381], where the optical transition occurs between the 5*d* excited state and 4*f* ground state. 4*f* states the spin is highly mixed with the orbital momentum. This enables spin-flip optical transitions between the 4*f* and the 5*d* levels with an optical lifetime of ~60 ns. Under dynamic decoupling, spin coherence lifetime reaches $T_2$ = 2 ms and is almost limited by the measured spin-lattice relaxation time $T_1$ = 4.5 ms. Another possibility to circumvent the intrinsically low emission of REI is to engineer them near high-quality optical resonator which can significantly increase the rate of spontaneous emission selectively at the resonator mode by harnessing the Purcell-effect. This has been recently achieved in $Er^{3+}$ doped YSO[382] and ytterbium doped $YVO_4$ or YVO[383] that made possible to detect single photon emission from these ions. Furthermore, by careful choice of the direction of an external constant magnetic field, the enhancement of the emission has been tuned towards the spin-conserving excitation only with well-distinguishable optical transitions for the different spin states[382,383]. If the laser and the cavity mode were tuned only one of the spin-conserving excitation energies then emission is expected only for the resonant bright spin state and no emission is expected for the off-resonant dark spin state. During the optical cycles a spontaneous decay from the bright spin state to the dark state can occur with some low probability that remains dark. Thus, the observation of the emission is correlated to the spin state of the REI in its ground state. Beside the optical readout of the REI spin state, spin dephasing and the spin dephasing limits in YVO have been observed[383].

Theory here faces with several challenges: highly correlated atomic like orbitals, large spin-orbit interaction and possibly non-negligible electron-phonon interaction for the case of quasi degenerate orbitals. One obvious choice is to apply post Hartree-Fock based methods where the correlation energy between the orbitals can be systematically taken into account by raising the complexity of the approach including relativistic effects (c.f. a review in Ref.[384]). The crystals are modeled by finite small clusters (10s of atoms like ultrasmall quantum dots) in the post Hartree-Fock based methods. To our knowledge, no systematic theoretical study has been conducted how the size of the clusters would affect the results. Alternatively, Kohn-Sham DFT theory may be used but it may fail due to the complex correlation effects of the atomic like orbitals. Recently, a computationally tractable solution has been developed that combines the HSE hybrid density functional theory with orbital dependent exchange interaction that was demonstrated for cerium dioxide[385]. Another solution can be to embed CI methods into DFT[203] or to very efficiently calculate the CI method in medium size clusters with periodic boundary conditions by means of density matrix renormalization group (DMRG) algorithms[179].

#### 6.1.8. Lithium fluoride

Using time-dependent density functional theory embedded-cluster simulations, the electronic and optical properties of two adjacent singlet-coupled F color centers (anion-vacancy) in lithium fluoride (LiF) were investigated[386], where LiF has extremely large band gap over 10 eV. It was found that to accurately simulate the entangled-defect system, it is necessary to consider the dynamical correlations between the defect electrons and the adjacent ionic lattice. To our knowledge, no single photon emitter has been observed so far in this insulator.

### 6.2. Potential moderate/small bandgap semiconductors

Diamond is already proven to be a promising host material for quantum information technology that builds up from carbon atoms. The graphite, including the single sheet of graphite, i.e., graphene, is a zero-gap semiconductor, therefore it cannot host color centers. However, some forms of carbon nanotubes with relatively small diameter can introduce small band gap which opens the door for{Citation} hosting NIR emitters. By considering small band gap materials for hosting qubits, silicon crystal is an obvious choice which is an elementary semiconductor. Silicon is the most used semiconductor for integrated circuits which has a band gap at 1.17 eV at cryogenic temperatures. Beside using dopant atoms for realizing qubits, i.e. Kane quantum computer (see below), NIR color centers have been introduced for quantum optics studies (see Table I). Considering its ideal interface compatibility with conventional Si-based technology, the application of silicon in quantum information technology has attracted a great attention in recent years. In this sub-section, we discuss the characteristic, performance and perspective of these two promising materials for quantum information technology.

#### 6.2.1. Carbon nanotube

The development of a solid state photon source based on carbon-related nano materials has received a lot of attention[387,388]. The photoluminescence of semiconducting single-wall carbon nanotubes was first observed in 2002[389]. The bandgap of carbon nanotubes can vary from zero to 2 eV, depending significantly on the tube chirality. The advantages of carbon nanotubes include the straightforward tuning of the emission wavelength, between 850 nm and 2 μm, by changing the chiral species. Unfortunately, in principle, single-photon emission requires a quantum mechanical quasi-two-level system, yet the one-dimensional (1D) band structure of carbon nanotubes conflicts with this major requirement.

One scheme towards a single photon source is the strong exciton–exciton interaction in 1D systems[390]. Via this strong exciton–exciton interaction, the multiple excitons created in a carbon nanotube can be annihilated until only a single exciton remains to emit a single photon, namely, the exciton–exciton annihilation effect. As it does not rely on exciton localization, room temperature single photon generation is possible. In 2015, room-temperature partial single-photon emission with purity of about 50% was observed[391].

Another strategy for obtaining reliable single-photon emission from carbon nanotubes relies on exciton localization. Deep potential trapping (much greater than $k_BT$) is required for localization at room temperature. However, because the trapping potential in a pristine carbon nanotube is very shallow, single photon emission in a pristine carbon nanotube is limited to cryogenic temperatures. Chemical functionalization of carbon nanotubes provides one way to strengthen the exciton localization and generate localized excitons with well depths of 100 meV or greater, which are higher than $k_BT$ at room temperature[392–395]. The physical mechanism is the strong exciton trapping at the $sp^3$ defect centers[393,396,397]. The zero-dimensional nature of trapped excitons also brings novel physical phenomenon, such as the interaction of 1D excitons with the 1D phonon modes in carbon nanotubes[396,397]. Furthermore, exciton localization can lead to significantly longer decay times, and with diameter decreases of (7, 5) to (5, 4) of the nanotubes, the decay time increases from 75 ps to 600 ps, demonstrating strong chirality dependence[398].

### 6.2.2. Silicon – Kane quantum computer

A completely different approach towards building quantum computers was proposed by Bruce Kane in 1998[399], which is based on the combination of classical semiconductor technology and electron-nuclear magnetic resonance techniques. As the most successful material applied in semiconductor device technology, silicon (Si) has an indirect bandgap of about 1.17 eV at cryogenic temperatures that reduces to about 1.12 eV at room temperature. The isotopically pure $^{28}$Si has a nuclear spin of 0 that can be a perfect host for defect qubit spins. When the silicon substrate is doped with isotopically pure $^{31}$P (phosphorus), with providing a donor electron with an electron spin and also nuclear spin of ½, the nuclear spin of the phosphorous donors can realize the function to encode qubits with extremely long coherence times and low error rates, because the donors are extremely separated from environment. This design combining the advantage of tunable quantum well for localizing the donor electron using gate electrodes – called quantum dots[400] – and nuclear magnetic resonance has the advantage of scalability[399]. In this system, also named the Kane quantum computer, the nuclear spins of the phosphorous donors perform single-qubit operations. Two nuclear spins can interact mediated by the extended electron wave packet. As the electron wave packet is sensitive to external electric field, the nuclear spin interaction can be controlled by gate voltage, which is essential for realizing the operation of quantum computation. With an array of such donors embedded beneath the surface of a pure silicon wafer, a silicon-based nuclear spin quantum computer was expected. Indeed, using electron-nuclear double resonance technique the electron spin could be successfully encoded into the nuclear spin of $^{31}$P with an overall store-readout fidelity of 90 per cent. The coherence lifetime of the quantum memory element at 5.5 K exceeded 1 s in $^{28}$Si enriched Si[401].

The idea of Kane quantum computer faces at least two challenges: i) efficient readout of single qubits; ii) precise alignment of the donor atoms in Si. By engineering the P donors between two electrodes, i.e. field-effect transistor (FET) configuration, the electrical detection of single P donor electron spins was demonstrated via spin-selective recombination between a deep paramagnetic defect acting as a trap for the P donor electron at the silicon and $SiO_2$ interface. This result was published[402] after 6 years of Kane's publication[399]. The mechanism can be described as a spin-to-charge conversion where the high purity of Si material makes possible to detect single electrons in the FET device. With using pulsed electrical detected magnetic resonance, the Rabi nutation of single P donor electrons (the coherent state) was observed in the Si FET device with using the same mechanism[403]. The disadvantage of this method is that the spin-to-charge conversion probability between the states of the deep defect (presumably a Si dangling bond at the $Si/SiO_2$ interface) and the P donor heavily depends on their locations, and it is extremely complicated to control the formation and location of the deep defect. A breakthrough in the single-shot readout of P donors was achieved in a single electron transistor (SET) architecture with using 1.5 Tesla magnetic field for efficient Zeeman splitting at 40 mK temperature, in order to achieve a spin-selective tunneling of the P donor electron from the SET region towards the electrodes[404,405]. Morello *at al.* observed a spin lifetime of 6 seconds at a magnetic field of 1.5 tesla, and achieved a spin readout fidelity better than 90 per cent[405]. Although, this method operates at ultralow temperatures compared to the spin-to-charge readout mechanism but here only the location of the P donor should be precisely controlled. Later, a top-gated nanostructure, fabricated on an isotopically engineered $^{28}$Si substrate, was used to increase the electron spin readout contrast to about 97 per cent and $^{31}$P nuclear spin readout contrast to 99.995 per cent[406]. Although, the control fidelity approached 99.99 per cent[406], the coherence time of the $^{31}$P nuclear spin can be much extended by removing the donor electron spin by illumination, i.e., charging the donor. It was demonstrated that the coherence time of the $^{31}$P nuclear spin is 39 minutes at room temperature and 3 hours at cryogenic temperatures[407]. We note that the control and readout of the qubit still occurs at low temperatures. These favorable qubit properties make feasible to study the scalability of the qubits and their interaction[408]. To this end, the tight control on the formation and placement of the donors is required.

A breakthrough in the precise alignment of P donors was achieved by a combination of scanning tunneling microscopy (STM) and hydrogen-resist lithography[409]. Simmons group demonstrated a single-atom transistor in which an individual phosphorus dopant atom was deterministically placed with a spatial accuracy of one lattice site in SET. The SET operated at liquid helium temperatures, and millikelvin electron transport measurements confirmed the presence of discrete quantum levels in the energy spectrum of the P atom[409]. Furthermore, the coherent control of single P donor qubit was demonstrated[410]. This technology has been recently applied to fabricate the two-qubit exchange gate between phosphorus donor electron spin qubits in silicon using independent single-shot spin readout with a readout fidelity of about 94 per cent which operates very fast (about 800 picosecond) [411]. In the two-qubit exchange mechanism the precise alignment of the two P donors was inevitable. This is a crucial step with respect to the previous breakthroughs achieved in the demonstration of various two-qubit gates in Si[411–415].

In addition to phosphorous dopant, arsenic in silicon is another attractive platform for quantum computing, since arsenic dopants have many advantages over phosphorus, include a higher solid solubility in bulk silicon and a lower diffusivity than phosphorus. Moreover, atomic spin−orbit interaction strength of arsenic is twice of that of phosphorus, and triple times of nuclear spin value[416]. The high nuclear spin value present opportunities for simplifications in physical implementations of quantum gate structures. Very recently, using a scanning tunneling microscope tip, Stock *et al.* reported the successful fabrication of atomic-precision of arsenic in silicon[417]. Obviously, further studies on the quantum properties of the arsenic in silicon are desired. Arsenic has an advantage of the relatively large quadrupole of the nuclear spins that can be harnessed to control the nuclear spin states of the ionized arsenic donors via strain[418]. Obviously, further studies on the qubit properties of the arsenic in silicon are desired. A very interesting donor alternative in Si is bismuth (Bi). Although, Bi has much lower solubility than P in Si but $^{209}$Bi offers a 20-dimensional Hilbert space rather than a 4-dimensional Hilbert space of $^{31}$P which has a great potential in quantum operations. It was experimentally demonstrated that Bi donor has similar coherence times like P donor's, and the nuclear spins can be coherently manipulated[419]. Notably, group-III shallow acceptors in silicon, e.g., boron, which have strong spin-orbit coupling, exhibit ultra-long coherence times of 10 ms that can rival the best electron-spin qubits[420]. The large spin-orbit coupling may help to manipulate the qubit with electric fields instead of magnetic fields which is technologically much friendlier.

We note that it was a common belief that the relatively shallow donors and acceptors in Si can be well understood by using the effective mass theory with some fitting parameters for the central cell correction and related empirical tight binding theory[421,422]. In particular, the electrostatic tuning the hyperfine interaction between the electron spin and the P donor nuclear spin in an FET[423] or the large valley-orbit splitting in a silicon gated nanowire device[424] could be well understood with this picture. *Ab initio* theory predicted that the former effect can be even significantly greater in a silicon nanowire where the physical dimension of the silicon nanowire is reduced to the quantum confinement regime[425] where the strain acting on the qubit depends on the diameter and surface termination of the Si nanowire.

Recently, strain was applied to various donors in Si to tune the hyperfine coupling between the corresponding nuclear spin and the donor spin[426]. Although, empirical tight binding theory seemed to mostly reproduce the experimental data but *ab initio* theory may be needed for accurate prediction, in particular, for the large donor ions. The challenge for *ab initio* periodic cluster calculations is to accommodate the defect (wavefunction) in a sufficiently large supercell, or to apply the appropriate scaling method to approach the isolated donor limit.

The spin-to-photon interface to create flying qubits in Si may be realized by deep donors like chalcogen donors[427]. For the prototypical $^{77}Se^{+}$ donor, lower bounds on the transition dipole moment and excited-state lifetime were measured with long-lived spin at cryogenic temperatures, enabling access to the strong coupling limit of cavity quantum electrodynamics using known silicon photonic resonator technology and integrated silicon photonics[428,429].

Another approach is to use deep fluorescent centers in Si to realize quantum emitters. Erbium in Si emits light in the telecom wavelength which is a REI impurity in Si with optical transition between the split *4f* orbitals (e.g., Ref.[107]). A hybrid approach was demonstrated for the readout of the electron spin in which optical excitation is used to change the charge state (conditional on its spin state) of an erbium defect center in a SET, and this change is then detected electrically[430]. Recently, ensembles of G-centers and G-center-related single photon emitters were created in silicon by carbon implantation and annealing[159,431]. A G-center mostly emits at 1280 nm (0.97 eV) wavelength in the near infrared region, as shown in Fig. 9, making it very compelling for quantum telecommunication[432]. The G-center has a known ODMR signal with an S = 1 metastable state with singlet ground and excited states (see Ref.[433] and references therein); thus, this quantum emitter may act as a qubit or quantum memory. A combined deep level transient spectroscopy, PL, EPR and ODMR spectroscopies study strongly argued that G-center is a bistable form of two carbon atoms sharing one Si site in the lattice[433], however, previous and recent *ab initio* calculations did not reach any satisfying consensus and conclusion about the microscopic origin of the center[434–437]. Additional *ab initio* studies are required for unambiguous identification and further characterization of the G-center in silicon. We note that the ZPL wavelength of C-center and W-center in silicon[438] are also very promising for quantum communication purposes (see Fig. 9). Further theoretical and experimental efforts are needed to harness these color centers in quantum technology applications.

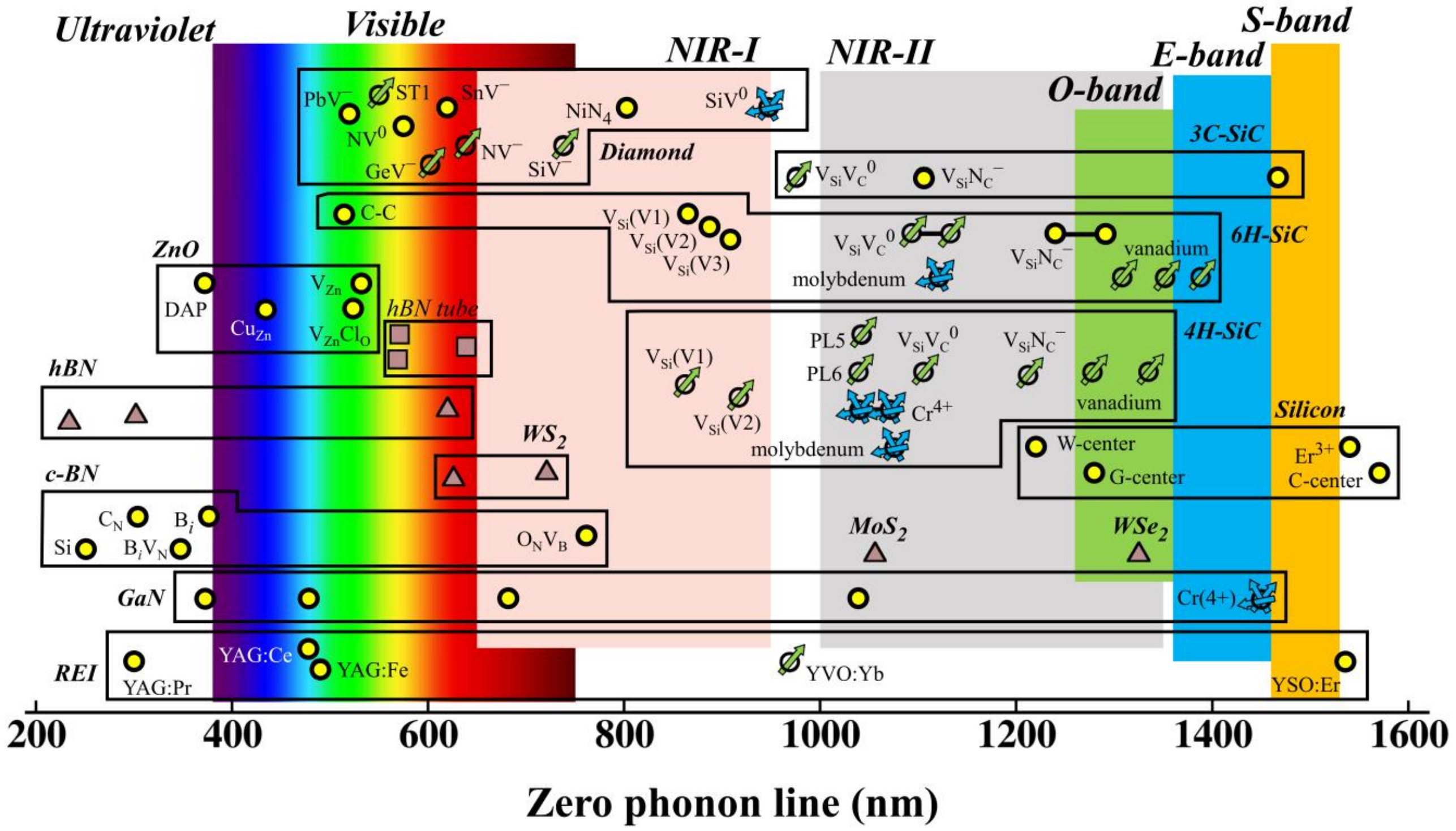

Figure 9. The plots of quantum-coherent materials versus zero-phonon line emission (unit in nm). In the region of near-infrared light (700 ~ 2500 nm), there are two windows presented as pink color (first biological window: 650 ~ 950 nm) and grey color (second biological window: 1000 ~ 1350 nm), that are important for *in vivo* imaging applications[439]. The three telecommunication operating wavelengths for fiber optic communication, namely O-band (original), E-band (extended), and S-band (short), are presented as green, blue, and orange colors, respectively. The ZPL values with unit in eV are shown in Table I. The color centers with unknown origin are not labeled. The symbol codes are the followings: 1D color center—purple square; 2D color center—purple triangle; 3D color center—yellow circle; coherent ensemble spins control—circle with blue arrows; coherent single spin control—circle with a single green arrow. In silicon, an oxygen related center, C-center, has a ZPL at 1570 nm and the corresponding DW factor is 0.100[438] beside a self-interstitials related color center, W-center, which has a ZPL energy at 1220 nm and the corresponding DW factor is ~0.40[157]. C-center and W-center in silicon were reported in neutron irradiated and annealed samples as ensembles[157], thus are not listed in Table I.

## 7. Summary and outlook

In this review, we have focused on color centers in solids, the emerging material platforms for quantum information technology with distinctly promising properties. A summary of the key parameters in experimental measurement for these promising materials, e.g., diamond, 4H-SiC, 3C-SiC, 6H-SiC, ZnO, GaN, *c*-BN, 2H-$MoS_2$, and for 2D materials, e.g., *h*-BN, $WSe_2$, $WS_2$, is given in Table I. Over the last decade, scientists have tried various materials, such as diamond, *h*-BN, SiC, and other wide bandgap semiconductors, for constructing the basic elements for quantum information processing devices[440]. Some startup companies have already begun to commercialize solid-state quantum systems, such as diamond NV centers, for quantum sensing[441]. Diamond NV center has a great potential to realize quantum computers operating at elevated temperatures when the technology is developed to imprint the qubits in a scalable 2D array arrangement. However, obviously, many challenges for industry-level application remain in any platform. Within the next few decades, one may expect that traditional and quantum computers will coexist, for each type has its own advantages[442], as predicted by Turing Award laureate Andrew Chi-Chih Yao. In 2016, The Quantum Manifesto[443] called upon Member States and the European Commission to launch a €1 billion Flagship-scale Initiative in Quantum Technology. Two years later, the National Science Foundation (USA) invested funds of up to $25 million in a new program, called Enabling Quantum Leap: Convergent Accelerated Discovery Foundries for Quantum Materials Science,

Engineering, and Information (Q-AMASE-i), to establish foundries with "mid-scale infrastructure for rapid prototyping and development of quantum materials and devices," according to the program solicitation[444]. Obviously, the above shows that the quantum computer will have a prominent part in the future. Although quantum computers may not supersede traditional computers in tasks for which traditional computers are already proven to be highly efficient, quantum computers could excel in many tasks, such as the design of new materials for health, energy storage and production, high temperature superconductors and new catalysts.

We briefly mentioned the challenges in this road in the paper that we perpetuate here. At limited temperature, the excitation of phonon will induce atoms in the solids displacing from their equilibrium positions. As a result, the electron/spin energy level have significant broadening because of the fluctuation of the atoms, leading to finite relaxation time for spin states. For example, significant reduction in spin coherence time was reported in isolated NV center at room temperature with respect to low temperature[46,195]. Furthermore, the excitation of phonons also may break time-reversal symmetry, result in a net magnetization in non-magnetic system[445]. In addition to its influence on spin coherence time, physicists also figured out how quantum computers can operate using coherent phonons[446–448]. In light of the aforementioned facts, therefore, we suggest that a better knowledge of spin-phonon coupling is highly important not only for understanding the fundamental physics, but also for the application of quantum technologies. The complex interplay between the phonons and spins can be monitored in such defect qubits in which the electron-phonon coupling and spin-orbit coupling are in the same order of magnitude[79]. Theory predicts that SnV and PbV defects in diamond are ideal platforms to this end[79].

Beside quantum computation, the development of quantum sensing and quantum communication devices is also a driving force in the field. In quantum sensor applications, the room temperature operation is a must for *in vivo* biological and medical studies, thus the search for room temperature defect qubits is of high importance in this context. Furthermore, the temperature dependence of materials properties such as high-temperature superconductivity can be only monitored by such quantum sensors that can operate in a broad range of temperatures. The use of light either in the manipulation or readout of the room temperature qubits poses restrictions on the wavelength region for biological studies. Two NIR wavelength regions are defined in the context where the absorption is minimal by typical biological systems (see NIR-I and NIR-II in Fig. 9 and Ref.[439]). Ideally, the color centers should be photo-excited and emit in the NIR-I region or rather in the NIR-II region to this end. We note that the phonon sideband of the emission may fall to the desired region which have longer wavelength than the marked ZPL wavelength in Fig. 9, as the optical readout of many color centers with low DW factor are carried out in the phonon sideband. The quest for the design of the optical emission is also evident for quantum communication applications. The efficiency of the quantum communication can be significantly increased if the critical ZPL emission of the color centers fall into the present telecommunication bands of the optical fibers (see Fig. 9), so it does not require any wavelength conversion that always leads to a loss of the signal intensity. The list of the presently observed single photon emitters and qubits in the context of technologically relevant wavelength regions are summarized in Fig. 9. It is obvious that the search for qubits with given fluorescence properties is required for optimization and efficient implementation of quantum technologies.

We note that the search of point defects for a given ZPL wavelength is an insufficient condition in the optimization of defect qubits. The interaction and coupling between different particles and quasi-particles in solids, including electrons, phonons, photons and spin, are important physical process in nature, and have direct impacts on performance of quantum devices. Considering the complex multi-particles interaction, it is still challenging in computational design of new material platform for quantum technology. Recently, with big data generated by theory and experiments, high-throughput calculation, screening and machine learning methodologies have been adopted to materials genome initiatives and materials informatics[449–451]. These methodologies have been exploited in discovery of various functional materials, including valleytronics materials[452], thermoelectric materials[453], thin film solar cell[454], organic-inorganic perovskites[455], and topological electronic materials[456]. Very recently, candidate materials for hosting defect qubits have been identified with the principles of constituting of the crystal by low-spin isotopes for securing long coherence times and of possessing >2 eV bandgap for robust optical control by means of machine learning techniques[457]. However, quantum-coherent materials are realized by point defects in the host that should be also identified. The first steps were taken to conduct this type of research to seek solid state defect qubit candidates[458] which is based on the computation of the key magneto-optical properties of the defects as list in Table I, i.e., database of defect properties based on the theoretical spectroscopy. Therefore, it is also expected that machine learning methodology can provide new insights and facilitate the development of new material platforms for quantum technology. A crucial issue that may stem to employ machine learning methodology to this end is the accuracy of the computation methods, in particular, for the excited states, which is critical in the credibility of the resulting database. The combination of the advantages of the density functional theory and configuration interaction seems to provide a good balance in terms of the tractable system size and the relatively good accuracy (about 0.1 eV) for highly correlated excited states[203].

It is exciting to note that, due to the enormous potential of quantum computers in the design of new materials and the importance of new material platforms for building scalable quantum computer networks with much better performance, the birth of the usable quantum computer will definitely speed up the development of new material platforms, resulting in "self-accelerating" progress in quantum information technology. The initial step has been recently taken in this direction[459]. In this context, theoretical simulation of experimental characterization is expected to play a more important role in future research.

Finally, we allude here an important point from Ref.[45]. We note that the full description of quantum bit cannot be separated from the description of the environment. The environment often is a source of noise for the qubits causing decoherence. This is clearly disadvantageous for quantum computer application but can be harnessed in quantum sensor protocols. We briefly mentioned some key quantities of the qubits such as readout contrasts, the longitudinal spin relaxation time, and the coherence time which are dependent on the environment such as strain, electric and magnetic fields, and temperature. By computing the coupling of the defect qubits properties to these key quantities, the sensitivity of quantum sensing protocols and optimization of quantum control can be designed and may guide future experimental studies. Only this comprehensive approach makes the *ab initio* search for alternative solid-state defect quantum bits reliable and powerful that might be superior for a given quantum technology application. One possible route along this direction is to combine the *ab initio* calculations and effective spin Hamiltonian approach to compute the readout contrasts, the longitudinal spin relaxation time, the coherence time, and other key quantities. A recent study has taken the first steps into this direction[460] which has been successfully applied to understand the optical response of qubits in a real environment[461]. Again, the usable quantum computer may result in a "self-accelerating" process by directly simulating the evolution of spin states of the defect qubit in the bath of external spins.

All-in-all, the future of color centers qubits looks bright in the joint effort of *ab initio* simulations – theoretical spectroscopy and qubit control – and experiments.

## Acknowledgements

A.G. acknowledges the support from the National Office of Research, Development and Innovation in Hungary for Quantum Technology Program (Grant No. 2017-1.2.1-NKP-2017-00001) and National Excellence Program (Grant No. KKP129866), and from the European Commission for the QuanTELCO project (Grant No. 862721). G.Z. and J.C. thank Dr. M. Sun for discussion and data collection.

## Data availability statement

Data sharing is not applicable to this article as no new data were created or analyzed in this study.

## References

[1] R.P. Feynman, International Journal of Theoretical Physics **21**, 467 (1982).

[2] A. Acín, I. Bloch, H. Buhrman, T. Calarco, C. Eichler, J. Eisert, D. Esteve, N. Gisin, S.J. Glaser, F. Jelezko, S. Kuhr, M. Lewenstein, M.F. Riedel, P.O. Schmidt, R. Thew, A. Wallraff, I. Walmsley, and F.K. Wilhelm, New J. Phys. **20**, 080201 (2018).

[3] D.P. DiVincenzo, Science **270**, 255 (1995).
[4] S. Lloyd, Science **273**, 1073 (1996).
[5] J.I. Cirac and P. Zoller, Science **301**, 176 (2003).
[6] L.-M. Duan and G.-C. Guo, Phys. Rev. Lett. **79**, 1953 (1997).
[7] J.-W. Pan, Z.-B. Chen, C.-Y. Lu, H. Weinfurter, A. Zeilinger, and M. Żukowski, Rev. Mod. Phys. **84**, 777 (2012).
[8] X. Rong, D. Lu, X. Kong, J. Geng, Y. Wang, F. Shi, C.-K. Duan, and J. Du, Advances in Physics: X **2**, 125 (2017).
[9] W.K. Wootters and W.H. Zurek, Nature **299**, 802 (1982).
[10] N. Gisin and R. Thew, Nature Photonics **1**, 165 (2007).
[11] D. Castelvecchi, Nature News **543**, 159 (2017).
[12] F. Arute, K. Arya, R. Babbush, D. Bacon, J.C. Bardin, R. Barends, R. Biswas, S. Boixo, F.G.S.L. Brandao, D.A. Buell, B. Burkett, Y. Chen, Z. Chen, B. Chiaro, R. Collins, W. Courtney, A. Dunsworth, E. Farhi, B. Foxen, A. Fowler, C. Gidney, M. Giustina, R. Graff, K. Guerin, S. Habegger, M.P. Harrigan, M.J. Hartmann, A. Ho, M. Hoffmann, T. Huang, T.S. Humble, S.V. Isakov, E. Jeffrey, Z. Jiang, D. Kafri, K. Kechedzhi, J. Kelly, P.V. Klimov, S. Knysh, A. Korotkov, F. Kostritsa, D. Landhuis, M. Lindmark, E. Lucero, D. Lyakh, S. Mandrà, J.R. McClean, M. McEwen, A. Megrant, X. Mi, K. Michielsen, M. Mohseni, J. Mutus, O. Naaman, M. Neeley, C. Neill, M.Y. Niu, E. Ostby, A. Petukhov, J.C. Platt, C. Quintana, E.G. Rieffel, P. Roushan, N.C. Rubin, D. Sank, K.J. Satzinger, V. Smelyanskiy, K.J. Sung, M.D. Trevithick, A. Vainsencher, B. Villalonga, T. White, Z.J. Yao, P. Yeh, A. Zalcman, H. Neven, and J.M. Martinis, Nature **574**, 505 (2019).
[13] M. Freedman, A. Kitaev, M. Larsen, and Z. Wang, Bull. Amer. Math. Soc. **40**, 31 (2003).
[14] B. Field and T. Simula, Quantum Science and Technology **3**, 045004 (2018).
[15] S. Charpentier, L. Galletti, G. Kunakova, R. Arpaia, Y. Song, R. Baghdadi, S.M. Wang, A. Kalaboukhov, E. Olsson, F. Tafuri, D. Golubev, J. Linder, T. Bauch, and F. Lombardi, Nature Communications **8**, (2017).
[16] P. Zhang, K. Yaji, T. Hashimoto, Y. Ota, T. Kondo, K. Okazaki, Z. Wang, J. Wen, G.D. Gu, H. Ding, and S. Shin, Science **360**, 182 (2018).
[17] M. Greiner and S. Fölling, Nature **453**, 736 (2008).
[18] D. Hanneke, J.P. Home, J.D. Jost, J.M. Amini, D. Leibfried, and D.J. Wineland, Nature Physics **6**, 13 (2010).
[19] B.P. Lanyon, C. Hempel, D. Nigg, M. Muller, R. Gerritsma, F. Zahringer, P. Schindler, J.T. Barreiro, M. Rambach, G. Kirchmair, M. Hennrich, P. Zoller, R. Blatt, and C.F. Roos, Science **334**, 57 (2011).
[20] X. Peng, J. Zhang, J. Du, and D. Suter, Phys. Rev. Lett. **103**, 140501 (2009).
[21] X. Peng and D. Suter, Front. Phys. China **5**, 1 (2010).
[22] Z. Li, M.-H. Yung, H. Chen, D. Lu, J.D. Whitfield, X. Peng, A. Aspuru-Guzik, and J. Du, Scientific Reports **1**, 88 (2011).
[23] J.Q. You and F. Nori, Physics Today **58**, 42 (2005).
[24] J. Clarke and F.K. Wilhelm, Nature **453**, 1031 (2008).
[25] J.Q. You and F. Nori, Nature **474**, 589 (2011).
[26] A.A. Houck, H.E. Türeci, and J. Koch, Nature Physics **8**, 292 (2012).
[27] P.I. Bunyk, E.M. Hoskinson, M.W. Johnson, E. Tolkacheva, F. Altomare, A.J. Berkley, R. Harris, J.P. Hilton, T. Lanting, A.J. Przybysz, and J. Whittaker, IEEE Trans. Appl. Supercond. **24**, 1 (2014).
[28] T.F. Rønnow, Z. Wang, J. Job, S. Boixo, S.V. Isakov, D. Wecker, J.M. Martinis, D.A. Lidar, and M. Troyer, Science **345**, 420 (2014).
[29] M. Greiner, O. Mandel, T. Esslinger, T.W. Hänsch, and I. Bloch, Nature **415**, 39 (2002).
[30] D. Leibfried, B. DeMarco, V. Meyer, M. Rowe, A. Ben-Kish, J. Britton, W.M. Itano, B. Jelenković, C. Langer, T. Rosenband, and D.J. Wineland, Phys. Rev. Lett. **89**, 247901 (2002).
[31] A. Friedenauer, H. Schmitz, J.T. Glueckert, D. Porras, and T. Schaetz, Nature Physics **4**, 757 (2008).
[32] M. Neeley, M. Ansmann, R.C. Bialczak, M. Hofheinz, E. Lucero, A.D. O'Connell, D. Sank, H. Wang, J. Wenner, A.N. Cleland, M.R. Geller, and J.M. Martinis, Science **325**, 722 (2009).
[33] R. Gerritsma, G. Kirchmair, F. Zähringer, E. Solano, R. Blatt, and C.F. Roos, Nature **463**, 68 (2010).
[34] K. Kim, M.-S. Chang, S. Korenblit, R. Islam, E.E. Edwards, J.K. Freericks, G.-D. Lin, L.-M. Duan, and C. Monroe, Nature **465**, 590 (2010).
[35] B.P. Lanyon, J.D. Whitfield, G.G. Gillett, M.E. Goggin, M.P. Almeida, I. Kassal, J.D. Biamonte, M. Mohseni, B.J. Powell, M. Barbieri, A. Aspuru-Guzik, and A.G. White, Nature Chemistry **2**, 106 (2010).
[36] R. Hanson and D.D. Awschalom, Nature **453**, 1043 (2008).

[37] T.D. Ladd, F. Jelezko, R. Laflamme, Y. Nakamura, C. Monroe, and J.L. O'Brien, Nature **464**, 45 (2010).
[38] I. Buluta, S. Ashhab, and F. Nori, Rep. Prog. Phys. **74**, 104401 (2011).
[39] I. Buluta and F. Nori, Science **326**, 108 (2009).
[40] D. Jaksch and P. Zoller, Annals of Physics **315**, 52 (2005).
[41] I. Bloch, J. Dalibard, and S. Nascimbène, Nature Physics **8**, 267 (2012).
[42] R. Blatt and C.F. Roos, Nature Physics **8**, 277 (2012).
[43] C. Schneider, D. Porras, and T. Schaetz, Rep. Prog. Phys. **75**, 024401 (2012).
[44] M.W. Doherty, N.B. Manson, P. Delaney, F. Jelezko, J. Wrachtrup, and L.C.L. Hollenberg, Physics Reports **528**, 1 (2013).
[45] Á. Gali, Nanophotonics **8**, 2192 (2019).
[46] G. Balasubramanian, P. Neumann, D. Twitchen, M. Markham, R. Kolesov, N. Mizuochi, J. Isoya, J. Achard, J. Beck, J. Tissler, V. Jacques, P.R. Hemmer, F. Jelezko, and J. Wrachtrup, Nat Mater **8**, 383 (2009).
[47] P.C. Maurer, G. Kucsko, C. Latta, L. Jiang, N.Y. Yao, S.D. Bennett, F. Pastawski, D. Hunger, N. Chisholm, M. Markham, D.J. Twitchen, J.I. Cirac, and M.D. Lukin, Science **336**, 1283 (2012).
[48] G.D. Fuchs, G. Burkard, P.V. Klimov, and D.D. Awschalom, Nature Physics **7**, 789 (2011).
[49] S. Yang, Y. Wang, D.D.B. Rao, T. Hien Tran, A.S. Momenzadeh, M. Markham, D.J. Twitchen, P. Wang, W. Yang, R. Stöhr, P. Neumann, H. Kosaka, and J. Wrachtrup, Nature Photonics **10**, 507 (2016).
[50] G. Waldherr, Y. Wang, S. Zaiser, M. Jamali, T. Schulte-Herbrüggen, H. Abe, T. Ohshima, J. Isoya, J.F. Du, P. Neumann, and J. Wrachtrup, Nature **506**, 204 (2014).
[51] P. Ovartchaiyapong, K.W. Lee, B.A. Myers, and A.C.B. Jayich, Nature Communications **5**, 4429 (2014).
[52] J.M. Taylor, P. Cappellaro, L. Childress, L. Jiang, D. Budker, P.R. Hemmer, A. Yacoby, R. Walsworth, and M.D. Lukin, Nat Phys **4**, 810 (2008).
[53] J.R. Maze, P.L. Stanwix, J.S. Hodges, S. Hong, J.M. Taylor, P. Cappellaro, L. Jiang, M.V.G. Dutt, E. Togan, A.S. Zibrov, A. Yacoby, R.L. Walsworth, and M.D. Lukin, Nature **455**, 644 (2008).
[54] F. Dolde, H. Fedder, M.W. Doherty, T. Nöbauer, F. Rempp, G. Balasubramanian, T. Wolf, F. Reinhard, L.C.L. Hollenberg, F. Jelezko, and J. Wrachtrup, Nature Physics **7**, 459 (2011).
[55] D.M. Toyli, C.F. de las Casas, D.J. Christle, V.V. Dobrovitski, and D.D. Awschalom, PNAS **110**, 8417 (2013).
[56] V.M. Acosta, E. Bauch, M.P. Ledbetter, A. Waxman, L.-S. Bouchard, and D. Budker, Phys. Rev. Lett. **104**, 070801 (2010).
[57] P. Neumann, I. Jakobi, F. Dolde, C. Burk, R. Reuter, G. Waldherr, J. Honert, T. Wolf, A. Brunner, J.H. Shim, D. Suter, H. Sumiya, J. Isoya, and J. Wrachtrup, Nano Letters **13**, 2738 (2013).
[58] F. Jelezko, T. Gaebel, I. Popa, M. Domhan, A. Gruber, and J. Wrachtrup, Phys. Rev. Lett. **93**, 130501 (2004).
[59] F. Jelezko and J. Wrachtrup, J. Phys.: Condens. Matter **16**, R1089 (2004).
[60] F. Jelezko and J. Wrachtrup, Phys. Stat. Sol. (a) **203**, 3207 (2006).
[61] J.R. Weber, W.F. Koehl, J.B. Varley, A. Janotti, B.B. Buckley, C.G.V. de Walle, and D.D. Awschalom, PNAS **107**, 8513 (2010).
[62] I. Aharonovich, S. Castelletto, D.A. Simpson, C.-H. Su, A.D. Greentree, and S. Prawer, Rep. Prog. Phys. **74**, 076501 (2011).
[63] L. Childress, R. Walsworth, and M. Lukin, Physics Today **67**, 38 (2014).
[64] L. Rondin, J.-P. Tetienne, T. Hingant, J.-F. Roch, P. Maletinsky, and V. Jacques, Rep. Prog. Phys. **77**, 056503 (2014).
[65] B.M. Maune, M.G. Borselli, B. Huang, T.D. Ladd, P.W. Deelman, K.S. Holabird, A.A. Kiselev, I. Alvarado-Rodriguez, R.S. Ross, A.E. Schmitz, M. Sokolich, C.A. Watson, M.F. Gyure, and A.T. Hunter, Nature **481**, 344 (2012).
[66] D.B. Szombati, S. Nadj-Perge, D. Car, S.R. Plissard, E.P. a. M. Bakkers, and L.P. Kouwenhoven, Nature Physics **12**, 568 (2016).
[67] H. Bluhm, S. Foletti, I. Neder, M. Rudner, D. Mahalu, V. Umansky, and A. Yacoby, Nature Physics **7**, 109 (2011).
[68] J.W. Britton, B.C. Sawyer, A.C. Keith, C.-C.J. Wang, J.K. Freericks, H. Uys, M.J. Biercuk, and J.J. Bollinger, Nature **484**, 489 (2012).
[69] J. Benhelm, G. Kirchmair, C.F. Roos, and R. Blatt, Nature Physics **4**, 463 (2008).
[70] A. Das, Y. Ronen, M. Heiblum, D. Mahalu, A.V. Kretinin, and H. Shtrikman, Nature Communications **3**, (2012).

[71] L.G. Herrmann, F. Portier, P. Roche, A.L. Yeyati, T. Kontos, and C. Strunk, Physical Review Letters **104**, (2010).
[72] E. Togan, Y. Chu, A.S. Trifonov, L. Jiang, J. Maze, L. Childress, M.V.G. Dutt, A.S. Sørensen, P.R. Hemmer, A.S. Zibrov, and M.D. Lukin, Nature **466**, 730 (2010).
[73] A.L. Falk, P.V. Klimov, B.B. Buckley, V. Ivády, I.A. Abrikosov, G. Calusine, W.F. Koehl, Á. Gali, and D.D. Awschalom, Phys. Rev. Lett. **112**, 187601 (2014).
[74] H. Zheng, A. Weismann, and R. Berndt, Nature Communications **5**, (2014).
[75] T.T. Tran, K. Bray, M.J. Ford, M. Toth, and I. Aharonovich, Nature Nanotechnology **11**, 37 (2016).
[76] J.H.N. Loubser and J.A. van Wyk, Rep. Prog. Phys. **41**, 1201 (1978).
[77] P. Siyushev, H. Pinto, M. Vörös, A. Gali, F. Jelezko, and J. Wrachtrup, Phys. Rev. Lett. **110**, 167402 (2013).
[78] A. Gali, Physica Status Solidi (b) **248**, 1337 (2011).
[79] G. Thiering and A. Gali, Physical Review X **8**, 021063 (2018).
[80] P. Narang, C.J. Ciccarino, J. Flick, and D. Englund, Adv. Funct. Mater. **29**, 1904557 (2019).
[81] N.T. Son, C.P. Anderson, A. Bourassa, K.C. Miao, C. Babin, M. Widmann, M. Niethammer, J. Ul Hassan, N. Morioka, I.G. Ivanov, F. Kaiser, J. Wrachtrup, and D.D. Awschalom, Appl. Phys. Lett. **116**, 190501 (2020).
[82] S. Castelletto and A. Boretti, J. Phys. Photonics **2**, 022001 (2020).
[83] A. Sajid, M.J. Ford, and J.R. Reimers, Rep. Prog. Phys. **83**, 044501 (2020).
[84] S. Castelletto, F.A. Inam, S. Sato, and A. Boretti, Beilstein J. Nanotechnol. **11**, 740 (2020).
[85] D.D. Awschalom, R. Hanson, J. Wrachtrup, and B.B. Zhou, Nature Photonics **12**, 516 (2018).
[86] C. Freysoldt, B. Grabowski, T. Hickel, J. Neugebauer, G. Kresse, A. Janotti, and C.G. Van de Walle, Rev. Mod. Phys. **86**, 253 (2014).
[87] C.E. Dreyer, A. Alkauskas, J.L. Lyons, A. Janotti, and C.G. Van de Walle, Annual Review of Materials Research **48**, 1 (2018).
[88] J.C. Phillips, Phys. Rev. **112**, 685 (1958).
[89] P.E. Blöchl, Phys. Rev. B **50**, 17953 (1994).
[90] S. Lany and A. Zunger, Phys. Rev. B **78**, 235104 (2008).
[91] G. Makov and M.C. Payne, Phys. Rev. B **51**, 4014 (1995).
[92] C. Freysoldt, J. Neugebauer, and C.G. Van de Walle, Phys. Rev. Lett. **102**, 016402 (2009).
[93] H.-P. Komsa, N. Berseneva, A.V. Krasheninnikov, and R.M. Nieminen, Phys. Rev. X **4**, 031044 (2014).
[94] D. Wang, D. Han, X.-B. Li, S.-Y. Xie, N.-K. Chen, W.Q. Tian, D. West, H.-B. Sun, and S.B. Zhang, Phys. Rev. Lett. **114**, 196801 (2015).
[95] M. Kaviani, P. Deák, B. Aradi, T. Frauenheim, J.-P. Chou, and A. Gali, Nano Lett. **14**, 4772 (2014).
[96] J.-P. Chou and A. Gali, MRS Communications **7**, 551 (2017).
[97] S.B. Zhang and J.E. Northrup, Phys. Rev. Lett. **67**, 2339 (1991).
[98] L. Gordon, A. Janotti, and C.G. Van de Walle, Phys. Rev. B **92**, 045208 (2015).
[99] G. Wolfowicz, C.P. Anderson, A.L. Yeats, S.J. Whiteley, J. Niklas, O.G. Poluektov, F.J. Heremans, and D.D. Awschalom, Nature Communications **8**, 1876 (2017).
[100] V. Ivády, I.A. Abrikosov, and A. Gali, Npj Comput Mater **4**, 76 (2018).
[101] Huang Kun, Rhys Avril, and Mott Nevill Francis, Proceedings of the Royal Society of London. Series A. Mathematical and Physical Sciences **204**, 406 (1950).
[102] A. Gali, E. Janzén, P. Deák, G. Kresse, and E. Kaxiras, Phys. Rev. Lett. **103**, 186404 (2009).
[103] A. Alkauskas, B.B. Buckley, D.D. Awschalom, and C.G.V. de Walle, New J. Phys. **16**, 073026 (2014).
[104] A. Gali, T. Demján, M. Vörös, G. Thiering, E. Cannuccia, and A. Marini, Nature Communications **7**, 11327 (2016).
[105] C. Monroe, Nature **416**, 238 (2002).
[106] M.H. Devoret and J.M. Martinis, in *Experimental Aspects of Quantum Computing*, edited by H.O. Everitt (Springer US, Boston, MA, 2005), pp. 163–203.
[107] S. Bertaina, S. Gambarelli, A. Tkachuk, I.N. Kurkin, B. Malkin, A. Stepanov, and B. Barbara, Nature Nanotechnology **2**, 39 (2007).
[108] A. Gruber, A. Dräbenstedt, C. Tietz, L. Fleury, J. Wrachtrup, and C. von Borczyskowski, Science **276**, 2012 (1997).
[109] G. Davies, J. Phys. C: Solid State Phys. **12**, 2551 (1979).

[110] Davies G., Hamer M. F., and Price William Charles, Proceedings of the Royal Society of London. A. Mathematical and Physical Sciences **348**, 285 (1976).
[111] U.F.S. D'Haenens-Johansson, A.M. Edmonds, B.L. Green, M.E. Newton, G. Davies, P.M. Martineau, R.U.A. Khan, and D.J. Twitchen, Phys. Rev. B **84**, 245208 (2011).
[112] T. Feng and B.D. Schwartz, Journal of Applied Physics **73**, 1415 (1993).
[113] Y.N. Palyanov, I.N. Kupriyanov, Y.M. Borzdov, and N.V. Surovtsev, Scientific Reports **5**, 14789 (2015).
[114] T. Iwasaki, Y. Miyamoto, T. Taniguchi, P. Siyushev, M.H. Metsch, F. Jelezko, and M. Hatano, Phys. Rev. Lett. **119**, 253601 (2017).
[115] J. Wrachtrup and F. Jelezko, Journal of Physics: Condensed Matter **18**, S807 (2006).
[116] M.E. Trusheim, N.H. Wan, K.C. Chen, C.J. Ciccarino, J. Flick, R. Sundararaman, G. Malladi, E. Bersin, M. Walsh, B. Lienhard, H. Bakhru, P. Narang, and D. Englund, Phys. Rev. B **99**, 075430 (2019).
[117] S.-Y. Lee, M. Widmann, T. Rendler, M.W. Doherty, T.M. Babinec, S. Yang, M. Eyer, P. Siyushev, B.J.M. Hausmann, M. Loncar, Z. Bodrog, A. Gali, N.B. Manson, H. Fedder, and J. Wrachtrup, Nat Nano **8**, 487 (2013).
[118] G.D. Fuchs, V.V. Dobrovitski, R. Hanson, A. Batra, C.D. Weis, T. Schenkel, and D.D. Awschalom, Phys. Rev. Lett. **101**, 117601 (2008).
[119] A.M. Edmonds, M.E. Newton, P.M. Martineau, D.J. Twitchen, and S.D. Williams, Phys. Rev. B **77**, 245205 (2008).
[120] C. Hepp, T. Müller, V. Waselowski, J.N. Becker, B. Pingault, H. Sternschulte, D. Steinmüller-Nethl, A. Gali, J.R. Maze, M. Atatüre, and C. Becher, Phys. Rev. Lett. **112**, 036405 (2014).
[121] G. Davies, Rep. Prog. Phys. **44**, 787 (1981).
[122] B.C. Rose, D. Huang, Z.-H. Zhang, P. Stevenson, A.M. Tyryshkin, S. Sangtawesin, S. Srinivasan, L. Loudin, M.L. Markham, A.M. Edmonds, D.J. Twitchen, S.A. Lyon, and N.P. de Leon, Science **361**, 60 (2018).
[123] E. Neu, D. Steinmetz, J. Riedrich-Möller, S. Gsell, M. Fischer, M. Schreck, and C. Becher, New J. Phys. **13**, 025012 (2011).
[124] A.L. Falk, B.B. Buckley, G. Calusine, W.F. Koehl, V.V. Dobrovitski, A. Politi, C.A. Zorman, P.X.-L. Feng, and D.D. Awschalom, Nature Communications **4**, 1819 (2013).
[125] W.F. Koehl, B.B. Buckley, F.J. Heremans, G. Calusine, and D.D. Awschalom, Nature **479**, 84 (2011).
[126] S.A. Zargaleh, B. Eble, S. Hameau, J.-L. Cantin, L. Legrand, M. Bernard, F. Margaillan, J.-S. Lauret, J.-F. Roch, H.J. von Bardeleben, E. Rauls, U. Gerstmann, and F. Treussart, Phys. Rev. B **94**, 060102 (2016).
[127] H.J. von Bardeleben, J.L. Cantin, A. Csóré, A. Gali, E. Rauls, and U. Gerstmann, Phys. Rev. B **94**, 121202 (2016).
[128] A. Csóré, H.J. von Bardeleben, J.L. Cantin, and A. Gali, Phys. Rev. B **96**, 085204 (2017).
[129] E. Sörman, N.T. Son, W.M. Chen, O. Kordina, C. Hallin, and E. Janzén, Phys. Rev. B **61**, 2613 (2000).
[130] W.F. Koehl, B. Diler, S.J. Whiteley, A. Bourassa, N.T. Son, E. Janzén, and D.D. Awschalom, Phys. Rev. B **95**, 035207 (2017).
[131] G. Wolfowicz, C.P. Anderson, B. Diler, O.G. Poluektov, F.J. Heremans, and D.D. Awschalom, Science Advances **6**, eaaz1192 (2020).
[132] D.J. Christle, P.V. Klimov, C.F. de las Casas, K. Szász, V. Ivády, V. Jokubavicius, J. Ul Hassan, M. Syväjärvi, W.F. Koehl, T. Ohshima, N.T. Son, E. Janzén, Á. Gali, and D.D. Awschalom, Physical Review X **7**, (2017).
[133] P. Udvarhelyi, G. Thiering, N. Morioka, C. Babin, F. Kaiser, D. Lukin, T. Ohshima, J. Ul-Hassan, N.T. Son, J. Vučković, J. Wrachtrup, and A. Gali, Phys. Rev. Applied **13**, 054017 (2020).
[134] S.A. Zargaleh, S. Hameau, B. Eble, F. Margaillan, H.J. von Bardeleben, J.L. Cantin, and W. Gao, Phys. Rev. B **98**, 165203 (2018).
[135] J. Wang, Y. Zhou, Z. Wang, A. Rasmita, J. Yang, X. Li, H.J. von Bardeleben, and W. Gao, Nature Communications **9**, 4106 (2018).
[136] A.L. Falk, P.V. Klimov, V. Ivády, K. Szász, D.J. Christle, W.F. Koehl, Á. Gali, and D.D. Awschalom, Phys. Rev. Lett. **114**, 247603 (2015).
[137] H. Kraus, V.A. Soltamov, F. Fuchs, D. Simin, A. Sperlich, P.G. Baranov, G.V. Astakhov, and V. Dyakonov, Scientific Reports **4**, 5303 (2014).
[138] N. Chejanovsky, Y. Kim, A. Zappe, B. Stuhlhofer, T. Taniguchi, K. Watanabe, D. Dasari, A. Finkler, J.H. Smet, and J. Wrachtrup, Scientific Reports **7**, (2017).
[139] J. Ahn, Z. Xu, J. Bang, A.E.L. Allcca, Y.P. Chen, and T. Li, Opt. Lett. **43**, 3778 (2018).

[140] A. Gottscholl, M. Kianinia, V. Soltamov, S. Orlinskii, G. Mamin, C. Bradac, C. Kasper, K. Krambrock, A. Sperlich, M. Toth, I. Aharonovich, and V. Dyakonov, Nature Materials **19**, 540 (2020).
[141] O. Ari, V. Fırat, N. Polat, O. Cakir, and S. Ates, in *Quantum Information and Measurement (QIM) V: Quantum Technologies* (OSA, Rome, 2019), p. S4A.2.
[142] S. Kumar, M. Brotóns-Gisbert, R. Al-Khuzheyri, A. Branny, G. Ballesteros-Garcia, J.F. Sánchez-Royo, and B.D. Gerardot, Optica **3**, 882 (2016).
[143] B. Schuler, D.Y. Qiu, S. Refaely-Abramson, C. Kastl, C.T. Chen, S. Barja, R.J. Koch, D.F. Ogletree, S. Aloni, A.M. Schwartzberg, J.B. Neaton, S.G. Louie, and A. Weber-Bargioni, Phys. Rev. Lett. **123**, 076801 (2019).
[144] L. Kulyuk, L. Charron, and E. Fortin, Phys. Rev. B **68**, 075314 (2003).
[145] V.D. Tkachev, V.B. Shepilo, and A.M. Zaitsev, Physica Status Solidi (b) **127**, K65 (1985).
[146] E.M. Shishonok and J.W. Steeds, Diamond and Related Materials **11**, 1774 (2002).
[147] E.M. Shishonok, J Appl Spectrosc **74**, 272 (2007).
[148] A. Tararan, S. di Sabatino, M. Gatti, T. Taniguchi, K. Watanabe, L. Reining, L.H.G. Tizei, M. Kociak, and A. Zobelli, Physical Review B **98**, (2018).
[149] A.J. Morfa, B.C. Gibson, M. Karg, T.J. Karle, A.D. Greentree, P. Mulvaney, and S. Tomljenovic-Hanic, Nano Letters **12**, 949 (2012).
[150] R. Dingle, Phys. Rev. Lett. **23**, 579 (1969).
[151] D.C. Reynolds, D.C. Look, B. Jogai, J.E. Van Nostrand, R. Jones, and J. Jenny, Solid State Communications **106**, 701 (1998).
[152] B.K. Meyer, H. Alves, D.M. Hofmann, W. Kriegseis, D. Forster, F. Bertram, J. Christen, A. Hoffmann, M. Straßburg, M. Dworzak, U. Haboeck, and A.V. Rodina, Physica Status Solidi (b) **241**, 231 (2004).
[153] R. Kuhnert and R. Helbig, Journal of Luminescence **26**, 203 (1981).
[154] M.R. Wagner, G. Callsen, J.S. Reparaz, J.-H. Schulze, R. Kirste, M. Cobet, I.A. Ostapenko, S. Rodt, C. Nenstiel, M. Kaiser, A. Hoffmann, A.V. Rodina, M.R. Phillips, S. Lautenschläger, S. Eisermann, and B.K. Meyer, Physical Review B **84**, (2011).
[155] Y. Zhou, Z. Wang, A. Rasmita, S. Kim, A. Berhane, Z. Bodrog, G. Adamo, A. Gali, I. Aharonovich, and W. Gao, Science Advances **4**, eaar3580 (2018).
[156] A. Kazimirov, N. Faleev, H. Temkin, M.J. Bedzyk, V. Dmitriev, and Yu. Melnik, Journal of Applied Physics **89**, 6092 (2001).
[157] G. Davies, Physics Reports **176**, 83 (1989).
[158] B. Zheng, J. Michel, F.Y.G. Ren, L.C. Kimerling, D.C. Jacobson, and J.M. Poate, Appl. Phys. Lett. **64**, 2842 (1994).
[159] W. Redjem, A. Durand, T. Herzig, A. Benali, S. Pezzagna, J. Meijer, A.Y. Kuznetsov, H.S. Nguyen, S. Cueff, J.-M. Gérard, I. Robert-Philip, B. Gil, D. Caliste, P. Pochet, M. Abbarchi, V. Jacques, A. Dréau, and G. Cassabois, ArXiv:2001.02136 [Cond-Mat, Physics:Physics, Physics:Quant-Ph] (2020).
[160] R. Kolesov, K. Xia, R. Reuter, R. Stöhr, A. Zappe, J. Meijer, P.R. Hemmer, and J. Wrachtrup, Nat Commun **3**, 1 (2012).
[161] P. Siyushev, K. Xia, R. Reuter, M. Jamali, N. Zhao, N. Yang, C. Duan, N. Kukharchyk, A.D. Wieck, R. Kolesov, and J. Wrachtrup, Nature Communications **5**, 1 (2014).
[162] B. Car, L. Veissier, A. Louchet-Chauvet, J.-L. Le Gouët, and T. Chanelière, Phys. Rev. Lett. **120**, 197401 (2018).
[163] M. De Vido, A. Wojtusiak, and K. Ertel, Opt. Mater. Express **10**, 717 (2020).
[164] I. Nakamura, T. Yoshihiro, H. Inagawa, S. Fujiyoshi, and M. Matsushita, Scientific Reports **4**, 7364 (2014).
[165] V. Bachmann, C. Ronda, and A. Meijerink, Chem. Mater. **21**, 2077 (2009).
[166] D.O. Demchenko, I.C. Diallo, and M.A. Reshchikov, Journal of Applied Physics **119**, 035702 (2016).
[167] M.A. Reshchikov, M. Vorobiov, D.O. Demchenko, Ü. Özgür, H. Morkoç, A. Lesnik, M.P. Hoffmann, F. Hörich, A. Dadgar, and A. Strittmatter, Phys. Rev. B **98**, 125207 (2018).
[168] J.L. Lyons, A. Janotti, and C.G. Van de Walle, Appl. Phys. Lett. **97**, 152108 (2010).
[169] J.L. Lyons, A. Alkauskas, A. Janotti, and C.G.V. de Walle, Physica Status Solidi (b) **252**, 900 (2015).
[170] D.O. Demchenko, I.C. Diallo, and M.A. Reshchikov, Phys. Rev. Lett. **110**, 087404 (2013).
[171] M.A. Reshchikov, J.D. McNamara, F. Zhang, M. Monavarian, A. Usikov, H. Helava, Yu. Makarov, and H. Morkoç, Physical Review B **94**, (2016).

[172] A.M. Berhane, K.-Y. Jeong, Z. Bodrog, S. Fiedler, T. Schröder, N.V. Triviño, T. Palacios, A. Gali, M. Toth, D. Englund, and I. Aharonovich, Advanced Materials **29**, 1605092 (2017).
[173] S.A. Tawfik, S. Ali, M. Fronzi, M. Kianinia, T.T. Tran, C. Stampfl, I. Aharonovich, M. Toth, and M.J. Ford, Nanoscale **9**, 13575 (2017).
[174] X.Z. Du, J. Li, J.Y. Lin, and H.X. Jiang, Applied Physics Letters **106**, 021110 (2015).
[175] A. Katzir, J.T. Suss, A. Zunger, and A. Halperin, Physical Review B **11**, 2370 (1975).
[176] L. Weston, D. Wickramaratne, M. Mackoit, A. Alkauskas, and C.G. Van de Walle, Phys. Rev. B **97**, 214104 (2018).
[177] G. Cassabois, P. Valvin, and B. Gil, Phys. Rev. B **93**, 035207 (2016).
[178] X.Z. Du, M.R. Uddin, J. Li, J.Y. Lin, and H.X. Jiang, Appl. Phys. Lett. **110**, 092102 (2017).
[179] V. Ivády, G. Barcza, G. Thiering, S. Li, H. Hamdi, J.-P. Chou, Ö. Legeza, and A. Gali, Npj Computational Materials **6**, 1 (2020).
[180] M.A. Lourenço, M.M. Milošević, A. Gorin, R.M. Gwilliam, and K.P. Homewood, Scientific Reports **6**, 37501 (2016).
[181] A.M. Zaitsev, *Optical Properties of Diamond* (Springer Berlin Heidelberg, Berlin, Heidelberg, 2001).
[182] M.D. Anderson, S. Tarrago Velez, K. Seibold, H. Flayac, V. Savona, N. Sangouard, and C. Galland, Phys. Rev. Lett. **120**, 233601 (2018).
[183] M.-A. Lemonde, S. Meesala, A. Sipahigil, M.J.A. Schuetz, M.D. Lukin, M. Loncar, and P. Rabl, Phys. Rev. Lett. **120**, 213603 (2018).
[184] G. Zhang and B. Li, J. Chem. Phys. **123**, 114714 (2005).
[185] G. Zhang and B. Li, J. Chem. Phys. **123**, 014705 (2005).
[186] G. Zhang and Y.-W. Zhang, Physica Status Solidi (RRL) – Rapid Research Letters **7**, 754 (2013).
[187] G. Zhang and Y.-W. Zhang, Chinese Phys. B **26**, 034401 (2017).
[188] H. Bao, J. Chen, X. Gu, and B. Cao, ES Energy & Environment (2018).
[189] X. Wang, Z.-Z. Li, M.-P. Zhuo, Y. Wu, S. Chen, J. Yao, and H. Fu, Advanced Functional Materials **27**, 1703470 (2017).
[190] Q. Zheng, W.A. Saidi, Y. Xie, Z. Lan, O.V. Prezhdo, H. Petek, and J. Zhao, Nano Lett. **17**, 6435 (2017).
[191] G. Zhang and Y.-W. Zhang, Mechanics of Materials **91**, 382 (2015).
[192] W.L. Yang, Z.Q. Yin, Z.Y. Xu, M. Feng, and J.F. Du, Appl. Phys. Lett. **96**, 241113 (2010).
[193] H. Zhou, Y. Cai, G. Zhang, and Y.-W. Zhang, Npj 2D Materials and Applications **1**, 14 (2017).
[194] D. Li, J. He, G. Ding, Q. Tang, Y. Ying, J. He, C. Zhong, Y. Liu, C. Feng, Q. Sun, H. Zhou, P. Zhou, and G. Zhang, Advanced Functional Materials **28**, 1801685 (2018).
[195] I. Aharonovich, D. Englund, and M. Toth, Nature Photonics **10**, 631 (2016).
[196] W.F. Koehl, H. Seo, G. Galli, and D.D. Awschalom, MRS Bulletin **40**, 1146 (2015).
[197] L.J. Rogers, S. Armstrong, M.J. Sellars, and N.B. Manson, New J. Phys. **10**, 103024 (2008).
[198] M.L. Goldman, A. Sipahigil, M.W. Doherty, N.Y. Yao, S.D. Bennett, M. Markham, D.J. Twitchen, N.B. Manson, A. Kubanek, and M.D. Lukin, Phys. Rev. Lett. **114**, 145502 (2015).
[199] J.A. Larsson and P. Delaney, Phys. Rev. B **77**, 165201 (2008).
[200] S. Choi, M. Jain, and S.G. Louie, Phys. Rev. B **86**, 041202 (2012).
[201] A. Ranjbar, M. Babamoradi, M. Heidari Saani, M.A. Vesaghi, K. Esfarjani, and Y. Kawazoe, Phys. Rev. B **84**, 165212 (2011).
[202] J. Lischner, J. Deslippe, M. Jain, and S.G. Louie, Phys. Rev. Lett. **109**, 036406 (2012).
[203] M. Bockstedte, F. Schütz, T. Garratt, V. Ivády, and A. Gali, Npj Quantum Materials **3**, (2018).
[204] L. Shi, D. Yao, G. Zhang, and B. Li, Appl. Phys. Lett. **96**, 173108 (2010).
[205] W. Li, X. Dai, J. Morrone, G. Zhang, and R. Zhou, Nanoscale **9**, 12025 (2017).
[206] K.-M.C. Fu, C. Santori, P.E. Barclay, L.J. Rogers, N.B. Manson, and R.G. Beausoleil, Phys. Rev. Lett. **103**, 256404 (2009).
[207] H.A. Jahn, E. Teller, and F.G. Donnan, Proc. R. Soc. A **161**, 220 (1937).
[208] T.A. Abtew, Y.Y. Sun, B.-C. Shih, P. Dev, S.B. Zhang, and P. Zhang, Phys. Rev. Lett. **107**, 146403 (2011).
[209] G. Thiering and A. Gali, Phys. Rev. B **96**, 081115 (2017).
[210] G. Thiering and A. Gali, Phys. Rev. B **98**, 085207 (2018).
[211] J. Choi, S. Choi, G. Kucsko, P.C. Maurer, B.J. Shields, H. Sumiya, S. Onoda, J. Isoya, E. Demler, F. Jelezko, N.Y. Yao, and M.D. Lukin, Phys. Rev. Lett. **118**, 093601 (2017).
[212] J.-P. Chou, Z. Bodrog, and A. Gali, Phys. Rev. Lett. **120**, 136401 (2018).

[213] P. Kómár, E.M. Kessler, M. Bishof, L. Jiang, A.S. Sørensen, J. Ye, and M.D. Lukin, Nature Physics **10**, 582 (2014).
[214] N. Kalb, A.A. Reiserer, P.C. Humphreys, J.J.W. Bakermans, S.J. Kamerling, N.H. Nickerson, S.C. Benjamin, D.J. Twitchen, M. Markham, and R. Hanson, Science **356**, 928 (2017).
[215] J.-P. Chou, A. Retzker, and A. Gali, Nano Lett. **17**, 2294–2298 (2017).
[216] A. Stacey, K.M. O'Donnell, J.-P. Chou, A. Schenk, A. Tadich, N. Dontschuk, J. Cervenka, C. Pakes, A. Gali, A. Hoffman, and S. Prawer, Adv. Mater. Interfaces **2**, 1500079 (2015).
[217] S. Sangtawesin, B.L. Dwyer, S. Srinivasan, J.J. Allred, L.V.H. Rodgers, K. De Greve, A. Stacey, N. Dontschuk, K.M. O'Donnell, D. Hu, D.A. Evans, C. Jaye, D.A. Fischer, M.L. Markham, D.J. Twitchen, H. Park, M.D. Lukin, and N.P. de Leon, Phys. Rev. X **9**, 031052 (2019).
[218] M.H. Abobeih, J. Randall, C.E. Bradley, H.P. Bartling, M.A. Bakker, M.J. Degen, M. Markham, D.J. Twitchen, and T.H. Taminiau, Nature **576**, 411 (2019).
[219] C.E. Bradley, J. Randall, M.H. Abobeih, R.C. Berrevoets, M.J. Degen, M.A. Bakker, M. Markham, D.J. Twitchen, and T.H. Taminiau, Phys. Rev. X **9**, 031045 (2019).
[220] T. Lühmann, R. John, R. Wunderlich, J. Meijer, and S. Pezzagna, Nature Communications **10**, 1 (2019).
[221] E. Bourgeois, A. Jarmola, P. Siyushev, M. Gulka, J. Hruby, F. Jelezko, D. Budker, and M. Nesladek, Nature Communications **6**, 8577 (2015).
[222] P. Siyushev, M. Nesladek, E. Bourgeois, M. Gulka, J. Hruby, T. Yamamoto, M. Trupke, T. Teraji, J. Isoya, and F. Jelezko, Science **363**, 728 (2019).
[223] B. Smeltzer, L. Childress, and A. Gali, New J. Phys. **13**, 025021 (2011).
[224] A.P. Nizovtsev, S.Y. Kilin, A.L. Pushkarchuk, V.A. Pushkarchuk, S.A. Kuten, O.A. Zhikol, S. Schmitt, T. Unden, and F. Jelezko, New J. Phys. **20**, 023022 (2018).
[225] L.J. Rogers, K.D. Jahnke, M.H. Metsch, A. Sipahigil, J.M. Binder, T. Teraji, H. Sumiya, J. Isoya, M.D. Lukin, P. Hemmer, and F. Jelezko, Phys. Rev. Lett. **113**, 263602 (2014).
[226] B. Pingault, J.N. Becker, C.H.H. Schulte, C. Arend, C. Hepp, T. Godde, A.I. Tartakovskii, M. Markham, C. Becher, and M. Atatüre, Phys. Rev. Lett. **113**, 263601 (2014).
[227] K.D. Jahnke, A. Sipahigil, J.M. Binder, M.W. Doherty, M. Metsch, L.J. Rogers, N.B. Manson, M.D. Lukin, and F. Jelezko, New J. Phys. **17**, 043011 (2015).
[228] L.J. Rogers, K.D. Jahnke, M.W. Doherty, A. Dietrich, L.P. McGuinness, C. Müller, T. Teraji, H. Sumiya, J. Isoya, N.B. Manson, and F. Jelezko, Physical Review B **89**, (2014).
[229] A. Gali and J.R. Maze, Phys. Rev. B **88**, 235205 (2013).
[230] C.D. Clark, H. Kanda, I. Kiflawi, and G. Sittas, Phys. Rev. B **51**, 16681 (1995).
[231] L.J. Rogers, K.D. Jahnke, T. Teraji, L. Marseglia, C. Müller, B. Naydenov, H. Schauffert, C. Kranz, J. Isoya, L.P. McGuinness, and F. Jelezko, Nature Communications **5**, 1 (2014).
[232] B. Pingault, D.-D. Jarausch, C. Hepp, L. Klintberg, J.N. Becker, M. Markham, C. Becher, and M. Atatüre, Nature Communications **8**, 15579 (2017).
[233] D.D. Sukachev, A. Sipahigil, C.T. Nguyen, M.K. Bhaskar, R.E. Evans, F. Jelezko, and M.D. Lukin, Phys. Rev. Lett. **119**, 223602 (2017).
[234] J.N. Becker, B. Pingault, D. Groß, M. Gündoğan, N. Kukharchyk, M. Markham, A. Edmonds, M. Atatüre, P. Bushev, and C. Becher, Phys. Rev. Lett. **120**, 053603 (2018).
[235] A. Sipahigil, K.D. Jahnke, L.J. Rogers, T. Teraji, J. Isoya, A.S. Zibrov, F. Jelezko, and M.D. Lukin, Phys. Rev. Lett. **113**, 113602 (2014).
[236] A. Sipahigil, R.E. Evans, D.D. Sukachev, M.J. Burek, J. Borregaard, M.K. Bhaskar, C.T. Nguyen, J.L. Pacheco, H.A. Atikian, C. Meuwly, R.M. Camacho, F. Jelezko, E. Bielejec, H. Park, M. Lončar, and M.D. Lukin, Science **354**, 847 (2016).
[237] S. Maity, L. Shao, S. Bogdanović, S. Meesala, Y.-I. Sohn, N. Sinclair, B. Pingault, M. Chalupnik, C. Chia, L. Zheng, K. Lai, and M. Lončar, Nat Commun **11**, (2020).
[238] B.L. Green, S. Mottishaw, B.G. Breeze, A.M. Edmonds, U.F.S. D'Haenens-Johansson, M.W. Doherty, S.D. Williams, D.J. Twitchen, and M.E. Newton, Physical Review Letters **119**, 096402 (2017).
[239] B.C. Rose, G. Thiering, A.M. Tyryshkin, A.M. Edmonds, M.L. Markham, A. Gali, S.A. Lyon, and N.P. de Leon, Phys. Rev. B **98**, 235140 (2018).
[240] G. Thiering and A. Gali, Npj Comput Mater **5**, 18 (2019).
[241] B.L. Green, M.W. Doherty, E. Nako, N.B. Manson, U.F.S. D'Haenens-Johansson, S.D. Williams, D.J. Twitchen, and M.E. Newton, Phys. Rev. B **99**, 161112 (2019).

[242] Z.-H. Zhang, P. Stevenson, G. Thiering, B.C. Rose, D. Huang, A.M. Edmonds, M.L. Markham, S.A. Lyon, A. Gali, and N.P. de Leon, ArXiv:2004.12544 [Cond-Mat, Physics:Quant-Ph] (2020).
[243] A. Dietrich, K.D. Jahnke, J.M. Binder, T. Teraji, J. Isoya, L.J. Rogers, and F. Jelezko, New J. Phys. **16**, 113019 (2014).
[244] A. Szenes, B. Bánhelyi, L.Z. Szabó, G. Szabó, T. Csendes, and M. Csete, Scientific Reports **7**, 13845 (2017).
[245] T. Iwasaki, F. Ishibashi, Y. Miyamoto, Y. Doi, S. Kobayashi, T. Miyazaki, K. Tahara, K.D. Jahnke, L.J. Rogers, B. Naydenov, F. Jelezko, S. Yamasaki, S. Nagamachi, T. Inubushi, N. Mizuochi, and M. Hatano, Scientific Reports **5**, 12882 (2015).
[246] J.P. Goss, P.R. Briddon, M.J. Rayson, S.J. Sque, and R. Jones, Phys. Rev. B **72**, 035214 (2005).
[247] P. Siyushev, M.H. Metsch, A. Ijaz, J.M. Binder, M.K. Bhaskar, D.D. Sukachev, A. Sipahigil, R.E. Evans, C.T. Nguyen, M.D. Lukin, P.R. Hemmer, Y.N. Palyanov, I.N. Kupriyanov, Y.M. Borzdov, L.J. Rogers, and F. Jelezko, Phys. Rev. B **96**, 081201 (2017).
[248] A.E. Rugar, C. Dory, S. Sun, and J. Vučković, Phys. Rev. B **99**, 205417 (2019).
[249] S. Ditalia Tchernij, T. Lühmann, T. Herzig, J. Küpper, A. Damin, S. Santonocito, M. Signorile, P. Traina, E. Moreva, F. Celegato, S. Pezzagna, I.P. Degiovanni, P. Olivero, M. Jakšić, J. Meijer, P.M. Genovese, and J. Forneris, ACS Photonics **5**, 4864 (2018).
[250] J. Görlitz, D. Herrmann, G. Thiering, P. Fuchs, M. Gandil, T. Iwasaki, T. Taniguchi, M. Kieschnick, J. Meijer, M. Hatano, A. Gali, and C. Becher, New J. Phys. **22**, 013048 (2020).
[251] M.E. Trusheim, B. Pingault, N.H. Wan, M. Gündoğan, L. De Santis, R. Debroux, D. Gangloff, C. Purser, K.C. Chen, M. Walsh, J.J. Rose, J.N. Becker, B. Lienhard, E. Bersin, I. Paradeisanos, G. Wang, D. Lyzwa, A.R.-P. Montblanch, G. Malladi, H. Bakhru, A.C. Ferrari, I.A. Walmsley, M. Atatüre, and D. Englund, Phys. Rev. Lett. **124**, 023602 (2020).
[252] A.E. Rugar, H. Lu, C. Dory, S. Sun, P.J. McQuade, Z.-X. Shen, N.A. Melosh, and J. Vučković, Nano Lett. **20**, 1614 (2020).
[253] M.T. Westerhausen, A.T. Trycz, C. Stewart, M. Nonahal, B. Regan, M. Kianinia, and I. Aharonovich, ACS Appl. Mater. Interfaces **12**, 29700 (2020).
[254] A. Gali, Materials Science Forum **679–680**, 225 (2011).
[255] K. Szász, V. Ivády, I.A. Abrikosov, E. Janzén, M. Bockstedte, and A. Gali, Phys. Rev. B **91**, 121201 (2015).
[256] N.T. Son, P. Carlsson, J. ul Hassan, E. Janzén, T. Umeda, J. Isoya, A. Gali, M. Bockstedte, N. Morishita, T. Ohshima, and H. Itoh, Phys. Rev. Lett. **96**, 055501 (2006).
[257] S. Castelletto, B.C. Johnson, V. Ivády, N. Stavrias, T. Umeda, A. Gali, and T. Ohshima, Nature Materials **13**, 151 (2014).
[258] A. Lohrmann, B.C. Johnson, J.C. McCallum, and S. Castelletto, Reports on Progress in Physics **80**, 034502 (2017).
[259] K.C. Miao, A. Bourassa, C.P. Anderson, S.J. Whiteley, A.L. Crook, S.L. Bayliss, G. Wolfowicz, G. Thiering, P. Udvarhelyi, V. Ivády, H. Abe, T. Ohshima, Á. Gali, and D.D. Awschalom, Science Advances **5**, eaay0527 (2019).
[260] V. Ivády, J. Davidsson, N. Delegan, A.L. Falk, P.V. Klimov, S.J. Whiteley, S.O. Hruszkewycz, M.V. Holt, F.J. Heremans, N.T. Son, D.D. Awschalom, I.A. Abrikosov, and A. Gali, Nature Communications **10**, 1 (2019).
[261] D.J. Christle, A.L. Falk, P. Andrich, P.V. Klimov, J.U. Hassan, N.T. Son, E. Janzén, T. Ohshima, and D.D. Awschalom, Nature Materials **14**, 160 (2015).
[262] G. Calusine, A. Politi, and D.D. Awschalom, Appl. Phys. Lett. **105**, 011123 (2014).
[263] P. Udvarhelyi and A. Gali, Phys. Rev. Applied **10**, 054010 (2018).
[264] H. Seo, A.L. Falk, P.V. Klimov, K.C. Miao, G. Galli, and D.D. Awschalom, Nature Communications **7**, 12935 (2016).
[265] S.J. Whiteley, G. Wolfowicz, C.P. Anderson, A. Bourassa, H. Ma, M. Ye, G. Koolstra, K.J. Satzinger, M.V. Holt, F.J. Heremans, A.N. Cleland, D.I. Schuster, G. Galli, and D.D. Awschalom, Nature Physics **15**, 490 (2019).
[266] P.V. Klimov, A.L. Falk, B.B. Buckley, and D.D. Awschalom, Phys. Rev. Lett. **112**, 087601 (2014).
[267] C.P. Anderson, A. Bourassa, K.C. Miao, G. Wolfowicz, P.J. Mintun, A.L. Crook, H. Abe, J.U. Hassan, N.T. Son, T. Ohshima, and D.D. Awschalom, Science **366**, 1225 (2019).

[268] V. Ivády, K. Szász, A.L. Falk, P.V. Klimov, D.J. Christle, E. Janzén, I.A. Abrikosov, D.D. Awschalom, and A. Gali, Phys. Rev. B **92**, 115206 (2015).
[269] V. Ivády, P.V. Klimov, K.C. Miao, A.L. Falk, D.J. Christle, K. Szász, I.A. Abrikosov, D.D. Awschalom, and A. Gali, Phys. Rev. Lett. **117**, 220503 (2016).
[270] Mt. Wagner, B. Magnusson, W.M. Chen, E. Janzén, E. Sörman, C. Hallin, and J.L. Lindström, Phys. Rev. B **62**, 16555 (2000).
[271] V. Ivády, J. Davidsson, N.T. Son, T. Ohshima, I.A. Abrikosov, and A. Gali, Phys. Rev. B **96**, 161114 (2017).
[272] N. Mizuochi, S. Yamasaki, H. Takizawa, N. Morishita, T. Ohshima, H. Itoh, and J. Isoya, Phys. Rev. B **68**, 165206 (2003).
[273] Z. Shang, A. Hashemi, Y. Berencén, H.-P. Komsa, P. Erhart, S. Zhou, M. Helm, A.V. Krasheninnikov, and G.V. Astakhov, Phys. Rev. B **101**, 144109 (2020).
[274] V.A. Soltamov, B.V. Yavkin, D.O. Tolmachev, R.A. Babunts, A.G. Badalyan, V.Yu. Davydov, E.N. Mokhov, I.I. Proskuryakov, S.B. Orlinskii, and P.G. Baranov, Phys. Rev. Lett. **115**, 247602 (2015).
[275] J. Heyd, G.E. Scuseria, and M. Ernzerhof, The Journal of Chemical Physics **118**, 8207 (2003).
[276] J. Heyd, G.E. Scuseria, and M. Ernzerhof, J. Chem. Phys. **124**, 219906 (2006).
[277] J.P. Perdew, K. Burke, and M. Ernzerhof, Phys. Rev. Lett. **77**, 3865 (1996).
[278] Ö.O. Soykal, P. Dev, and S.E. Economou, Phys. Rev. B **93**, 081207 (2016).
[279] T. Biktagirov, W.G. Schmidt, U. Gerstmann, B. Yavkin, S. Orlinskii, P. Baranov, V. Dyakonov, and V. Soltamov, Phys. Rev. B **98**, 195204 (2018).
[280] J. Davidsson, V. Ivády, R. Armiento, T. Ohshima, N.T. Son, A. Gali, and I.A. Abrikosov, Appl. Phys. Lett. **114**, 112107 (2019).
[281] F. Fuchs, B. Stender, M. Trupke, D. Simin, J. Pflaum, V. Dyakonov, and G.V. Astakhov, Nature Communications **6**, 7578 (2015).
[282] M. Widmann, S.-Y. Lee, T. Rendler, N.T. Son, H. Fedder, S. Paik, L.-P. Yang, N. Zhao, S. Yang, I. Booker, A. Denisenko, M. Jamali, S.A. Momenzadeh, I. Gerhardt, T. Ohshima, A. Gali, E. Janzén, and J. Wrachtrup, Nat Mater **14**, 164 (2015).
[283] M. Niethammer, M. Widmann, T. Rendler, N. Morioka, Y.-C. Chen, R. Stöhr, J.U. Hassan, S. Onoda, T. Ohshima, S.-Y. Lee, A. Mukherjee, J. Isoya, N.T. Son, and J. Wrachtrup, Nature Communications **10**, 1 (2019).
[284] F. Fuchs, V.A. Soltamov, S. Väth, P.G. Baranov, E.N. Mokhov, G.V. Astakhov, and V. Dyakonov, Scientific Reports **3**, 1 (2013).
[285] H. Kraus, V.A. Soltamov, D. Riedel, S. Väth, F. Fuchs, A. Sperlich, P.G. Baranov, V. Dyakonov, and G.V. Astakhov, Nature Physics **10**, 157 (2014).
[286] E. Janzén, A. Gali, P. Carlsson, A. Gällström, B. Magnusson, and N.T. Son, Physica B: Condensed Matter **404**, 4354 (2009).
[287] A. Gali, Journal of Materials Research **27**, 897 (2012).
[288] R. Nagy, M. Niethammer, M. Widmann, Y.-C. Chen, P. Udvarhelyi, C. Bonato, J.U. Hassan, R. Karhu, I.G. Ivanov, N.T. Son, J.R. Maze, T. Ohshima, Ö.O. Soykal, Á. Gali, S.-Y. Lee, F. Kaiser, and J. Wrachtrup, Nature Communications **10**, 1 (2019).
[289] P. Udvarhelyi, R. Nagy, F. Kaiser, S.-Y. Lee, J. Wrachtrup, and A. Gali, Phys. Rev. Applied **11**, 044022 (2019).
[290] N. Morioka, C. Babin, R. Nagy, I. Gediz, E. Hesselmeier, D. Liu, M. Joliffe, M. Niethammer, D. Dasari, V. Vorobyov, R. Kolesov, R. Stöhr, J. Ul-Hassan, N.T. Son, T. Ohshima, P. Udvarhelyi, G. Thiering, A. Gali, J. Wrachtrup, and F. Kaiser, Nature Communications **11**, 2516 (2020).
[291] H. Ozawa, K. Tahara, H. Ishiwata, M. Hatano, and T. Iwasaki, Appl. Phys. Express **10**, 045501 (2017).
[292] F. Pan, M. Zhao, and L. Mei, Journal of Applied Physics **108**, 043917 (2010).
[293] J.R. Weber, W.F. Koehl, J.B. Varley, A. Janotti, B.B. Buckley, C.G. Van de Walle, and D.D. Awschalom, Journal of Applied Physics **109**, 102417 (2011).
[294] J.-F. Wang, F.-F. Yan, Q. Li, Z.-H. Liu, H. Liu, G.-P. Guo, L.-P. Guo, X. Zhou, J.-M. Cui, J. Wang, Z.-Q. Zhou, X.-Y. Xu, J.-S. Xu, C.-F. Li, and G.-C. Guo, Phys. Rev. Lett. **124**, 223601 (2020).
[295] N.T. Son, A. Ellison, B. Magnusson, M.F. MacMillan, W.M. Chen, B. Monemar, and E. Janzén, Journal of Applied Physics **86**, 4348 (1999).
[296] L. Spindlberger, A. Csóré, G. Thiering, S. Putz, R. Karhu, J.U. Hassan, N.T. Son, T. Fromherz, A. Gali, and M. Trupke, Phys. Rev. Applied **12**, 014015 (2019).

[297] B. Kaufmann, A. Dörnen, and F.S. Ham, in *Volume 196–201, Materials Science Forum* (1995), pp. 707–712.
[298] T. Bosma, G.J.J. Lof, C.M. Gilardoni, O.V. Zwier, F. Hendriks, B. Magnusson, A. Ellison, A. Gällström, I.G. Ivanov, N.T. Son, R.W.A. Havenith, and C.H. van der Wal, Npj Quantum Information **4**, 1 (2018).
[299] A. Csóré and A. Gali, ArXiv:1909.11587 [Cond-Mat, Physics:Quant-Ph] (2019).
[300] M. Widmann, M. Niethammer, D.Yu. Fedyanin, I.A. Khramtsov, T. Rendler, I.D. Booker, J. Ul Hassan, N. Morioka, Y.-C. Chen, I.G. Ivanov, N.T. Son, T. Ohshima, M. Bockstedte, A. Gali, C. Bonato, S.-Y. Lee, and J. Wrachtrup, Nano Lett. **19**, 7173 (2019).
[301] M.E. Levinshteĭn, S.L. Rumyantsev, and M. Shur, *Properties of Advanced Semiconductor Materials: GaN, AlN, InN, BN, SiC, SiGe* (Wiley, New York, 2001).
[302] Q. Li, J.-F. Wang, F.-F. Yan, J.-Y. Zhou, H.-F. Wang, H. Liu, L.-P. Guo, X. Zhou, A. Gali, Z.-H. Liu, Z.-Q. Wang, K. Sun, G.-P. Guo, J.-S. Tang, J.-S. Xu, C.-F. Li, and G.-C. Guo, ArXiv:2005.07876 [Cond-Mat, Physics:Quant-Ph] (2020).
[303] D.M. Toyli, D.J. Christle, A. Alkauskas, B.B. Buckley, C.G. Van de Walle, and D.D. Awschalom, Phys. Rev. X **2**, 031001 (2012).
[304] B.C. Johnson, J. Woerle, D. Haasmann, C.T.-K. Lew, R.A. Parker, H. Knowles, B. Pingault, M. Atature, A. Gali, S. Dimitrijev, M. Camarda, and J.C. McCallum, Phys. Rev. Applied **12**, 044024 (2019).
[305] T.T. Tran, C. Zachreson, A.M. Berhane, K. Bray, R.G. Sandstrom, L.H. Li, T. Taniguchi, K. Watanabe, I. Aharonovich, and M. Toth, Phys. Rev. Applied **5**, 034005 (2016).
[306] N.R. Jungwirth, B. Calderon, Y. Ji, M.G. Spencer, M.E. Flatté, and G.D. Fuchs, Nano Letters **16**, 6052 (2016).
[307] N. Chejanovsky, M. Rezai, F. Paolucci, Y. Kim, T. Rendler, W. Rouabeh, F. Fávaro de Oliveira, P. Herlinger, A. Denisenko, S. Yang, I. Gerhardt, A. Finkler, J.H. Smet, and J. Wrachtrup, Nano Letters **16**, 7037 (2016).
[308] T.T. Tran, C. Elbadawi, D. Totonjian, C.J. Lobo, G. Grosso, H. Moon, D.R. Englund, M.J. Ford, I. Aharonovich, and M. Toth, ACS Nano **10**, 7331 (2016).
[309] N.V. Proscia, Z. Shotan, H. Jayakumar, P. Reddy, C. Cohen, M. Dollar, A. Alkauskas, M. Doherty, C.A. Meriles, and V.M. Menon, Optica **5**, 1128 (2018).
[310] M. Ye, H. Seo, and G. Galli, Npj Computational Materials **5**, 44 (2019).
[311] N.L. McDougall, J.G. Partridge, R.J. Nicholls, S.P. Russo, and D.G. McCulloch, Physical Review B **96**, (2017).
[312] G.D. Cheng, Y.G. Zhang, L. Yan, H.F. Huang, Q. Huang, Y.X. Song, Y. Chen, and Z. Tang, Computational Materials Science **129**, 247 (2017).
[313] M. Abdi, J.-P. Chou, A. Gali, and M.B. Plenio, ACS Photonics **5**, 1967 (2018).
[314] Z.-Q. Xu, C. Elbadawi, T.T. Tran, M. Kianinia, X. Li, D. Liu, T.B. Hoffman, M. Nguyen, S. Kim, J.H. Edgar, X. Wu, L. Song, S. Ali, M. Ford, M. Toth, and I. Aharonovich, Nanoscale **10**, 7957 (2018).
[315] F. Karsai, M. Humer, E. Flage-Larsen, P. Blaha, and G. Kresse, Phys. Rev. B **98**, 235205 (2018).
[316] M.A. Feldman, A. Puretzky, L. Lindsay, E. Tucker, D.P. Briggs, P.G. Evans, R.F. Haglund, and B.J. Lawrie, Phys. Rev. B **99**, 020101 (2019).
[317] Y. Luo, N. Liu, X. Li, J.C. Hone, and S. Strauf, 2D Mater. **6**, 035017 (2019).
[318] M. Kianinia, B. Regan, S.A. Tawfik, T.T. Tran, M.J. Ford, I. Aharonovich, and M. Toth, ACS Photonics **4**, 768 (2017).
[319] N.R. Jungwirth and G.D. Fuchs, Physical Review Letters **119**, 057401 (2017).
[320] T.T. Tran, D. Wang, Z.-Q. Xu, A. Yang, M. Toth, T.W. Odom, and I. Aharonovich, Nano Lett. **17**, 2634 (2017).
[321] R. Bourrellier, S. Meuret, A. Tararan, O. Stéphan, M. Kociak, L.H.G. Tizei, and A. Zobelli, Nano Lett. **16**, 4317 (2016).
[322] A. Sajid, J.R. Reimers, and M.J. Ford, Phys. Rev. B **97**, 064101 (2018).
[323] F. Wu, A. Galatas, R. Sundararaman, D. Rocca, and Y. Ping, Physical Review Materials **1**, (2017).
[324] M. Toth and I. Aharonovich, Annual Review of Physical Chemistry **70**, 123 (2019).
[325] C. Jin, F. Lin, K. Suenaga, and S. Iijima, Phys. Rev. Lett. **102**, 195505 (2009).
[326] L.J. Martínez, T. Pelini, V. Waselowski, J.R. Maze, B. Gil, G. Cassabois, and V. Jacques, Phys. Rev. B **94**, 121405 (2016).
[327] G. Noh, D. Choi, J.-H. Kim, D.-G. Im, Y.-H. Kim, H. Seo, and J. Lee, Nano Lett. **18**, 4710 (2018).

[328] O.L. Krivanek, M.F. Chisholm, V. Nicolosi, T.J. Pennycook, G.J. Corbin, N. Dellby, M.F. Murfitt, C.S. Own, Z.S. Szilagyi, M.P. Oxley, S.T. Pantelides, and S.J. Pennycook, Nature **464**, 571 (2010).
[329] M.R. Uddin, J. Li, J.Y. Lin, and H.X. Jiang, Appl. Phys. Lett. **110**, 182107 (2017).
[330] M.R. Ahmadpour Monazam, U. Ludacka, H.-P. Komsa, and J. Kotakoski, Appl. Phys. Lett. **115**, 071604 (2019).
[331] F. Wu, A. Galatas, R. Sundararaman, D. Rocca, and Y. Ping, Phys. Rev. Materials **1**, 071001 (2017).
[332] T. Korona and M. Chojecki, Int J Quantum Chem **119**, e25925 (2019).
[333] N. Chejanovsky, A. Mukherjee, Y. Kim, A. Denisenko, A. Finkler, T. Taniguchi, K. Watanabe, D.B.R. Dasari, J.H. Smet, and J. Wrachtrup, ArXiv:1906.05903 [Cond-Mat, Physics:Physics] (2019).
[334] M. Mackoit-Sinkevičienė, M. Maciaszek, C.G. Van de Walle, and A. Alkauskas, Appl. Phys. Lett. **115**, 212101 (2019).
[335] M.E. Turiansky, A. Alkauskas, L.C. Bassett, and C.G. Van de Walle, Phys. Rev. Lett. **123**, 127401 (2019).
[336] J.R. Reimers, A. Sajid, R. Kobayashi, and M.J. Ford, Journal of Chemical Theory and Computation **14**, 1602 (2018).
[337] S. Li, J.-P. Chou, A. Hu, M.B. Plenio, P. Udvarhelyi, G. Thiering, M. Abdi, and A. Gali, arXiv:2001.02749 (2020).
[338] Z. Yu, Y. Pan, Y. Shen, Z. Wang, Z.-Y. Ong, T. Xu, R. Xin, L. Pan, B. Wang, L. Sun, J. Wang, G. Zhang, Y.W. Zhang, Y. Shi, and X. Wang, Nature Communications **5**, 5290 (2014).
[339] W. Li, G. Zhang, M. Guo, and Y.-W. Zhang, Nano Res. **7**, 518 (2014).
[340] Y. Liu, H. Wu, H.-C. Cheng, S. Yang, E. Zhu, Q. He, M. Ding, D. Li, J. Guo, N.O. Weiss, Y. Huang, and X. Duan, Nano Lett. **15**, 3030 (2015).
[341] Y. Cui, R. Xin, Z. Yu, Y. Pan, Z.-Y. Ong, X. Wei, J. Wang, H. Nan, Z. Ni, Y. Wu, T. Chen, Y. Shi, B. Wang, G. Zhang, Y.-W. Zhang, and X. Wang, Advanced Materials **27**, 5230 (2015).
[342] Y. Cheng, L.-D. Koh, D. Li, B. Ji, Y. Zhang, J. Yeo, G. Guan, M.-Y. Han, and Y.-W. Zhang, ACS Appl. Mater. Interfaces **7**, 21787 (2015).
[343] Y. Cheng, G. Zhang, Y. Zhang, T. Chang, Q.-X. Pei, Y. Cai, and Y.-W. Zhang, Nanoscale **10**, 1660 (2018).
[344] G. Chen, N. Matsuhisa, Z. Liu, D. Qi, P. Cai, Y. Jiang, C. Wan, Y. Cui, W.R. Leow, Z. Liu, S. Gong, K.-Q. Zhang, Y. Cheng, and X. Chen, Advanced Materials **30**, 1800129 (2018).
[345] G. Zhang and Y.-W. Zhang, Journal of Materials Chemistry C **5**, 7684 (2017).
[346] Y.-Y. Zhang, Q.-X. Pei, C.-M. Wang, Y. Cheng, and Y.-W. Zhang, Journal of Applied Physics **114**, 073504 (2013).
[347] Y. Jiang, Z. Liu, N. Matsuhisa, D. Qi, W.R. Leow, H. Yang, J. Yu, G. Chen, Y. Liu, C. Wan, Z. Liu, and X. Chen, Advanced Materials **30**, 1706589 (2018).
[348] W. Zhou, X. Zou, S. Najmaei, Z. Liu, Y. Shi, J. Kong, J. Lou, P.M. Ajayan, B.I. Yakobson, and J.-C. Idrobo, Nano Letters **13**, 2615 (2013).
[349] W. Li, Y. Yang, J.K. Weber, G. Zhang, and R. Zhou, ACS Nano **10**, 1829 (2016).
[350] M. Koperski, K. Nogajewski, A. Arora, V. Cherkez, P. Mallet, J.-Y. Veuillen, J. Marcus, P. Kossacki, and M. Potemski, Nature Nanotechnology **10**, 503 (2015).
[351] Y.-M. He, G. Clark, J.R. Schaibley, Y. He, M.-C. Chen, Y.-J. Wei, X. Ding, Q. Zhang, W. Yao, X. Xu, C.-Y. Lu, and J.-W. Pan, Nature Nanotechnology **10**, 497 (2015).
[352] C. Chakraborty, L. Kinnischtzke, K.M. Goodfellow, R. Beams, and A.N. Vamivakas, Nature Nanotechnology **10**, 507 (2015).
[353] A. Srivastava, M. Sidler, A.V. Allain, D.S. Lembke, A. Kis, and A. Imamoğlu, Nature Nanotechnology **10**, 491 (2015).
[354] P. Tonndorf, R. Schmidt, R. Schneider, J. Kern, M. Buscema, G.A. Steele, A. Castellanos-Gomez, H.S.J. van der Zant, S.M. de Vasconcellos, and R. Bratschitsch, Optica, OPTICA **2**, 347 (2015).
[355] Y.-M. He, O. Iff, N. Lundt, V. Baumann, M. Davanco, K. Srinivasan, S. Höfling, and C. Schneider, Nature Communications **7**, 13409 (2016).
[356] Y. Ye, X. Dou, K. Ding, Y. Chen, D. Jiang, F. Yang, and B. Sun, Phys. Rev. B **95**, 245313 (2017).
[357] A. Branny, S. Kumar, R. Proux, and B.D. Gerardot, Nature Communications **8**, 15053 (2017).
[358] C. Palacios-Berraquero, D.M. Kara, A.R.-P. Montblanch, M. Barbone, P. Latawiec, D. Yoon, A.K. Ott, M. Loncar, A.C. Ferrari, and M. Atatüre, Nature Communications **8**, 15093 (2017).
[359] C. Su, M. Tripathi, Q.-B. Yan, Z. Wang, Z. Zhang, C. Hofer, H. Wang, L. Basile, G. Su, M. Dong, J.C. Meyer, J. Kotakoski, J. Kong, J.-C. Idrobo, T. Susi, and J. Li, Science Advances **5**, eaav2252 (2019).

[360] H. Zeng, G. Duan, Y. Li, S. Yang, X. Xu, and W. Cai, Advanced Functional Materials **20**, 561 (2010).
[361] N.R. Jungwirth, H.-S. Chang, M. Jiang, and G.D. Fuchs, ACS Nano **10**, 1210 (2016).
[362] A. Janotti and C.G.V. de Walle, Rep. Prog. Phys. **72**, 126501 (2009).
[363] J.L. Lyons, J.B. Varley, D. Steiauf, A. Janotti, and C.G. Van de Walle, Journal of Applied Physics **122**, 035704 (2017).
[364] S. Choi, B.C. Johnson, S. Castelletto, C. Ton-That, M.R. Phillips, and I. Aharonovich, Appl. Phys. Lett. **104**, 261101 (2014).
[365] K. Chung, T.J. Karle, A. Khalid, A.N. Abraham, R. Shukla, B.C. Gibson, D.A. Simpson, A.B. Djurišic, H. Amekura, and S. Tomljenovic-Hanic, Nanophotonics **6**, 269 (2017).
[366] C. Stewart, M. Kianinia, R. Previdi, T.T. Tran, I. Aharonovich, C. Bradac, and C. Bradac, Opt. Lett., OL **44**, 4873 (2019).
[367] J.B. Varley and V. Lordi, Appl. Phys. Lett. **103**, 102103 (2013).
[368] K. Chung, Y.H. Leung, C.H. To, A.B. Djurišić, and S. Tomljenovic-Hanic, Beilstein J. Nanotechnol. **9**, 1085 (2018).
[369] F.L. Roux, K. Gao, M. Holmes, S. Kako, M. Arita, and Y. Arakawa, Scientific Reports **7**, 16107 (2017).
[370] D. Gammon, E.S. Snow, B.V. Shanabrook, D.S. Katzer, and D. Park, Science **273**, 87 (1996).
[371] M. Arita, F. Le Roux, M.J. Holmes, S. Kako, and Y. Arakawa, Nano Letters **17**, 2902 (2017).
[372] A. Szállás, K. Szász, X.T. Trinh, N.T. Son, E. Janzén, and A. Gali, Journal of Applied Physics **116**, 113702 (2014).
[373] S. Wei and A. Zunger, Appl. Phys. Lett. **69**, 2719 (1996).
[374] H. Okumura, T. Kimoto, and J. Suda, Appl. Phys. Express **4**, 025502 (2011).
[375] N.T. Son, A. Gali, Á. Szabó, M. Bickermann, T. Ohshima, J. Isoya, and E. Janzén, Appl. Phys. Lett. **98**, 242116 (2011).
[376] J.B. Varley, A. Janotti, and C.G. Van de Walle, Phys. Rev. B **93**, 161201 (2016).
[377] H. Seo, H. Ma, M. Govoni, and G. Galli, Phys. Rev. Materials **1**, 075002 (2017).
[378] Y. Xue, H. Wang, N. Xie, Q. Yang, F. Xu, B. Shen, J. Shi, D. Jiang, X. Dou, T. Yu, and B. Sun, J. Phys. Chem. Lett. **11**, 2689 (2020).
[379] M. Afzelius, N. Gisin, and H. de Riedmatten, Physics Today **68**, 42 (2015).
[380] M. Zhong, M.P. Hedges, R.L. Ahlefeldt, J.G. Bartholomew, S.E. Beavan, S.M. Wittig, J.J. Longdell, and M.J. Sellars, Nature **517**, 177 (2015).
[381] R. Kolesov, K. Xia, R. Reuter, M. Jamali, R. Stöhr, T. Inal, P. Siyushev, and J. Wrachtrup, Phys. Rev. Lett. **111**, 120502 (2013).
[382] M. Raha, S. Chen, C.M. Phenicie, S. Ourari, A.M. Dibos, and J.D. Thompson, Nature Communications **11**, 1605 (2020).
[383] J.M. Kindem, A. Ruskuc, J.G. Bartholomew, J. Rochman, Y.Q. Huan, and A. Faraon, Nature **580**, 201 (2020).
[384] J. Luzon and R. Sessoli, Dalton Trans. **41**, 13556 (2012).
[385] V. Ivády, A. Gali, and I.A. Abrikosov, J. Phys.: Condens. Matter **29**, 454002 (2017).
[386] B.G. Janesko, Phys. Rev. B **97**, 085138 (2018).
[387] H.S. Knowles, D.M. Kara, and M. Atatüre, Nat Mater **13**, 21 (2014).
[388] X. He, H. Htoon, S.K. Doorn, W.H.P. Pernice, F. Pyatkov, R. Krupke, A. Jeantet, Y. Chassagneux, and C. Voisin, Nature Materials **17**, 663 (2018).
[389] M.J. O'Connell, S.M. Bachilo, C.B. Huffman, V.C. Moore, M.S. Strano, E.H. Haroz, K.L. Rialon, P.J. Boul, W.H. Noon, C. Kittrell, J. Ma, R.H. Hauge, R.B. Weisman, and R.E. Smalley, Science **297**, 593 (2002).
[390] A. Högele, C. Galland, M. Winger, and A. Imamoğlu, Phys. Rev. Lett. **100**, 217401 (2008).
[391] X. Ma, O. Roslyak, J.G. Duque, X. Pang, S.K. Doorn, A. Piryatinski, D.H. Dunlap, and H. Htoon, Phys. Rev. Lett. **115**, 017401 (2015).
[392] S. Ghosh, S.M. Bachilo, R.A. Simonette, K.M. Beckingham, and R.B. Weisman, Science **330**, 1656 (2010).
[393] Y. Piao, B. Meany, L.R. Powell, N. Valley, H. Kwon, G.C. Schatz, and Y. Wang, Nature Chemistry **5**, 840 (2013).
[394] X. Ma, N.F. Hartmann, J.K.S. Baldwin, S.K. Doorn, and H. Htoon, Nature Nanotechnology **10**, 671 (2015).
[395] X. He, N.F. Hartmann, X. Ma, Y. Kim, R. Ihly, J.L. Blackburn, W. Gao, J. Kono, Y. Yomogida, A. Hirano, T. Tanaka, H. Kataura, H. Htoon, and S.K. Doorn, Nature Photonics **11**, 577 (2017).

[396] C. Galland, A. Högele, H.E. Türeci, and A. Imamoğlu, Phys. Rev. Lett. **101**, 067402 (2008).
[397] X. Ma, L. Adamska, H. Yamaguchi, S.E. Yalcin, S. Tretiak, S.K. Doorn, and H. Htoon, ACS Nano **8**, 10782 (2014).
[398] N.F. Hartmann, K.A. Velizhanin, E.H. Haroz, M. Kim, X. Ma, Y. Wang, H. Htoon, and S.K. Doorn, ACS Nano **10**, 8355 (2016).
[399] B.E. Kane, Nature **393**, 133 (1998).
[400] S.J. Angus, A.J. Ferguson, A.S. Dzurak, and R.G. Clark, Nano Lett. **7**, 2051 (2007).
[401] J.J.L. Morton, A.M. Tyryshkin, R.M. Brown, S. Shankar, B.W. Lovett, A. Ardavan, T. Schenkel, E.E. Haller, J.W. Ager, and S.A. Lyon, Nature **455**, 1085 (2008).
[402] M. Xiao, I. Martin, E. Yablonovitch, and H.W. Jiang, Nature **430**, 435 (2004).
[403] A.R. Stegner, C. Boehme, H. Huebl, M. Stutzmann, K. Lips, and M.S. Brandt, Nature Physics **2**, 835 (2006).
[404] A. Morello, C.C. Escott, H. Huebl, L.H. Willems van Beveren, L.C.L. Hollenberg, D.N. Jamieson, A.S. Dzurak, and R.G. Clark, Phys. Rev. B **80**, 081307 (2009).
[405] A. Morello, J.J. Pla, F.A. Zwanenburg, K.W. Chan, K.Y. Tan, H. Huebl, M. Möttönen, C.D. Nugroho, C. Yang, J.A. van Donkelaar, A.D.C. Alves, D.N. Jamieson, C.C. Escott, L.C.L. Hollenberg, R.G. Clark, and A.S. Dzurak, Nature **467**, 687 (2010).
[406] J.T. Muhonen, J.P. Dehollain, A. Laucht, F.E. Hudson, R. Kalra, T. Sekiguchi, K.M. Itoh, D.N. Jamieson, J.C. McCallum, A.S. Dzurak, and A. Morello, Nature Nanotechnology **9**, 986 (2014).
[407] K. Saeedi, S. Simmons, J.Z. Salvail, P. Dluhy, H. Riemann, N.V. Abrosimov, P. Becker, H.-J. Pohl, J.J.L. Morton, and M.L.W. Thewalt, Science **342**, 830 (2013).
[408] S. Simmons, R.M. Brown, H. Riemann, N.V. Abrosimov, P. Becker, H.-J. Pohl, M.L.W. Thewalt, K.M. Itoh, and J.J.L. Morton, Nature **470**, 69 (2011).
[409] M. Fuechsle, J.A. Miwa, S. Mahapatra, H. Ryu, S. Lee, O. Warschkow, L.C.L. Hollenberg, G. Klimeck, and M.Y. Simmons, Nature Nanotechnology **7**, 242 (2012).
[410] J.J. Pla, K.Y. Tan, J.P. Dehollain, W.H. Lim, J.J.L. Morton, D.N. Jamieson, A.S. Dzurak, and A. Morello, Nature **489**, 541 (2012).
[411] Y. He, S.K. Gorman, D. Keith, L. Kranz, J.G. Keizer, and M.Y. Simmons, Nature **571**, 371 (2019).
[412] M. Veldhorst, C.H. Yang, J.C.C. Hwang, W. Huang, J.P. Dehollain, J.T. Muhonen, S. Simmons, A. Laucht, F.E. Hudson, K.M. Itoh, A. Morello, and A.S. Dzurak, Nature **526**, 410 (2015).
[413] T.F. Watson, S.G.J. Philips, E. Kawakami, D.R. Ward, P. Scarlino, M. Veldhorst, D.E. Savage, M.G. Lagally, M. Friesen, S.N. Coppersmith, M.A. Eriksson, and L.M.K. Vandersypen, Nature **555**, 633 (2018).
[414] D.M. Zajac, A.J. Sigillito, M. Russ, F. Borjans, J.M. Taylor, G. Burkard, and J.R. Petta, Science **359**, 439 (2018).
[415] W. Huang, C.H. Yang, K.W. Chan, T. Tanttu, B. Hensen, R.C.C. Leon, M.A. Fogarty, J.C.C. Hwang, F.E. Hudson, K.M. Itoh, A. Morello, A. Laucht, and A.S. Dzurak, Nature **569**, 532 (2019).
[416] R. Lo Nardo, G. Wolfowicz, S. Simmons, A.M. Tyryshkin, H. Riemann, N.V. Abrosimov, P. Becker, H.-J. Pohl, M. Steger, S.A. Lyon, M.L.W. Thewalt, and J.J.L. Morton, Phys. Rev. B **92**, 165201 (2015).
[417] T.J.Z. Stock, O. Warschkow, P.C. Constantinou, J. Li, S. Fearn, E. Crane, E.V.S. Hofmann, A. Kölker, D.R. McKenzie, S.R. Schofield, and N.J. Curson, ACS Nano **14**, 3316 (2020).
[418] D.P. Franke, F.M. Hrubesch, M. Künzl, H.-W. Becker, K.M. Itoh, M. Stutzmann, F. Hoehne, L. Dreher, and M.S. Brandt, Phys. Rev. Lett. **115**, 057601 (2015).
[419] G.W. Morley, M. Warner, A.M. Stoneham, P.T. Greenland, J. van Tol, C.W.M. Kay, and G. Aeppli, Nature Materials **9**, 725 (2010).
[420] T. Kobayashi, J. Salfi, C. Chua, J. van der Heijden, M.G. House, D. Culcer, W.D. Hutchison, B.C. Johnson, J.C. McCallum, H. Riemann, N.V. Abrosimov, P. Becker, H.-J. Pohl, M.Y. Simmons, and S. Rogge, Nat. Mater. (2020).
[421] Y.M. Niquet, D. Rideau, C. Tavernier, H. Jaouen, and X. Blase, Phys. Rev. B **79**, 245201 (2009).
[422] M. Usman, R. Rahman, J. Salfi, J. Bocquel, B. Voisin, S. Rogge, G. Klimeck, and L.L.C. Hollenberg, J. Phys.: Condens. Matter **27**, 154207 (2015).
[423] L. Dreher, T.A. Hilker, A. Brandlmaier, S.T.B. Goennenwein, H. Huebl, M. Stutzmann, and M.S. Brandt, Phys. Rev. Lett. **106**, 037601 (2011).
[424] B. Roche, E. Dupont-Ferrier, B. Voisin, M. Cobian, X. Jehl, R. Wacquez, M. Vinet, Y.-M. Niquet, and M. Sanquer, Phys. Rev. Lett. **108**, 206812 (2012).
[425] B. Yan, R. Rurali, and Á. Gali, Nano Lett. **12**, 3460 (2012).

[426] J. Mansir, P. Conti, Z. Zeng, J.J. Pla, P. Bertet, M.W. Swift, C.G. Van de Walle, M.L.W. Thewalt, B. Sklenard, Y.M. Niquet, and J.J.L. Morton, Phys. Rev. Lett. **120**, 167701 (2018).
[427] M. Steger, A. Yang, M.L.W. Thewalt, M. Cardona, H. Riemann, N.V. Abrosimov, M.F. Churbanov, A.V. Gusev, A.D. Bulanov, I.D. Kovalev, A.K. Kaliteevskii, O.N. Godisov, P. Becker, H.-J. Pohl, E.E. Haller, and J.W. Ager, Physical Review B **80**, (2009).
[428] K.J. Morse, R.J.S. Abraham, A. DeAbreu, C. Bowness, T.S. Richards, H. Riemann, N.V. Abrosimov, P. Becker, H.-J. Pohl, M.L.W. Thewalt, and S. Simmons, Science Advances **3**, e1700930 (2017).
[429] A. DeAbreu, C. Bowness, R.J.S. Abraham, A. Medvedova, K.J. Morse, H. Riemann, N.V. Abrosimov, P. Becker, H.-J. Pohl, M.L.W. Thewalt, and S. Simmons, Phys. Rev. Applied **11**, 044036 (2019).
[430] C. Yin, M. Rancic, G.G. de Boo, N. Stavrias, J.C. McCallum, M.J. Sellars, and S. Rogge, Nature **497**, 91 (2013).
[431] M. Hollenbach, Y. Berencén, U. Kentsch, M. Helm, and G.V. Astakhov, Opt. Express, OE **28**, 26111 (2020).
[432] C. Beaufils, W. Redjem, E. Rousseau, V. Jacques, A.Yu. Kuznetsov, C. Raynaud, C. Voisin, A. Benali, T. Herzig, S. Pezzagna, J. Meijer, M. Abbarchi, and G. Cassabois, Phys. Rev. B **97**, 035303 (2018).
[433] L.W. Song, X.D. Zhan, B.W. Benson, and G.D. Watkins, Phys. Rev. B **42**, 5765 (1990).
[434] A. Mattoni, F. Bernardini, and L. Colombo, Phys. Rev. B **66**, 195214 (2002).
[435] A. Docaj and S.K. Estreicher, Physica B: Condensed Matter **407**, 2981 (2012).
[436] H. Wang, A. Chroneos, C.A. Londos, E.N. Sgourou, and U. Schwingenschlögl, Journal of Applied Physics **115**, 183509 (2014).
[437] D. Timerkaeva, C. Attaccalite, G. Brenet, D. Caliste, and P. Pochet, Journal of Applied Physics **123**, 161421 (2018).
[438] C.E. Jones and W.D. Compton, Radiation Effects **9**, 83 (1971).
[439] A.M. Smith, M.C. Mancini, and S. Nie, Nature Nanotech **4**, 710 (2009).
[440] M. Atatüre, D. Englund, N. Vamivakas, S.-Y. Lee, and J. Wrachtrup, Nature Reviews Materials **3**, 38 (2018).
[441] J. Palmer, The Economist, https://www.economist.com/technology-quarterly/2017-03-09/quantum-devices (2017).
[442] M. Poo and L. Wang, Natl Sci Rev **5**, 598 (2018).
[443] See http://qurope.eu/manifesto for the full of Quantum Manifesto and the list of the endorses, (2016).
[444] See https://www.nsf.gov/pubs/2018/nsf18578/nsf18578.htm for more details of the program solicitation of the National Science Foundation 2018, (2018).
[445] D. Shin, H. Hübener, U. De Giovannini, H. Jin, A. Rubio, and N. Park, Nat Commun **9**, 638 (2018).
[446] R. Ruskov and C. Tahan, Phys. Rev. B **88**, 064308 (2013).
[447] C.M. Reinke and I. El-Kady, AIP Advances **6**, 122002 (2016).
[448] A. Macridin, P. Spentzouris, J. Amundson, and R. Harnik, Phys. Rev. Lett. **121**, 110504 (2018).
[449] R. Ramprasad, R. Batra, G. Pilania, A. Mannodi-Kanakkithodi, and C. Kim, Npj Computational Materials **3**, 1 (2017).
[450] K.T. Butler, D.W. Davies, H. Cartwright, O. Isayev, and A. Walsh, Nature **559**, 547 (2018).
[451] T. Wang, C. Zhang, H. Snoussi, and G. Zhang, Advanced Functional Materials **30**, 1906041 (2020).
[452] K. Tanaka, K. Hachiya, W. Zhang, K. Matsuda, and Y. Miyauchi, ACS Nano **13**, 12687 (2019).
[453] L. Xi, S. Pan, X. Li, Y. Xu, J. Ni, X. Sun, J. Yang, J. Luo, J. Xi, W. Zhu, X. Li, D. Jiang, R. Dronskowski, X. Shi, G.J. Snyder, and W. Zhang, J. Am. Chem. Soc. **140**, 10785 (2018).
[454] N. Majeed, M. Saladina, M. Krompiec, S. Greedy, C. Deibel, and R.C.I. MacKenzie, Advanced Functional Materials **30**, 1907259 (2020).
[455] S. Lu, Q. Zhou, Y. Ouyang, Y. Guo, Q. Li, and J. Wang, Nature Communications **9**, 1 (2018).
[456] T. Zhang, Y. Jiang, Z. Song, H. Huang, Y. He, Z. Fang, H. Weng, and C. Fang, Nature **566**, 475 (2019).
[457] A.M. Ferrenti, N.P. de Leon, J.D. Thompson, and R.J. Cava, Npj Computational Materials **6**, 1 (2020).
[458] J. Davidsson, V. Ivády, R. Armiento, N.T. Son, A. Gali, and I.A. Abrikosov, New J. Phys. **20**, 023035 (2018).
[459] H. Ma, M. Govoni, and G. Galli, Npj Comput Mater **6**, 85 (2020).
[460] V. Ivády, Phys. Rev. B **101**, 155203 (2020).
[461] V. Ivády, H. Zheng, A. Wickenbrock, L. Bougas, G. Chatzidrosos, K. Nakamura, H. Sumiya, T. Ohshima, J. Isoya, D. Budker, I.A. Abrikosov, and A. Gali, ArXiv:2006.05085 [Cond-Mat] (2020).